\documentclass[11pt]{article}    
\usepackage{setspace}
\usepackage{hyperref}
\usepackage{graphics}
\usepackage{epsfig}
\usepackage{url}
\usepackage{amsmath}
\usepackage{mathrsfs}
\usepackage[utf8]{inputenc}
\usepackage{graphicx}
\usepackage{fancyhdr}
\usepackage[flushleft]{threeparttable}
\usepackage[nottoc]{tocbibind}
\usepackage[graphicx]{realboxes}
\usepackage{pdflscape}
\usepackage{hypcap}
\usepackage{amssymb}
\usepackage{setspace}
\usepackage{hyperref}
\usepackage{graphics}
\usepackage{epsfig}
\usepackage{url}
\usepackage{amsmath}
\usepackage{mathrsfs}
\usepackage{amssymb}
\usepackage{ rotating, graphicx}
\usepackage{float}
\usepackage[caption = false]{subfig}

\usepackage{amsmath}
\usepackage{bm}
\usepackage{amsfonts}
\usepackage{amsmath}
\usepackage{mathrsfs}
\usepackage{multirow}
\numberwithin{equation}{section}
\usepackage{tablefootnote}
\usepackage{booktabs}
\usepackage{multirow}
\usepackage{siunitx}
\usepackage[flushleft]{threeparttable}
\usepackage{color}
\usepackage{hyperref}
\usepackage{graphicx}
\usepackage{graphics}
\usepackage{adjustbox}
\usepackage{titlesec}
\usepackage{geometry}
\usepackage{lipsum}     
\usepackage{amsmath}
\usepackage{amssymb}

\newcommand{\bfg}[1]{\mbox{\boldmath $#1$\unboldmath}}
\newcommand{\fraca}[2]{\displaystyle\frac{#1}{#2}}
\def \R {{\rm I\kern -2.2pt R\hskip 1pt}}

\newtheorem{definition}{Definition}

\newtheorem{remark}{Remark}
 \newtheorem{example}{Example}

\usepackage{color}
\begin{document}
	
	\begin{center} { \large \sc  Goodness of fit tests for the pseudo-Poisson distribution}\\
	
		\vskip 0.1in {\bf {\bf Banoth Veeranna}\footnote{veerusukya40@gmail.com},  B.G. Manjunath }\footnote{bgmanjunath@gmail.com} and {\bf B. Shobha }\footnote{bsstat@uohyd.ac.in} 
	\\
	\vskip 0.1in 
	School of Mathematics and Statistics, University of Hyderabad, Hyderabad, India. \\
	
	\bigskip

	\end{center}
	
	\begin{abstract}
	 Bivariate count models having one marginal and the other conditionals being of the Poissons form are called pseudo-Poisson distributions.  Such models have simple  flexible dependence structures,  possess fast computation algorithms and generate a sufficiently large  number of parametric families. It has been strongly  argued  that the pseudo-Poisson model will be the first choice to consider in modelling bivariate over-dispersed data with positive correlation and having one of the marginal equi-dispersed.  Yet, before we start fitting,  it is necessary to test  whether the given data is compatible with the assumed pseudo-Poisson model. Hence, in the present note we derive and propose a few goodness-of-fit tests  for the bivariate pseudo-Poisson distribution. Also we emphasize two tests, a lesser known test based on the supremes of the absolute difference between the estimated probability generating function and its empirical counterpart. A new test has been proposed based on the difference between the estimated bivariate Fisher dispersion index and its empirical indices. However, we also consider the potential of applying the bivariate tests that depend on the generating function (like the Kocherlakota and Kocherlakota and Mu\~noz and Gamero tests) and the univariate goodness-of-fit tests (like the Chi-square test) to the pseudo-Poisson data. However, for each of the tests considered we analyse finite, large and asymptotic properties. Nevertheless, we compare the power (bivariate classical Poisson and Com-Max bivariate Poisson as alternatives) of each of the tests suggested and also include examples of application to real-life data. In a nutshell we are developing an R package which includes a test for compatibility of the data with the bivariate pseudo-Poisson model.

	\end{abstract}
	
	\bigskip

	\section{Introduction}
	Indeed, goodness-of-fit (GoF) is a statistical procedure to test whether the given data is compatible with the assumed distribution. Any GoF test required the following three steps: $(1)$ identifying the unique characteristic of the assumed model (examples: distribution function, generating function or density function); $(2)$ compute the empirical version of the assumed characteristic; $(3)$ with the pre-assumed  measure(examples: L$_1$ - or L$_2$-space), measure the distance between assumed item in Step $(1)$ and its empirical one, in  Step $(2)$. A rejection region can be computed with a given level and the cut-off value for the distance measure determined. However, if the rejection region can not be derived explicitly then one can use Bootstrapping technique to generate a critical region.  
	The general steps requried to simulate a rejection region using Bootstrapping are  discussed more in Section $4$. We refer to Meintanis \cite{m16} and Nikitin \cite{n19} for detailed discussion of the GoF tests which involve the aforementioned steps.  Besides, there do exist or can be constructed tests which are not  based on a unique characteristic of the assumed distribution.  For example,  consider the univariate Poisson distribution, there exists a GoF test which depends  on the Fisher index of dispression.   We also know that the Poisson distribution belongs to the class of  equi-dispersed models but this property does not characterize the Poisson distribution.  Hence,  such tests, which are not based on a 
	unique characteristic of the assumed distributions  are not consistent tests.

	The literature on GoF tests for bivariate count data is sparse. For the classical bivariate and multivariate Poisson distributions a GoF test using the probability generating function is discussed by Mu\~noz and Gamero \cite{nj14} and Mu\~noz and Gamero \cite{nj16}. For a recent review on the available bivariate GoF tests and also a new test using  the differentiation of the probability generating function, see Mu\~noz \cite{n21}.

	In the following sections we are starting with a test defined in Kocherlakota and Kocherlakota \cite{kk92} and a few
	 bivariate GoF tests reviewed in Mu\~noz \cite{n21}.  In addition to  the classical GoF tests using probability generating function (p.g.f.), we considered a less known test which will be supremum of the absolute difference between estimated p.g.f. and empirical ones.  In addition, we are introducing a non-consistent tests which are based on the moments, in particular, defining test taking difference of estimated bivariate Fisher index and its empirical counterpart. We examine each test's finite, large, and asymptotic properties and recommending a few tests based on their power and robustness analysis.

	Before we start discussion on GoF tests we would like to make a few remarks on the bivariate pseudo-Poisson model and its relevance in the literature.  Finally, we refer to Arnold and Manjunath \cite{am21} and Arnold \cite{amsv22} et. al. for classical inferential aspects, characterization, Bayesian analysis and also an example of applications of the bivariate pseudo-Poisson model.
	
	\section{Bivariate pseudo-Poisson models}  
	In the following we will be discussing the bivariate pseudo-Poisson model, see Arnold and Manjunath \cite{am21} page 2307.

	\begin{definition}
		A $2$-dimensional random variable $(X,Y)$ is said to have a bivariate Pseudo-Poisson distribution if there exists a positive constant $\lambda_1$ such that
		$$X \sim \mathscr{P}(\lambda_1)$$
		and a function $ \lambda_2: \{0,1,2,...\} \rightarrow (0, \infty)$ such that, for every non-negative integer $x$,
		$$Y |X = x \sim \mathscr{P}(\lambda_2(x)).$$
	\end{definition}
	
	Here we restrict the form of the function $\lambda_2(x)$ to be a polynomial with unknown coefficients. In particularly the simple form we assume is that $\lambda_2(x) = \lambda_2 + \lambda_3 x$, then the above bivariate distribution will be of the form

	\begin{equation} \label{pseudo-P-lin-regress}
		X \sim \mathscr{P}(\lambda_1) 
	\end{equation}
	and
	\begin{equation}
		Y|X=x \sim \mathscr{P}(\lambda_2+\lambda_3 x).
	\end{equation}
	
	The parameter space for this model is $\{(\lambda_1,\lambda_2,\lambda_3):\lambda_1>0,\lambda_2 > 0,\lambda_3\geq 0\}$. The case in which the variables are independent corresponds to  the choice $\lambda_3=0$.  The probabililty generating function (p.g.f) for this bivariate Pseudo-Poisson distribution is given by

	\bigskip
	
	\begin{eqnarray}
		G(t_1,t_2) =  e^{\lambda_2 (t_2-1)} e^{ \lambda_1[t_1  e^{\lambda_3(t_2-1)}-1]};  \mbox{      }  t_1,t_2 \in \mathbb{R}. 
	\end{eqnarray}
	
	\begin{remark}
		As noted in Arnold and Manjunath \cite{am21}, for the case $\lambda_2 =0$, the bivariate pseudo-Poisson distribution reduces to the bivariate Poisson-Poisson distribution.  The corresponding Poisson-Poisson distribution was originally introduced by Leiter and Hamdani \cite{lh73} in modelling traffic accidents and fatalities count data. 	We make a remark that the bivariate pseudo-Poisson  model is a generalization of the Poisson-Poisson distribution. 
		
		The joint p.g.f in equation (2.3) deduces to 
		\begin{eqnarray}
			G_{II}(t_1,t_2) =   e^{ \lambda_1[t_1  e^{\lambda_3(t_2-1)}-1]};  \mbox{      }  t_1,t_2 \in \mathbb{R}.
		\end{eqnarray}
		
	\end{remark}
	
	Now, the marginal p.g.f of $Y$ is

	\begin{eqnarray}
		G(1,t_2) = G_{Y}(t_2) 
		= e^{\lambda_2 (t_2-1)} e^{\lambda_1 [ e^{\lambda_3(t_2-1)}-1 ]}; \mbox{     } t_2 \in \mathbb{R} .
	\end{eqnarray}
	Note that in general the p.g.f in equation (2.4) can not be simplified to compute all marginal probabilities. Yet, we can use equation (2.4) to derive a few marginal probabilities of $Y$. The derivation of marginal probability of $Y$ is demonstrated for   $Y=0,1,2,3$ in Appendix A.1   and one can still extend the mentioned  procedure to get albeit complicated values  for the probability that $Y$ assumes any positive value. Besides, the derivation of the other conditional distribution of the bivariate pseudo-Poisson, i.e., $f(x|y)$, has been included in Appendices A.2.

	In the following sections we discuss a few one dimensional  distributions which are 
	derived from the bivariate pseudo-Poisson for the case $\lambda_2=0$. Moreover, the derived univariate distributions has classical relevance to the two parameter Neyman Type A and Thomas distribution.

	\subsection{Neyman Type A distribution}
	As noted in Arnold and Manjunath \cite{am21}, in the case in which $\lambda_2=0$ the marginal distribution is a  Neyman Type A distribution with $\lambda_3$ being the index of  clumping (see page 403 of Johnson, Kemp and Kotz \cite{jkk05}). It can also be recognized as a Poisson mixture of Poisson distributions.  Now, the marginal mass function of $Y$ is given by
	\begin{eqnarray}
		P(Y=y) = \frac{e^{-\lambda_1}\lambda_3^{y}}{y!} \sum_{j=0}^{\infty} \frac{(\lambda_1 e^{-\lambda_3})^j j^{y}}{j!};  \mbox{              } y=0,1,2,... .
	\end{eqnarray}
	i.e. $Y$ has a Poisson distribution with the parameter $\lambda_1$ while $\lambda_1$ is also a  Poisson distribution with the parameter $\lambda_3$.  
	We refer to Glesson and Douglas \cite{gd75} and Johnson, Kemp and Kotz \cite{jkk05} Section 9.6 for applications and inferential aspects of the Neyman Type A distribution.

	\subsection{Thomas distribution}
	Consider the joint probability generating function defined in equation (2.4), i.e.,

	\begin{eqnarray}
		G_{II}(t_1,t_2) =   e^{ \lambda_1[t_1  e^{\lambda_3(t_2-1)}-1]};  \mbox{      }  t_1,t_2 \in \mathbb{R}.
	\end{eqnarray}

	Take $t_1=t_2: =t$ and the above p.g.f. deduces to 
	
	\begin{eqnarray}
		G^*(t)=G(t,t) =   e^{ \lambda_1[t  e^{\lambda_3(t-1)}-1]};  \mbox{      }  t \in \mathbb{R}.
	\end{eqnarray}
	
	Note that the above univariate p.g.f. is the p.g.f. of the Thomas distribution with parameter $\lambda_1$ and $\lambda_3$.
	The probability mass function of the Thomas distribution is given as 
	
	\begin{eqnarray}
		P(Z=z) = \frac{e^{-\lambda_1}}{z!} \sum_{j=1}^{z} {z \choose j} (\lambda_1 e^{-\lambda_3})^j (j\lambda_3)^{z-j}, \mbox{     }  z= 0,1,2,... .
	\end{eqnarray}
	
	For further, applications and inferential aspects of the Thomas distribution we refer to Glesson and Douglas \cite{gd75} and Johnson, Kemp and Kotz \cite{jkk05}) Section 9.10.
	
	\begin{remark}
		The Neyman Type A and the Thomas distribution has a historical relevance in modeling plant and animal populations. For example: suppose that the number of clusters of eggs an insect lays and  number of eggs per each cluster have specified probability distributions. Then for the Neyman Type A distribution and Thomas distributions the number of clusters  of eggs laid by the insect is follow a Poisson distribution with parameter $\lambda_1$: for the Neyman Type A the number of eggs per cluster is also a Poisson distribution with parameter $\lambda_3$.  But for the Thomas distribution  the parent of the cluster is always to be present with the number of eggs(offspring) and  which has a shifted Poisson distribution with support $\{1,2,3,...\}$ and the parameter $\lambda_3$.  
		Note that Neyman Type A and Thomas distributions can be generated by a  mixture of distributions and also a  random sum of random variables .

		Consider that the  mixing distribution is a Poisson with parameter $\lambda_1$  with mixture has a Poisson with parameter $\lambda_3$ then the resultant random variable has a Neyman Type A distribution. In sequel, if the mixing distribution is a Poisson with parameter $\lambda_1$ and the $j$th distribution in the  mixture has a distribution of the form $j+Y(j)$, where $Y(j)$ has a Poisson with parameter $j\lambda_3$ then the resultant random variable has a Thomas distribution.

		However, for a random sum of random variables  (also known as Stopped-Sum distributions): let us consider that the size $N$ of the initial generation is a random variable and that each individual $i$ of this generation independetly gives a random variable $Y_i$, where $Y_1,Y_2,...$ has a common distribution.
		Then the total number of individuals is $S_N = Y_1 + ...+Y_N$.
		For the case that  $N$ is a Poisson random variable with parameter $\lambda_1$ and  $Y_i$ is a Poisson random variable with parameter $\lambda_3$ then the random sum $S_N$ has a Neyman Type A distribution.  However, if $Y_i$ is a shifted Poisson with parameter $\lambda_3$ and support $\{1,2,3,...\}$ then the random sum $S_N$ has a Thomas distribution. 
	\end{remark}

\begin{remark}
	The other conditional mass function, i.e., conditional mass function of $X$ given $Y=y$ is recognized in Section 5 of Leiter and Hamdan \cite{lh73} and in Appendix 3 of Arnold and Manjunath \cite{am21}. However,  in Appendix A.2 in the current note we have derived the conditional mass function  and  identified that the expression as a Stirling number of the second kind.

\end{remark}	
	\section{GoF tests}
	In the following section we discuss GoF tests which are based  on the moments (non-consistent tests), on unique characteristics (consistent tests) and  a simple classical  $\chi^2$ goodness of fit test.
	
		\subsection{New test based on moments}
	In the following we will be extending an univariate GoF test based on the Fisher index to the bivariate case.
	We know that for a multivariate distribution the Fisher index of dispersion is not uniquely defined.  However, in the following we use the definition of the multivariate Fisher dispersion  given by  Kokonendji and Puig \cite{kp18} in  Section 3   as;  for any $d$-dimensional discrete random variable $\bf{Z}$ with mean vector $E(\bf{Z})$ and covariance matrix $Cov(\bf{Z})$ the generalized dispersion index  is 
	
	\begin{eqnarray}
		GDI(\bf{Z}) = \frac{ \sqrt{E (\bf{Z})}^T Cov(\bf{Z})\sqrt{E (\bf{Z}}) }{E (\bf{Z})^T E( \bf{Z})}.
	\end{eqnarray}

	For the bivariate pseudo-Poisson model, definte the random vector $\bfg{Z}= (X,Y)^T$ for  and the moments are (c.f. Arnold and Manjunath \cite{am21} page 2309--2310)
	\begin{eqnarray}
		E(\textbf{Z}) = (\lambda_1, \lambda_2 + \lambda_3 \lambda_1)^T
	\end{eqnarray}
	\[
	cov(\textbf{Z})=
	\begin{bmatrix}
		\lambda_1 &  \lambda_1 \lambda_3 \\
		\lambda_1 \lambda_3 & \lambda_2 + \lambda_3 \lambda_1 + \lambda^2_3 \lambda_1
	\end{bmatrix}.
	\]

	Now, using the  definition given in Kokonendji and Puig \cite{kp18} page 183, dispression index for the bivariate psuedo-Poisson is 
	
	\begin{eqnarray}
		GDI(\textbf{Z}) &=& \frac{\lambda_1^2 + 2 \lambda_1^{\frac{3}{2}} \lambda_3 \sqrt{\lambda_2 + \lambda_3 \lambda_1} + (\lambda_2 +  \lambda_3\lambda_1)(\lambda_2 + \lambda_3 \lambda_1 + \lambda_3^2 \lambda_1)}{\lambda_1^2 + (\lambda_2 + \lambda_3 \lambda_1)^2} \nonumber \\
		&=& 1 + \frac{ 2 \lambda_1^{\frac{3}{2}} \lambda_3 \sqrt{\lambda_2 + \lambda_3 \lambda_1} + (\lambda_2 +  \lambda_3\lambda_1)  \lambda_3^2 \lambda_1}{\lambda_1^2 + (\lambda_2 + \lambda_3 \lambda_1)^2}>1,
	\end{eqnarray}
	which indicates over-dispersion.

	For the corresponding sample version, consider the $n$ sample observations  \\ $\textbf{Z}_1 = (X_1,Y_1)^T$,..., $\textbf{Z}_n = (X_n,Y_n)^T$  from the bivariate pseudo-Poisson distribution. Now, denote $\overline{\textbf{Z}}_n= \frac{1}{n} \sum_{i=1}^{n} \textbf{Z}_i = (\overline{X}, \overline{Y})^T $ and $\widehat{cov(\textbf{Z})} =\frac{1}{n-1} \sum_{i=1}^{n} \textbf{Z}_i \textbf{Z}_i^T -  \overline{\textbf{Z}}_n \overline{\textbf{Z}}_n^T$ are sample mean vector and sample covariance matrix, respectively.  Then the empirical bivariate dispersion index is 
	
	\begin{eqnarray}
		\widehat{GDI(\textbf{Z})}_n = \frac{\sqrt{\overline{\textbf{Z}}^T_n } \widehat{cov(\textbf{Z})}\sqrt{\overline{\textbf{Z}}_n } }{\overline{\textbf{Z}}^T_n \overline{\textbf{Z}}_n}.
	\end{eqnarray}
	
	According to Theorem 1  in   Kokonendji and Puig \cite{kp18} page 184, as $n \rightarrow \infty$, \\
	$ \sqrt{n } \{\widehat{GDI(\textbf{Z})}_n -  GDI(\textbf{Z})\} \sim N(0, \sigma^2_{g})$,
	where $ \sigma^2_{g}= \Delta^T \Gamma \Delta$; 
	
	\begin{eqnarray*} 
		\Gamma = \begin{bmatrix}
			\Sigma & \textbf{0}\\
			\textbf{0} &  \textbf{0}
		\end{bmatrix} 
	\end{eqnarray*}
	and 
	\begin{eqnarray*} 
		\Sigma = \begin{bmatrix}
			var(X) & cov(X,Y)\\
			cov(X,Y) & var(Y)
		\end{bmatrix} .
	\end{eqnarray*}

 A new bivariate GoF test for the count data based on the Fisher dispersion index is 
	
	\begin{eqnarray}
		FI^{(.)}_{n} = \sqrt{n } \{\widehat{GDI(\textbf{Z})}_n -  GDI(\textbf{Z})\}
	\end{eqnarray}
	and the null hypothesis is rejected for large absolute values of $F^{(.)}_{n}$.
	The asymptotic distribution of the test statistic is
	
	\begin{eqnarray}
		\frac{ \widehat{GDI(\textbf{Z})}_n -  GDI(\textbf{Z})}{ \frac{\sigma_{g}}{\sqrt{n }}} \sim^{asy.} N(0,1), \mbox{   as   } n \rightarrow \infty.
	\end{eqnarray}
	For the detailed proof, c.f.  Theorem 1 in  Kokonendji and Puig \cite{kp18} page 184. 
However, for the two submodels of the bivariate pseudo-Poisson model, i.e., when $\lambda_2 = \lambda_3$ is Sub-Model I and when $\lambda_2 =0$  is Sub-Model II the new test statistics are 
	
		\begin{eqnarray}
		FI^{(SI)}_{n} = \sqrt{n } \{\widehat{GDI(\textbf{Z})}_n -  GDI^{(SI)}(\textbf{Z})\}
	\end{eqnarray}
	and
		\begin{eqnarray}
			FI^{(SI)}_{n} = \sqrt{n } \{\widehat{GDI(\textbf{Z})}_n -  GDI^{(SII)}(\textbf{Z})\}
		\end{eqnarray}
	  where 
		\begin{eqnarray}
			GDI^{(SI)}(\textbf{Z}) = 1 + \frac{ 2 \lambda_1^{\frac{3}{2}} \lambda_3^{\frac{3}{2}}   \sqrt{1 +  \lambda_1} + (1 + \lambda_1)  \lambda_3^3 \lambda_1}{\lambda_1^2 + \lambda_3^2 (1 +  \lambda_1)^2},
		\end{eqnarray}

 where 
\begin{eqnarray}
	GDI^{(SI)}(\textbf{Z}) = 1 + \frac{ 2 \lambda_1^{\frac{3}{2}} \lambda_3^{\frac{3}{2}}   \sqrt{\lambda_1} + \lambda_3^3 \lambda^2_1}{\lambda_1^2 + \lambda_3^2  \lambda_1^2}.
\end{eqnarray}
	one can derive test statistic 	
	$FI^{(SI)}_{n}$ and $FI^{(SII)}_{n}$.  The estimated dispersion index can be obtained by plugging in the m.l.e estimates of $\lambda_i$,$i=1,2,3$. Also, due to the invariance  and asymptotic properties of the m.l.e estimates the proposed test statistics are normal distributioned (with appropriate scaling).  For large sample sizes the null hypothesis is rejected whenever the test statistic absolute value is greater than the appropriate standard normal quantile value. In Section 4 we analyse finite, large and asymptotic behavior of the proposed test statistic.
	
	In addition, using bootstrapping techniques one can simulate the distribution of the above test and then testing for normality will also produce a robust GoF fit test.

	\subsection{Test based on the unique characteristic} 
	In the following we consider a few test statistics for the full, Sub-Model I and Sub-Model II. 
	
	\subsubsection{Mu\~noz and Gamero (M\&G) method}
	
	The GoF tests for a bivariate random variable based on the finite sample size is limited.  This is due to difficulty in deriving closed form expression for the critical region under finite sample size.  Yet,  in the following we use the finite sample size  test suggested  in  Mu\~noz and Gamero \cite{nj14}  for the classical bivariate Poisson distribution is  used to construct the GoF test for the  bivariate pseudo-Poisson distribution.  For a finite sample test based on the p.g.f to test GoF for the univariate Poisson, we refer to Rueda et al. \cite{uc91}.
	Furthermore,  using boostrapping  technique the critical region for the test is simulated and illustrated with an example in Section 4.
	
	Let  $(X,Y)$ be a bivariate random variable with p.g.f $G(t_1,t_2; \lambda_1,\lambda_2, \lambda_3)$, $(t_1,t_2)^T \in [0,1]^2$. For the given data set $(X_i,Y_i)$, $i=1,...,n$, we denote by $G_n(t_1,t_2) = \frac{1}{n}\sum_{i=1}^{n} t^{X_i}_1 t^{Y_i}_2 $ an  empirical counterpart of the bivariate p.g.f.   According to Mu\~noz and Gamero \cite{nj14}  a reasonable test for testing the compatiblity of the assumed density  should reject the null hypothesis for large values of  given statistic
	
	\begin{eqnarray}
		T^{(.)}_{P,n,w} (\hat{\lambda}_1,\hat{\lambda}_2, \hat{\lambda}_3) = \int_{0}^{1} \int_{0}^{1} g_n^2(t_1,t_2; \hat{\lambda}_1,\hat{\lambda}_2 \hat{\lambda}_3 ) w(t_1,t_2) dt_1 dt_2
	\end{eqnarray}
	where $\hat{\lambda}_1,\hat{\lambda}_2, \hat{\lambda}_3$ are consistent estimators of $\lambda$'s and  \\ $g_n(t_1,t_2; \hat{\lambda}_1,\hat{\lambda}_2 ,\hat{\lambda}_3 )$  $= \sqrt{n} \{  G_n(t_1,t_2) -G(t_1,t_2; \hat{\lambda}_1,\hat{\lambda}_2, \hat{ \lambda}_3) \}$ and also   $w(t_1,t_2) \geq 0$  is a measurable function satifying 
	
	\begin{eqnarray}
		\int_{0}^{1} \int_{0}^{1} w(t_1,t_2) dt_1 dt_2 < \infty.
	\end{eqnarray}
	
	The above condition implies that the test statistic  $T^{(.)}_{n,w}(\hat{\lambda}_1,\hat{\lambda}_2, \hat{\lambda}_3) $ is finite for the fixed sample size $n$.  Similarly for the Sub-model I \& II with approciate p.g.f one can derive test statistic 	
	$T^{(SI)}_{P,n,w}$ and $T^{(SII)}_{P,n,w}$.
	
	Due to the difficulty in obtaining explicit expression for the critical region, it has been argued in Mu\~noz and Gamero \cite{nj14} and in Mu\~noz \cite{n21}  the rejection regions can be simulated using bootstrapping methods. The general procedure to identify an appropriate weight function is difficult to argue. One can consider the weight functions which include a bigger family of functions. A few weights functions are considered in Appendix A.3 and also derived its test statistic. In Section 4 we analyzed the effect of weight functions and its on feasible parameter valueson on the critical region.

		\subsubsection{Kocherlakota and Kocherlakota(K\&K) method}
	Let $\textbf{Z}_1,...,\textbf{Z}_n$ be a random sample from the bivariate distribution $F(\textbf{z};\underline{\theta} )$, where $\underline{\theta}=(\theta_1,...,\theta_d)^T$ is the $d$-dimensional parameter vector. Let $G(t_1,t_2; \underline{\theta})$ be the p.g.f. of $\textbf{Z}=(X,Y)^T$, $t_1, t_2 \in \mathbb{R}^2$ and parameter vector $\underline{\theta}$ is  estimated by the maximum likelihood estimation (m.l.e) method and the estimator we denote by  $\widehat{\underline{\theta}}$. Let $G_n(t_1,t_2)= \frac{1}{n} \sum_{i=1}^{n} t_1^{X_i}t^{Y_i}_2$, $t \in \mathbb{R}$ be the empirical p.g.f. (e.p.g.f) then the test statistic

	\begin{eqnarray}
		T_{N} (t_1,t_2)  = \frac{G_n(t_1,t_2)- G(t_1,t_2; \widehat{\underline{\theta}})}{\sigma},   \mbox{      } |t_1|<1; |t_2|<1
	\end{eqnarray}
	
	asymptotically follows the standard normal distribution, where \\ $\sigma^2 = \frac{1}{n} [G(t_1^2,t_2^2; \underline{\theta}) -G^2(t_1,t_2; \underline{\theta})) ]- \sum_{i=1}^{d} \sum_{j=1}^{d} \sigma_{i,j} \frac{\partial G(t_1,t_2; \underline{\theta})}{\partial \theta_i} \frac{\partial G(t_1,t_2; \underline{\theta})}{\partial \theta_j}$, $((\sigma_{i,j}))$ is the inverse of the Fisher information matrix and $\sigma$ can be estimated by plugging in the m.l.e of $\underline{\theta}$.  We refer to Kocherlakota and Kocherlakota(K\&K) \cite{kk92}  for the asymptotic distribution of the test statistic.  Note that the computation of the Fisher information matrix for the full model is theoretically cumbersome, yet one can use numerical methods to evaluate the matrix.   However, we are considering the two submodels of the bivariate pseudo-Poisson and deriving their test statistics.

   Now, for the Sub-Model I the Fisher information matrix is 
	\[	I^{(SI)}(\lambda_1,\lambda_3) = n
	\begin{bmatrix} 
		E\Big( \frac{X}{\lambda^2_1} \Big)  & 0  \\
		0 & E\Big(  \frac{Y}{\lambda_3^2}\Big)  \\
	\end{bmatrix}  = \begin{bmatrix} 
		\frac{n}{ \lambda_1}   & 0  \\
		0 &  \frac{n(1+ \lambda_1)}{\lambda_3}  \\
	\end{bmatrix}
	.\] 
	
	Similarly,	for the  Sub-Model II, the Fisher information matrix is 
	\[	I^{(SII)}(\lambda_1,\lambda_3) = n
	\begin{bmatrix} 
		E\Big( \frac{X}{\lambda^2_1} \Big)  & 0  \\
		0 & E\Big(  \frac{Y}{\lambda_3^2}\Big)  \\
	\end{bmatrix} = \begin{bmatrix} 
		\frac{n}{\lambda_1}   & 0  \\
		0 &  \frac{n \lambda_1}{\lambda_3}  \\
	\end{bmatrix}
	.\] 
	
	The GoF test statistic is 
		\begin{eqnarray}
		T^{(SI)}_{PN} (t_1,t_2)  = \frac{G_n(t_1,t_2)- G_{I}(t_1,t_2; \hat{\lambda_1},\hat{\lambda_3} )}{\sigma^{(SI)} },   \mbox{      } |t_1|<1, |t_2|<1
	\end{eqnarray}
	where $G_n(.)$ is empirical p.g.f and $G_{I}(t_1,t_2; \hat{\lambda_1},\hat{\lambda_3} )$ is estimated p.g.f. of the Sub-Model I and  
	
	\begin{eqnarray}
		\sigma^{2(SI)} &=& \frac{1}{n} [G_{I}(t_1^2, t_2^2; \lambda_1, \lambda_3) -G_{I}^2(t_1, t_2; \lambda_1, \lambda_3)] -  \frac{\lambda_1}{n} \frac{\partial^2 G_{I}(t_1, t_2; \lambda_1, \lambda_3)}{\partial \lambda_1^2} \nonumber \\ && - \frac{ \lambda_3}{n(\lambda_1+1)} \frac{\partial^2 G_{I}(t_1, t_2; \lambda_1, \lambda_3)}{\partial \lambda_3^2}.
	\end{eqnarray}

	Similarly for the Sub-Model II, the GoF test statistic will be
	
\begin{eqnarray}
	T^{(SII)}_{PN}  = \frac{G_n(t_1,t_2)- G_{II}(t_1,t_2;\hat{\lambda_1},\hat{\lambda_3} )}{\sigma^{(SII)} },  \mbox{      } |t_1|<1, |t_2|<1
\end{eqnarray}
where $G_n(.)$ is empirical p.g.f and $G_{I}(t_1,t_2;\hat{\lambda_1},\hat{\lambda_3})$ is estimated p.g.f. of the Sub-Model II and  
	
	\begin{eqnarray}
	\sigma^{2(SII)}(t_1,t_2) &=& \frac{1}{n} [G(t_1^2, t_2^2; \lambda_1, \lambda_3) -G_{(II)}^2(t_1, t_2; \lambda_1, \lambda_3)] -  \frac{\lambda_1}{n} \frac{\partial^2 G_{(II)}(t_1, t_2; \lambda_1, \lambda_3)}{\partial \lambda_1^2} \nonumber \\ && - \frac{ \lambda_3}{n \lambda_1} \frac{\partial^2 G_{II}(t_1, t_2; \lambda_1, \lambda_3)}{\partial \lambda_3^2}.
	\end{eqnarray}
The boostrapped finite sample and asymptotic distributions of the GoF test statistic  of $T^{(.)}_{PN}$ are studied in Section 4.

  In the following we propose a  test procedure which will be supremum on the absolute value of the K\&K test statistic with $(t_1,  t_2)$ over $(-1,1) \times (-1,1) $. The reason behind proposing  such a test is exemplified in Section 4.
 The mentioned  GoF testing procedure for the K\&K method is originally discussed in Feiyan Chen \cite{cf13}  for the univariate and bivariate geometric models. Besides, Feiyan Chen (2013) also discusses K\&K method for the multiple $t$-values for the GoF test for geometric models, c.f. page 12 of Chen\cite{cf13} . However, in the present note we are interested in proposing  tests which are free from the choices of $t$-values hence the advantages or disadvantages  of considering multiple $t$-values are not discussed or illustrated in this note. 

The  GoF test statistic is

\begin{eqnarray}
	T^{(.)}_{SPN}  = \sup_{(t_1,t_2) \in \{(-1,1)\times(-1,1)\} } \Biggl |\frac{G_n(t_1,t_2)- G_{.}(t_1,t_2; \hat{\lambda_1},\hat{\lambda_3} )}{\sigma^{(.)} } \Biggl |
\end{eqnarray}
where $G_n(.)$, $G_{.}$ and $\sigma^{(.)}$ are defined in equation $(3.2)$ to $(3.5)$. Also note that deriving the  asymptotic distribution of the statistic $T^{(.)}_{SPN}$  is theoretically ambiguous.  Hence, in Section 4 the finite sample distribution of the test statistic $T^{(.)}_{SPN}$ is analyzed. 

	\begin{remark}
	For the Mu\~noz and Gamero (M\&G) and Kocherlakota and Kocherlakota(K\&K) the estimated p.g.f's can be obtained by plugging in the m.l.e estimates of $\lambda_i$, $i=1,2,3$.
	\end{remark}
	
	\begin{remark}
		In Meintanis \cite{m07} Theorem 2.3 characterizes the general p.g.f class of distributions called the CP-class, where  Thomas distribution  is a member of the CP-class. Which suggests a GoF test construction for the bivariate count models through the one dimensional p.g.f of particular form belongs to the CP-class,  see  Meintanis \cite{m07} page $23-25$. However, we have identified a missing link in the theorem and it is not clearly justified that knowing the form of the one dimensional p.g.f  assures in identifying the bivariate p.g.f. If we assume the theorem, a family of tests can be generated for the bivariate Poisson-Poisson distribution by testing only the one-dimensional Thomas distribution.  Since, we do not completely agree with the theorem, we are not recommending any test based on the one dimensional result to  conclude on the higher dimensional tests.
	\end{remark}

	\subsection{GoF test free from alternative}
	In the class of distribution free tests,  $\chi^2$ test is the commonly used even when there is no specific alternative hypothesis.  However, this also raises a difficulties in assessing the power of the test.  
	\subsubsection{$\chi^2$ GoF}

	In the following we are using classical $\chi^2$ GoF test and cell probabilites are computed upto $k$. The cell probability matrix is  given by

	\begin{center}
		\huge
		\resizebox{\columnwidth}{!}{%
			\begin{tabular} { | c | c | c | c | c | c | c |c|}
				\hline
				$X$ | $Y$& 0 & 1 & 2 & 3 & ... & k+ \\
				\hline
				0	& $p_{00}$  & $p_{01}$ & $p_{02}$ &  $p_{03}$& ... & $P(X=0) - \sum_{j=0}^{k-1} p_{0j} $\\
				\hline
				1 &  $p_{10}$  & $p_{11}$ & $p_{12}$ &  $p_{13}$& ...  &$P(X=1) -\sum_{j=0}^{k-1} p_{1j}$ \\
				\hline
				2 & $p_{20}$  & $p_{21}$ & $p_{22}$ &  $p_{23}$& ... &$P(X=2) -\sum_{j=0}^{k-1} p_{2j}$ \\
				\hline
				3&  $p_{30}$  & $p_{31}$ & $p_{32}$ &  $p_{33}$& ... &$P(X=3) -\sum_{j=0}^{k-1} p_{3j}$\\
				\hline
				... &... & ... & ... & ... & ...& ...\\
				\hline
				k+	& $P(Y=0) - \sum_{i=0}^{k-1} p_{i0}$ & $P(Y=1) - \sum_{i=0}^{k-1} p_{i1}$  & $P(Y=2) - \sum_{i=0}^{k-1} p_{i2}$ & $P(Y=3) - \sum_{i=0}^{k-1} p_{i3}$ & ... & $  1- \sum_{i=k}^{\infty}\sum_{j=k}^{\infty}p_{ij}$ \\
				\hline
			\end{tabular}
		}
	\end{center}
	where $p_{ij}= P(X=i,Y=j)$.  The test statistic is 
	\begin{eqnarray}
		T_{\chi^2} = \sum_{i=0}^{k} \sum_{j=0}^{k} \frac{ \Big( O_{i,j} - E_{i,j}\Big)^2}{E_{i,j}} 
	\end{eqnarray}
	
	where $k$ is the trucation point, $O_{i,j}$ is frequency of $(i,j)$ observation in the data of size $n$ and $E_{i,j}= n P(X=i,Y=j)$.  Hence, with Pearson theorem  $T_{\chi^2}$ follows  a $\chi^2$ distribution with  $[(k+1)\times(k+1)-1-3]$ degrees of freedom.

	 Similarly above two tests for  the Sub-Models I \& II  can be derived with appropriate cell probabilities $p_{ij}= P(X=i,Y=j)$.	
	In Section 4 we analyse finite sample and large sample behavior of the above two test statistics. 
	
\section{Examples}
	
\subsection{Simulation}	
	
	In the following we give a general procedure to analyse the finite sample distribution of the GoF test statistics with bootstrapping technique.

	\begin{enumerate}
		\item[Step 1] Simulate $n$ observations from the bivariate pseudo-Poisson with fixed parameter values.
		Otherwise,  estimate parameters by moment or m.l.e. method, say $\hat{\lambda_i}$.  Then compute GoF test statistics, say $T_{obs}$.
		\item[Step 2] Fix the number of bootstrapping  samples, say $B$ (ideal size is $5000$,$10000$) and sample $m(<n)$ observation from the above sample. , repeat Step 1 and compute  $T^b_{m,obs}$ for $b \in \{ 1,2,...,B\}$.
		\item [Step 3] From the frequency distribution of $T^b_{m,obs}$ obtain the quantile values and the empirical $p$-value is  $\frac{1}{B} \{ \mbox{Total no. of $T^b_{m,obs}$ greater than $T_{obs}$} \} $.
	\end{enumerate}

	\subsubsection{Test based on moments}
\begin{figure}[H]
		\subfloat[Full Model]{\includegraphics[width=6cm,height=3cm]{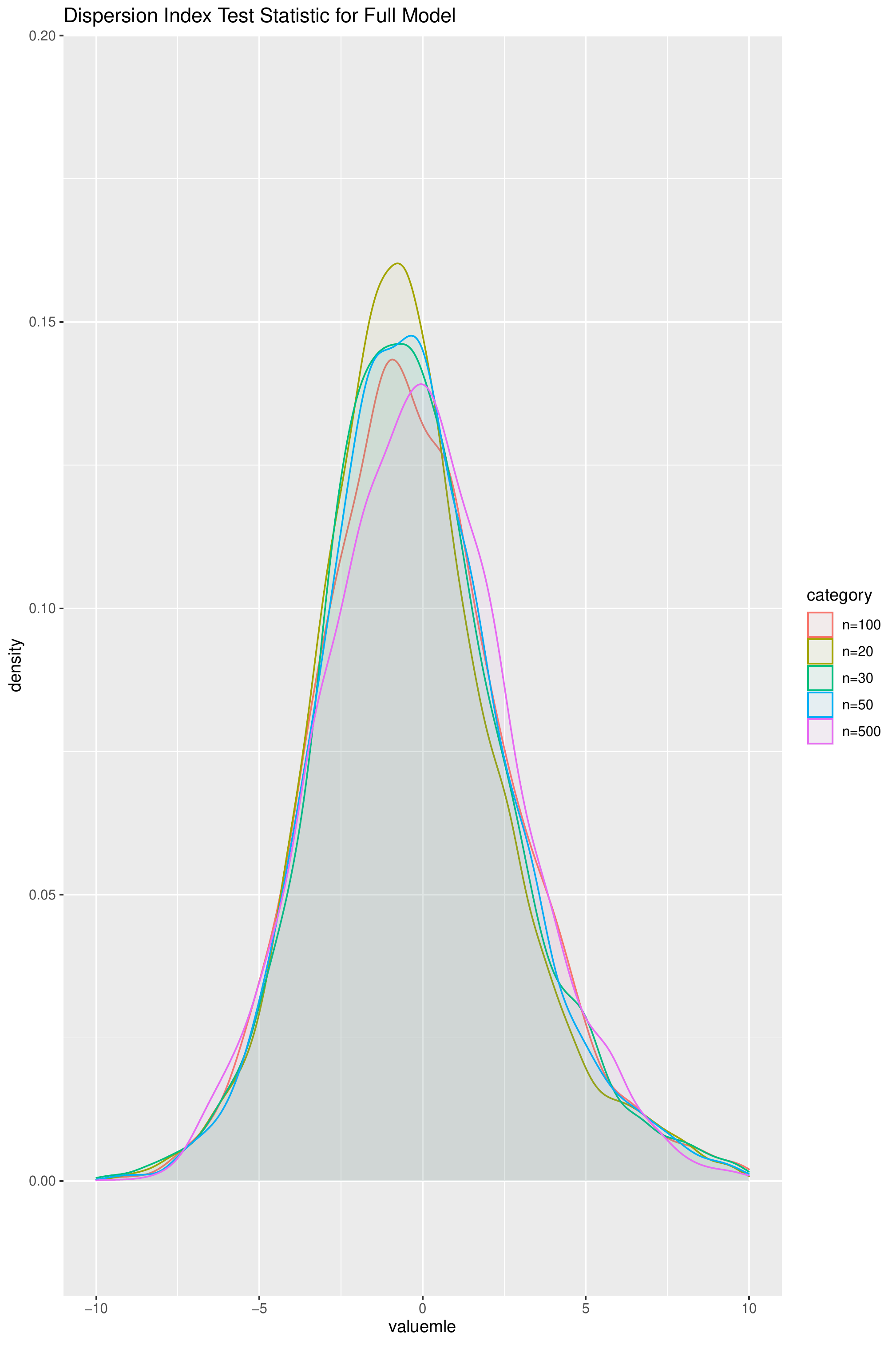}} 
		\subfloat[Sub-Model I]{\includegraphics[width=6cm,height=3cm]{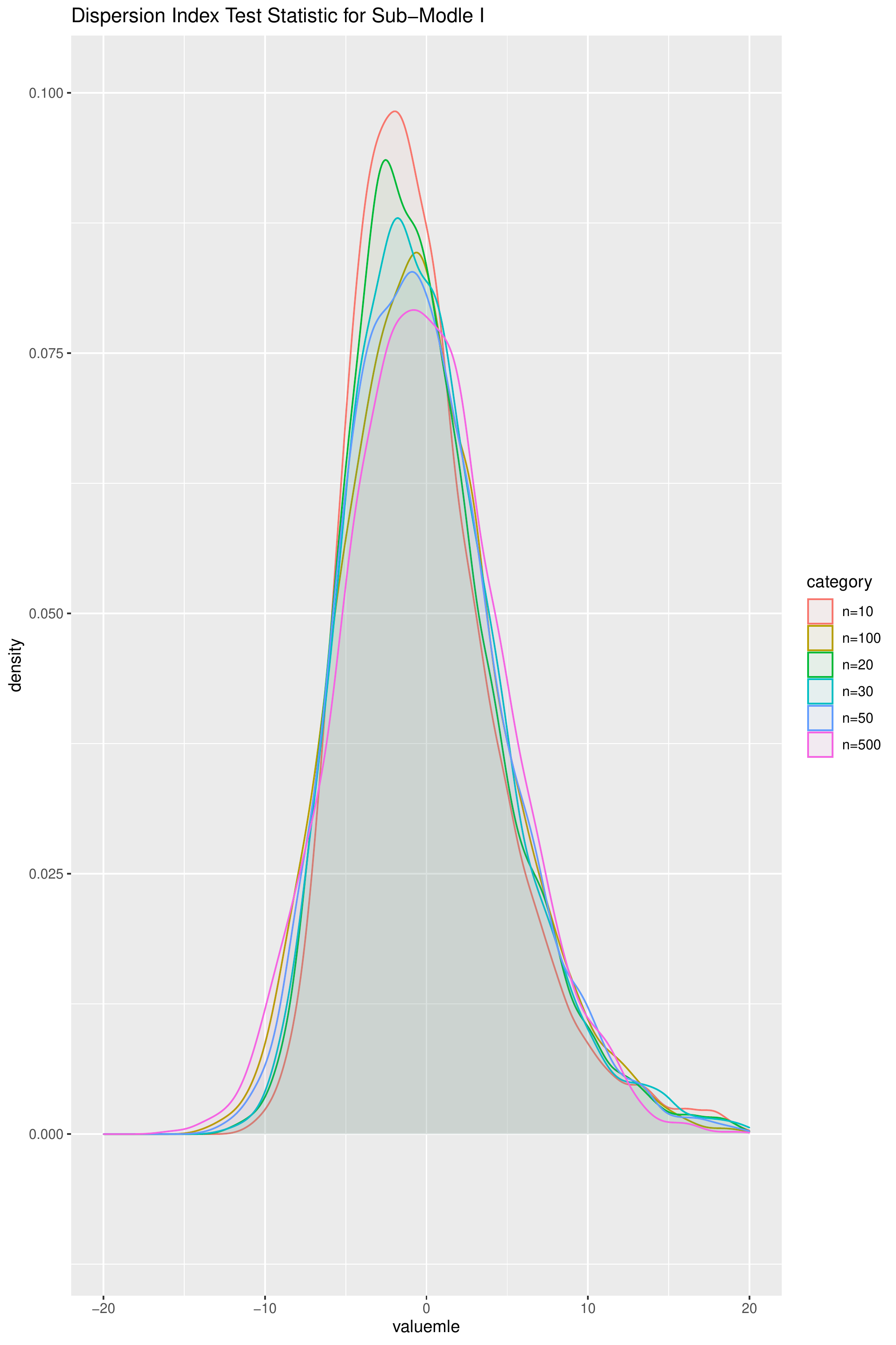}}
		\subfloat[Sub-Model II]{\includegraphics[width=6cm,height=3cm]{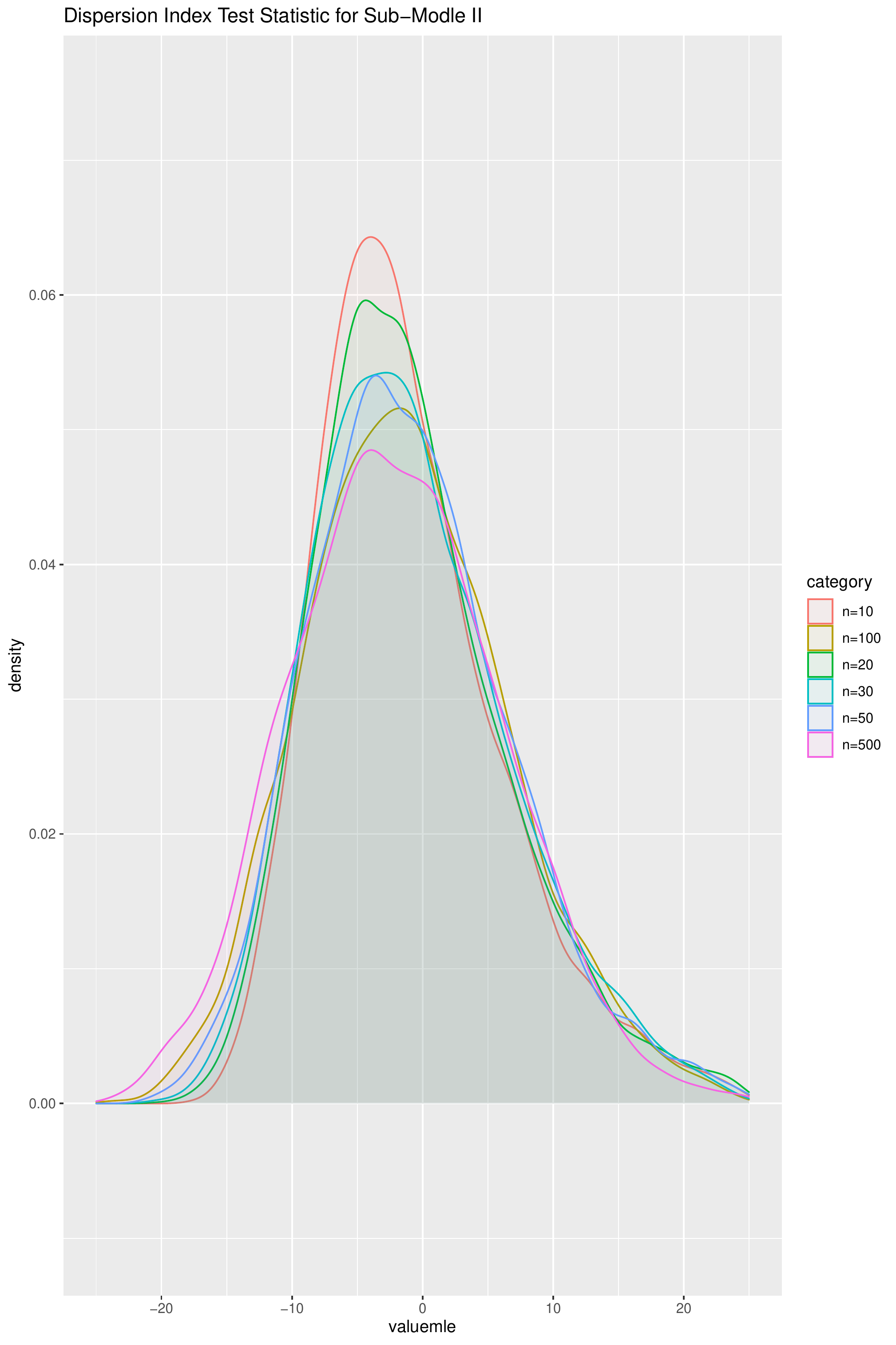}} 
		\caption{Distribution of the $FI^{(.)}_{n}$}
		\label{DIF} 
	\end{figure}

	In the current section we will be analysing the new non-consistent test defined in Section 3.1. 
	The finite sample distribution of the	$FI^{(.)}_{n}$, see Table \ref{DIT} and Figure \ref{DIF} for the  distribution  and its quantile values for the full and its sub-models. 
	From the simulation study it clearly showen that the distribution of test statistic showen to be standard normal behavior for increasing sample size. In addition, we make a note that for small and moderetely large sample sizes the test showen to be stable and consistent. 
	
	\subsubsection{Mu\~noz and Gamero(M\&G) method}
	
\begin{figure}[ht]
		\subfloat[]{\includegraphics[width=5cm,height=2.5cm]{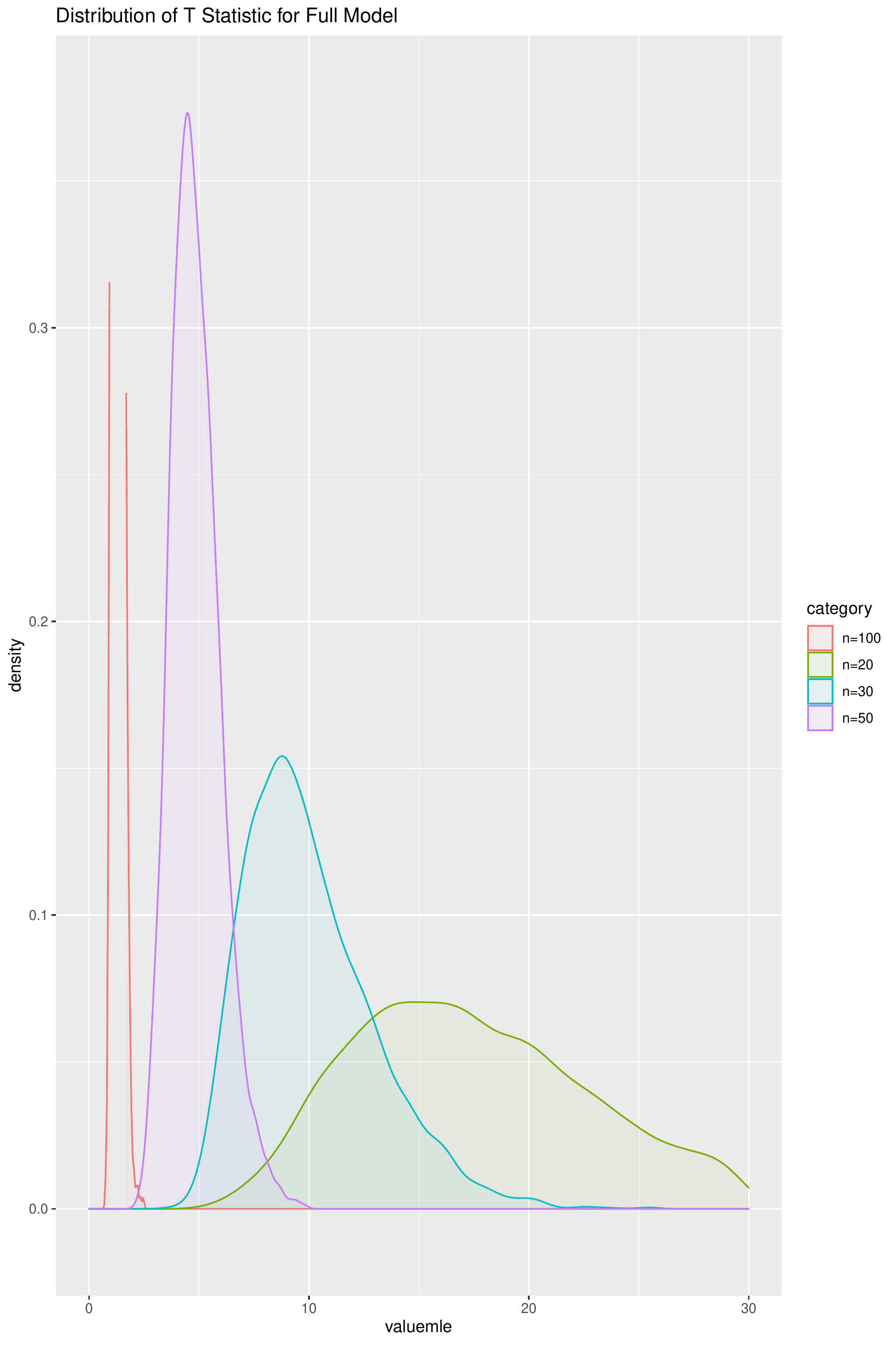}} 
		\subfloat[]{\includegraphics[width=5cm,height=2.5cm]{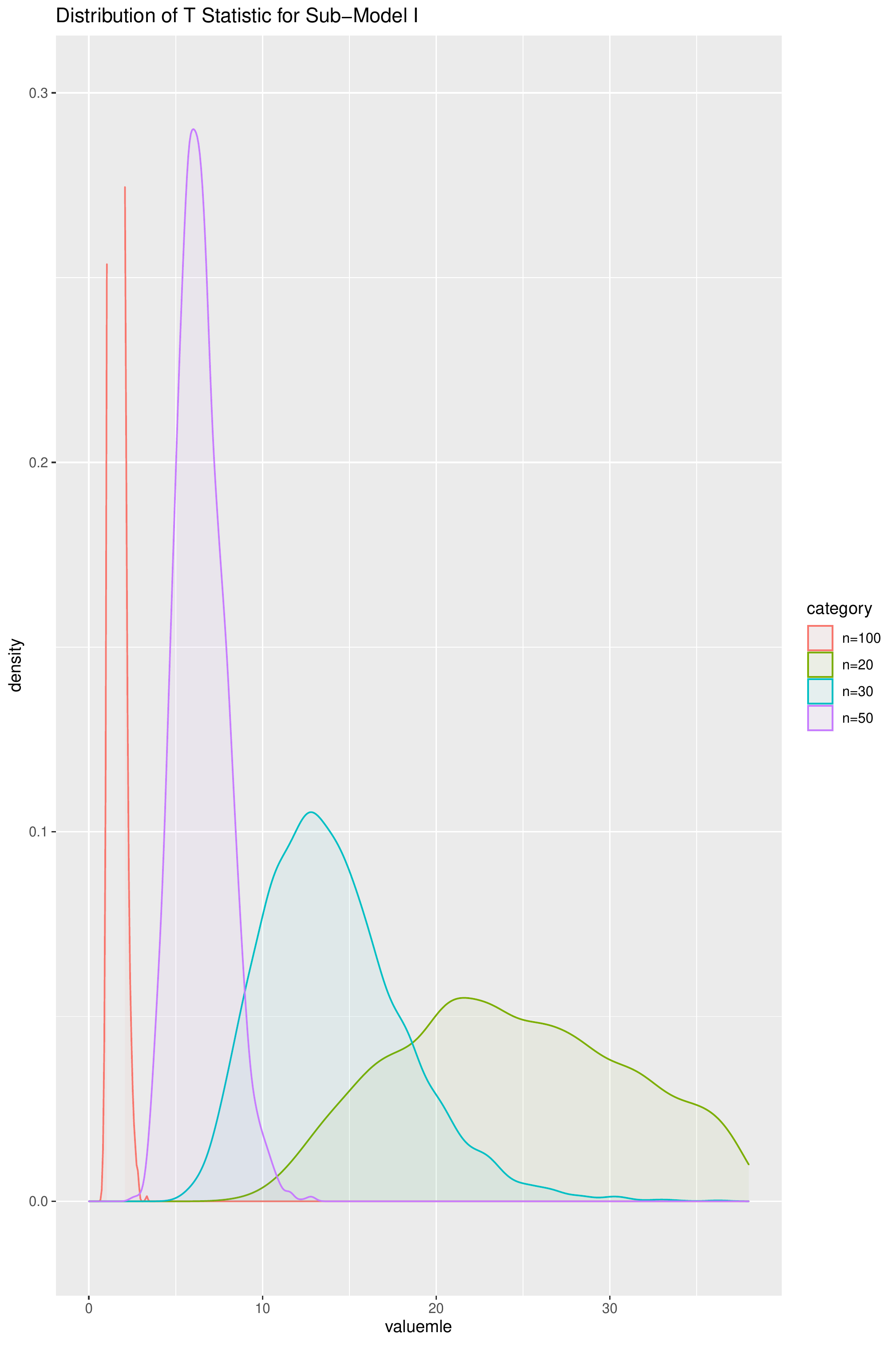}}
		\subfloat[]{\includegraphics[width=5cm,height=2.5cm]{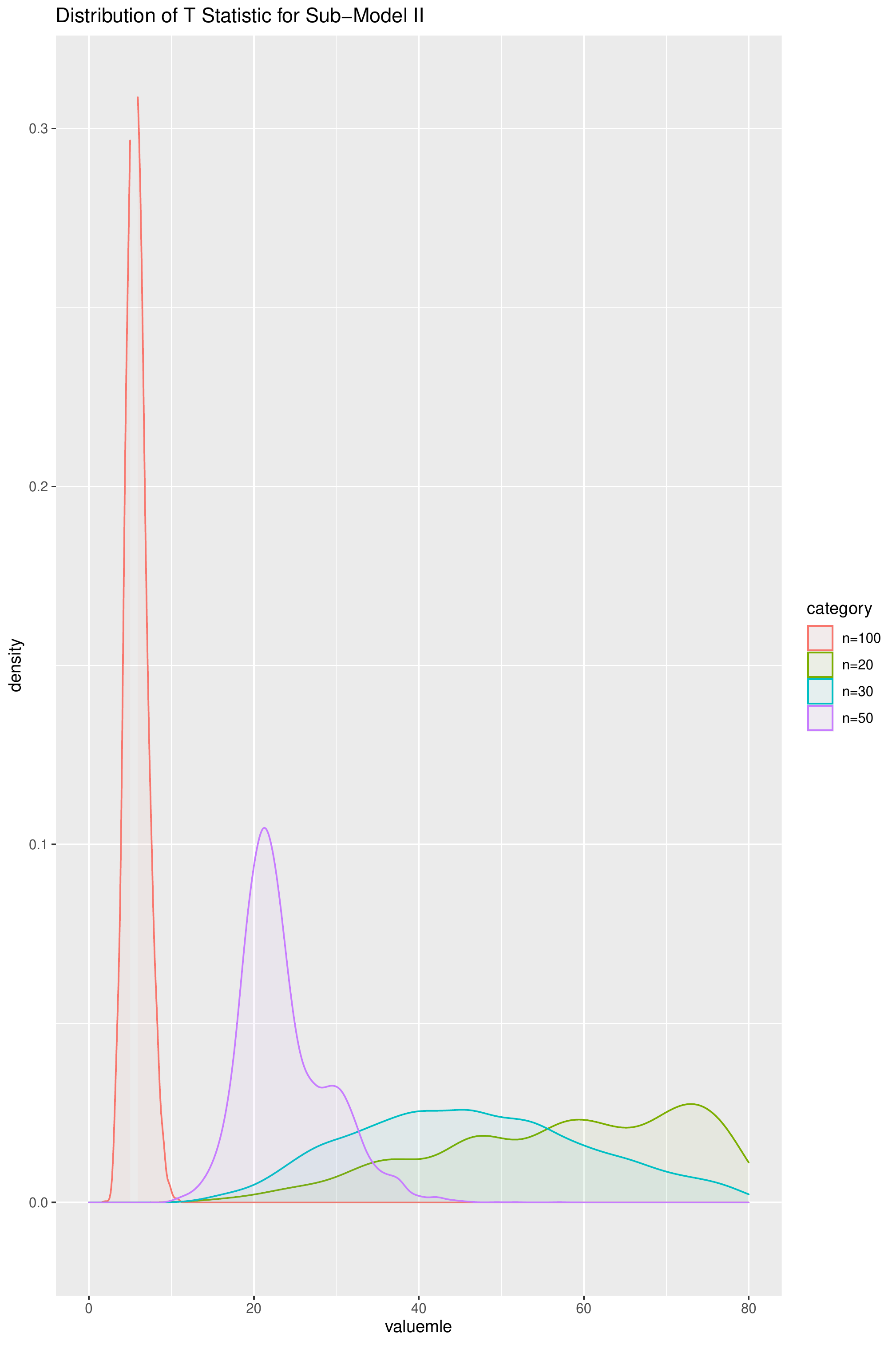}}
		\caption{Example 1}
		\label{EX1T}
	\end{figure}

\begin{figure}[H]
		\subfloat[]{\includegraphics[width=5cm,height=2.5cm]{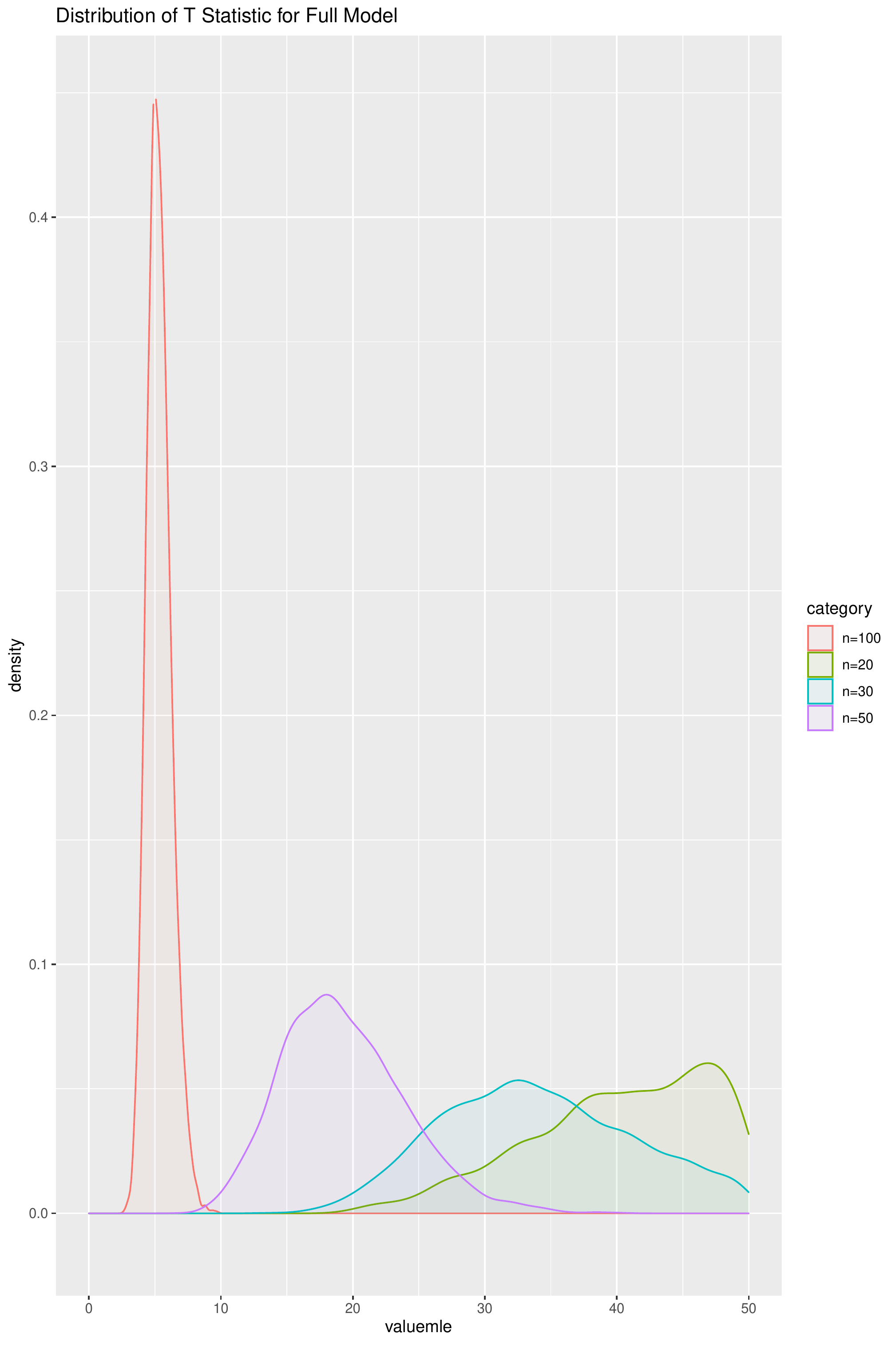}} 
		\subfloat[]{\includegraphics[width=5cm,height=2.5cm]{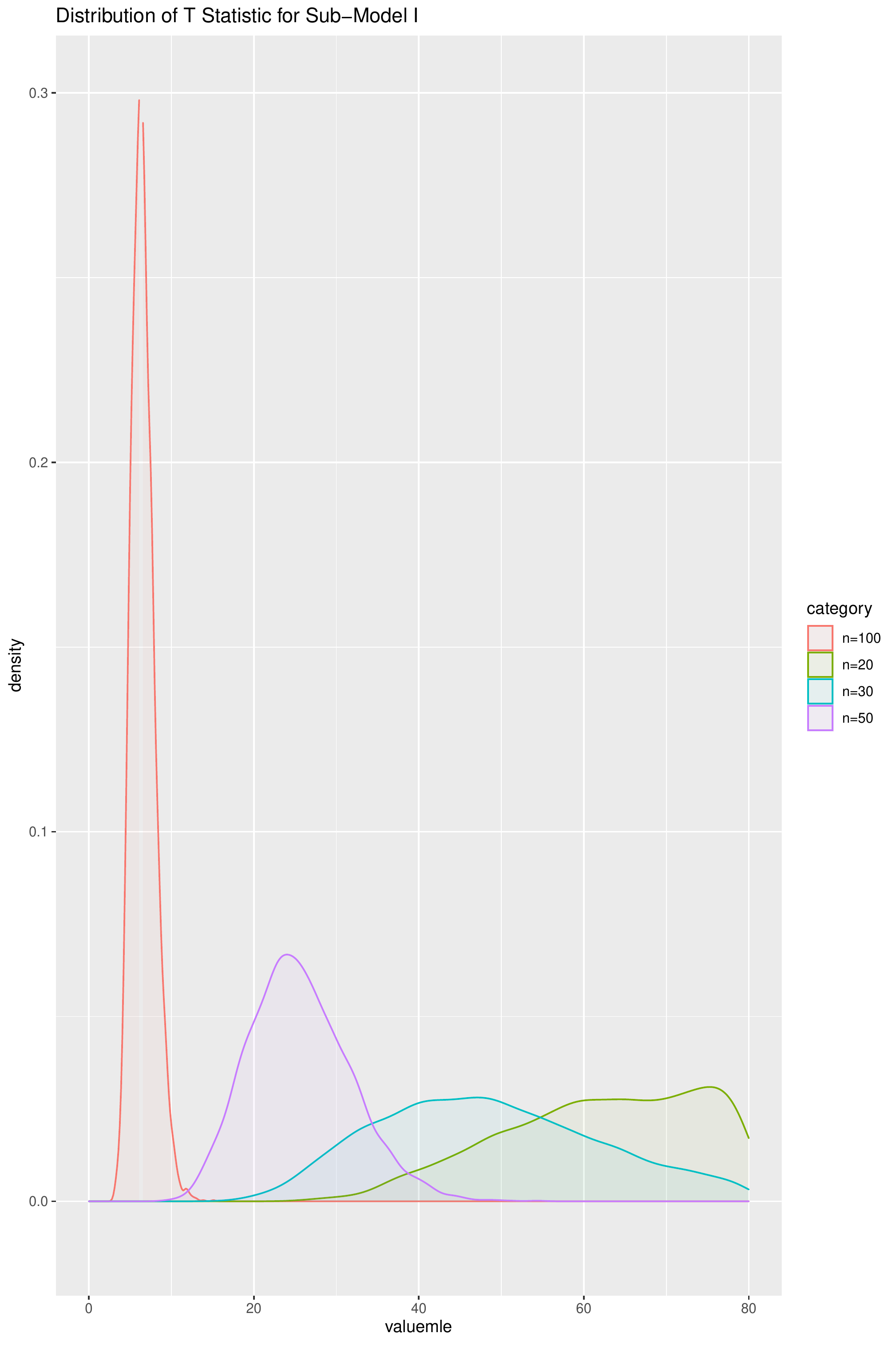}}
		\subfloat[]{\includegraphics[width=5cm,height=2.5cm]{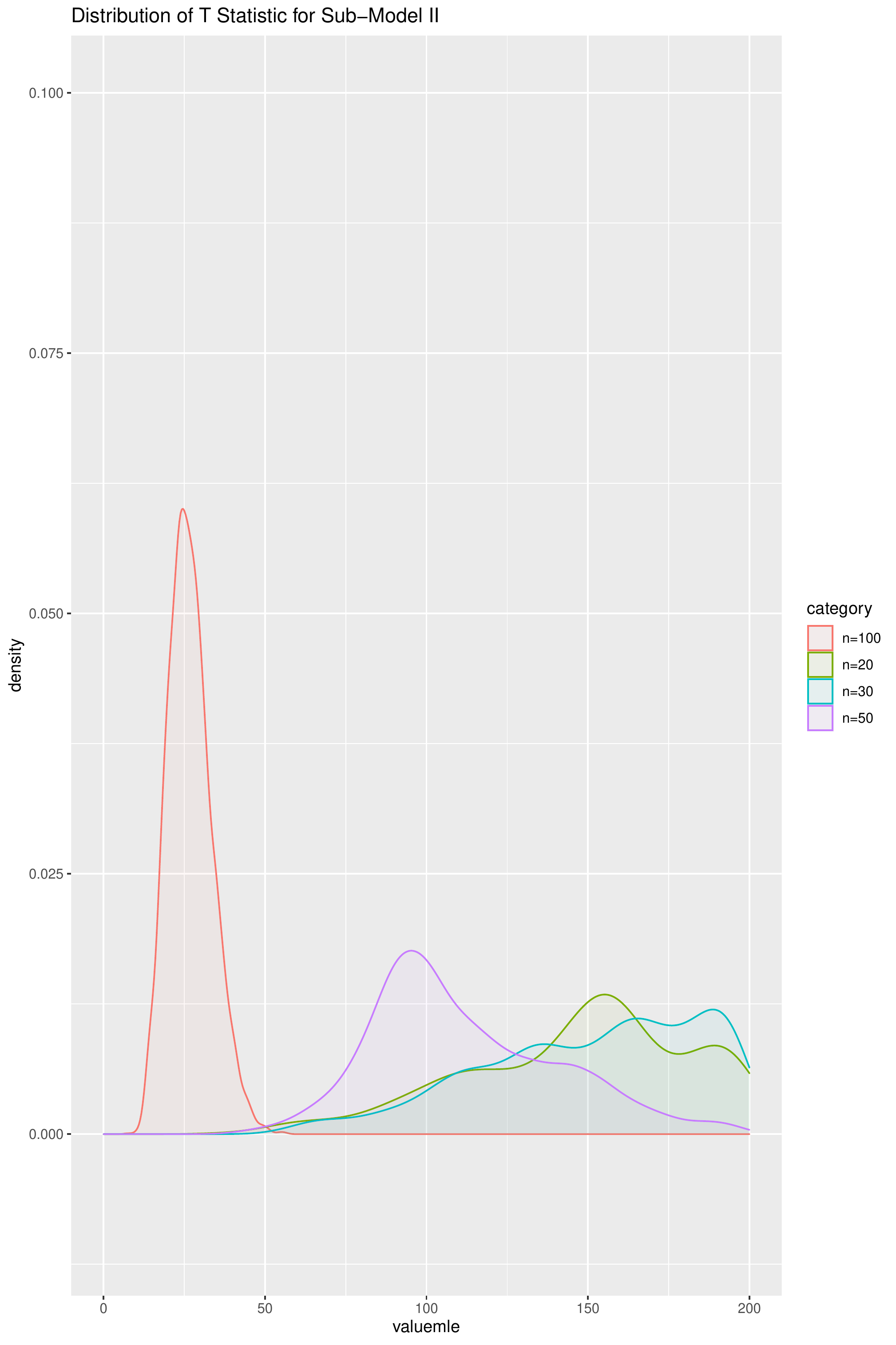}}
		\caption{Example 2}
		\label{EX2T}
	\end{figure}

\begin{figure}[H]
		\subfloat[]{\includegraphics[width=5cm,height=2.5cm]{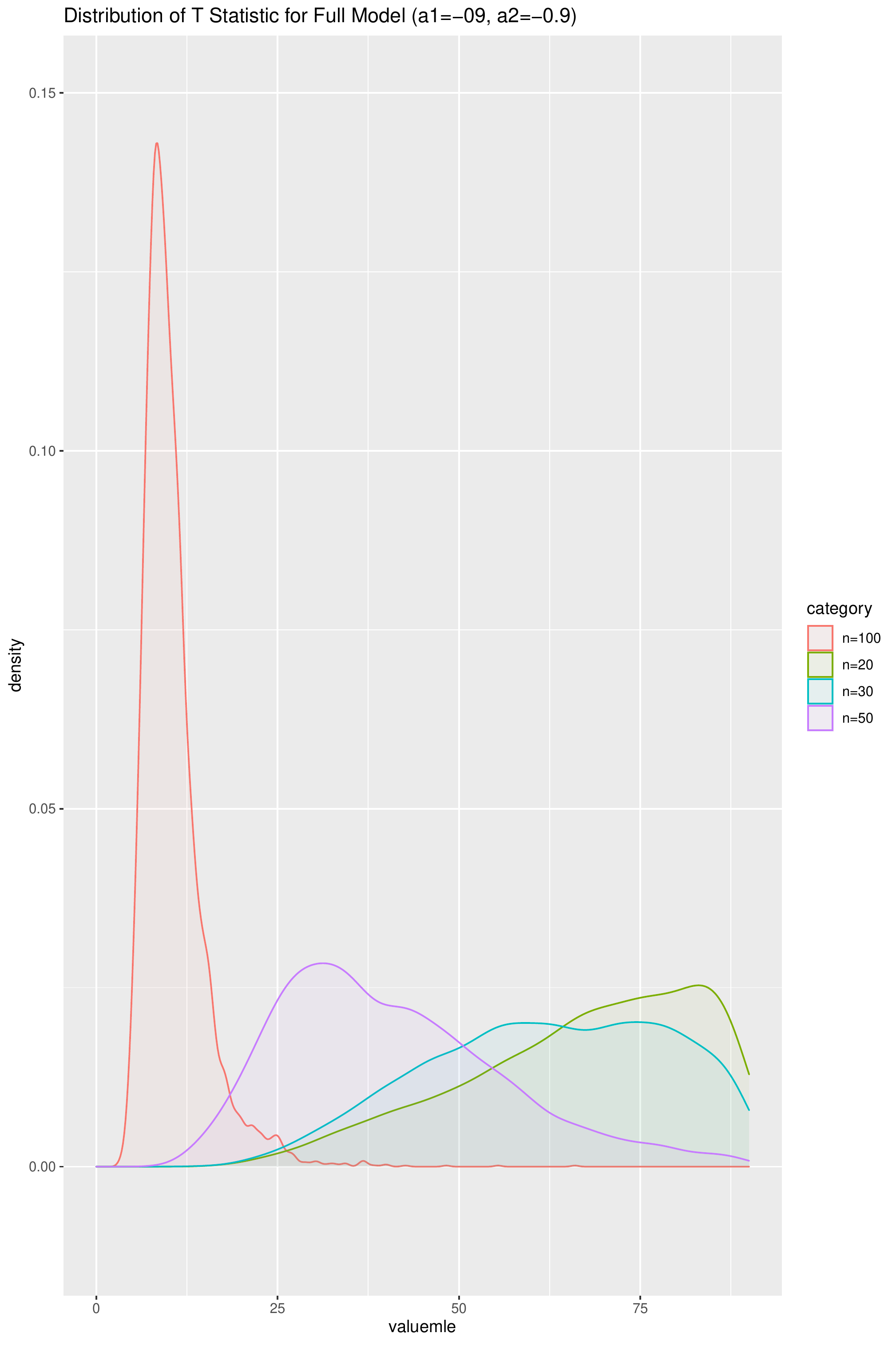}} 
		\subfloat[]{\includegraphics[width=5cm,height=2.5cm]{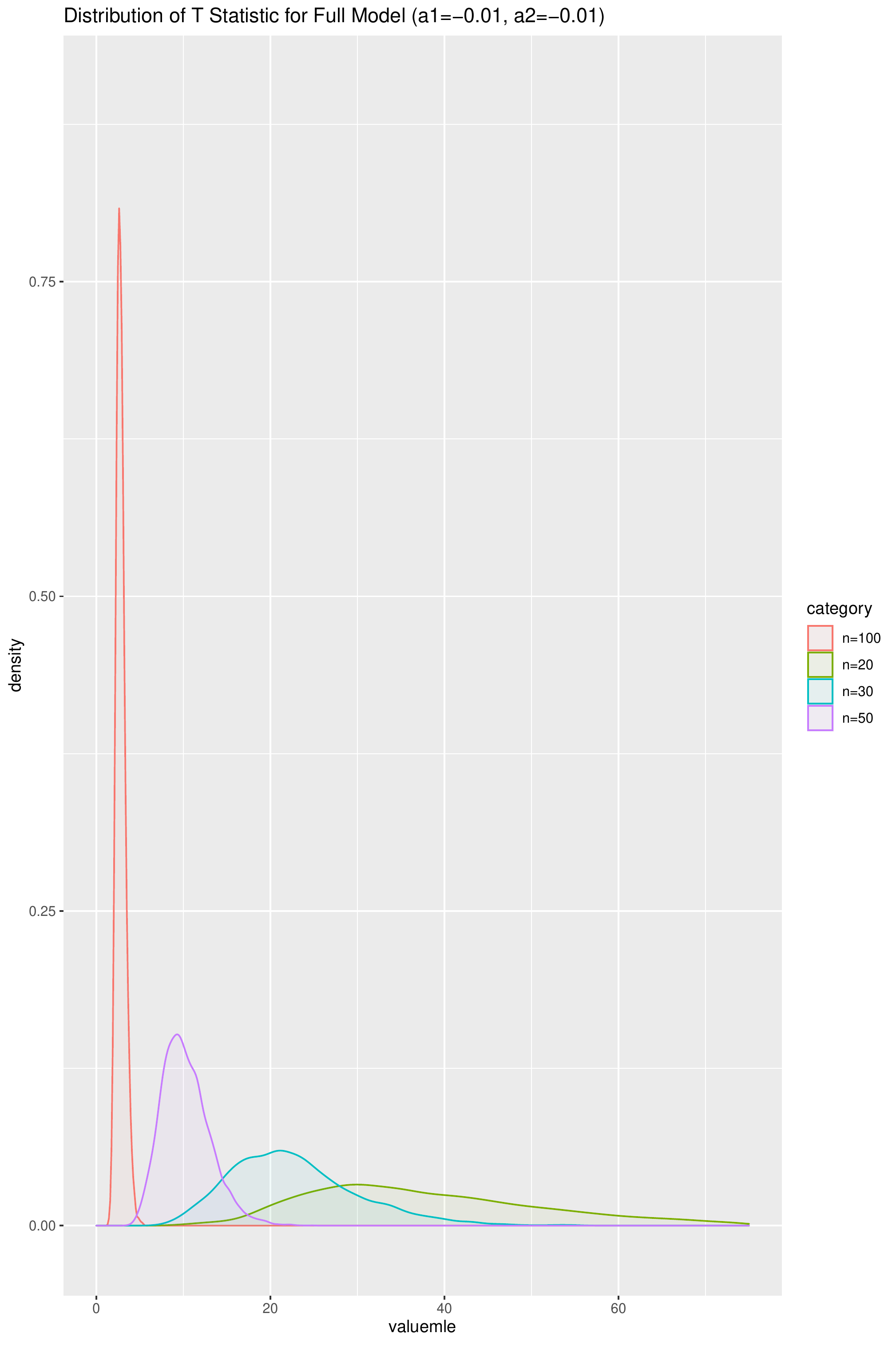}}
		\subfloat[]{\includegraphics[width=5cm,height=2.5cm]{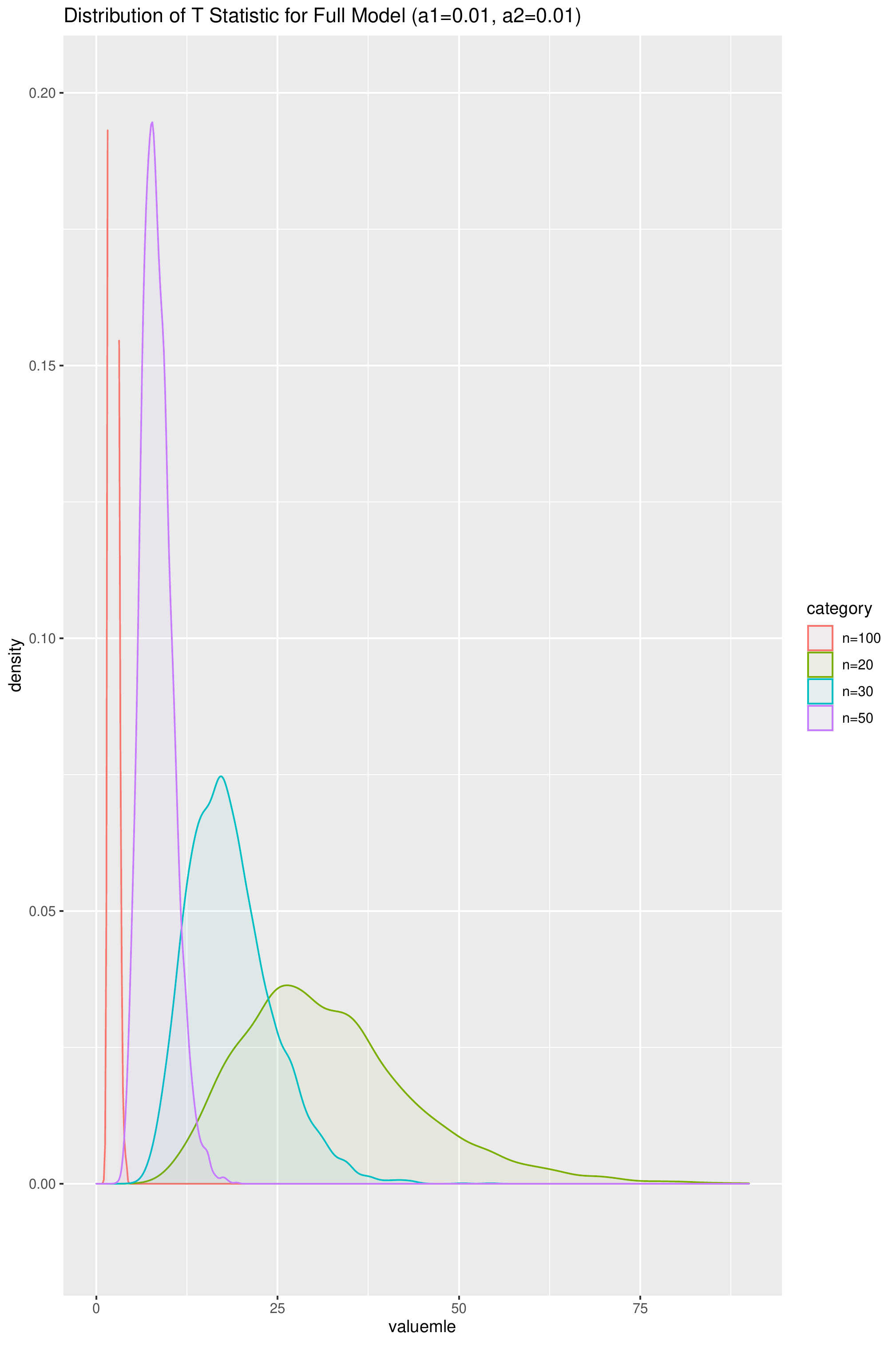}} \\
		\subfloat[]{\includegraphics[width=5cm,height=2.5cm]{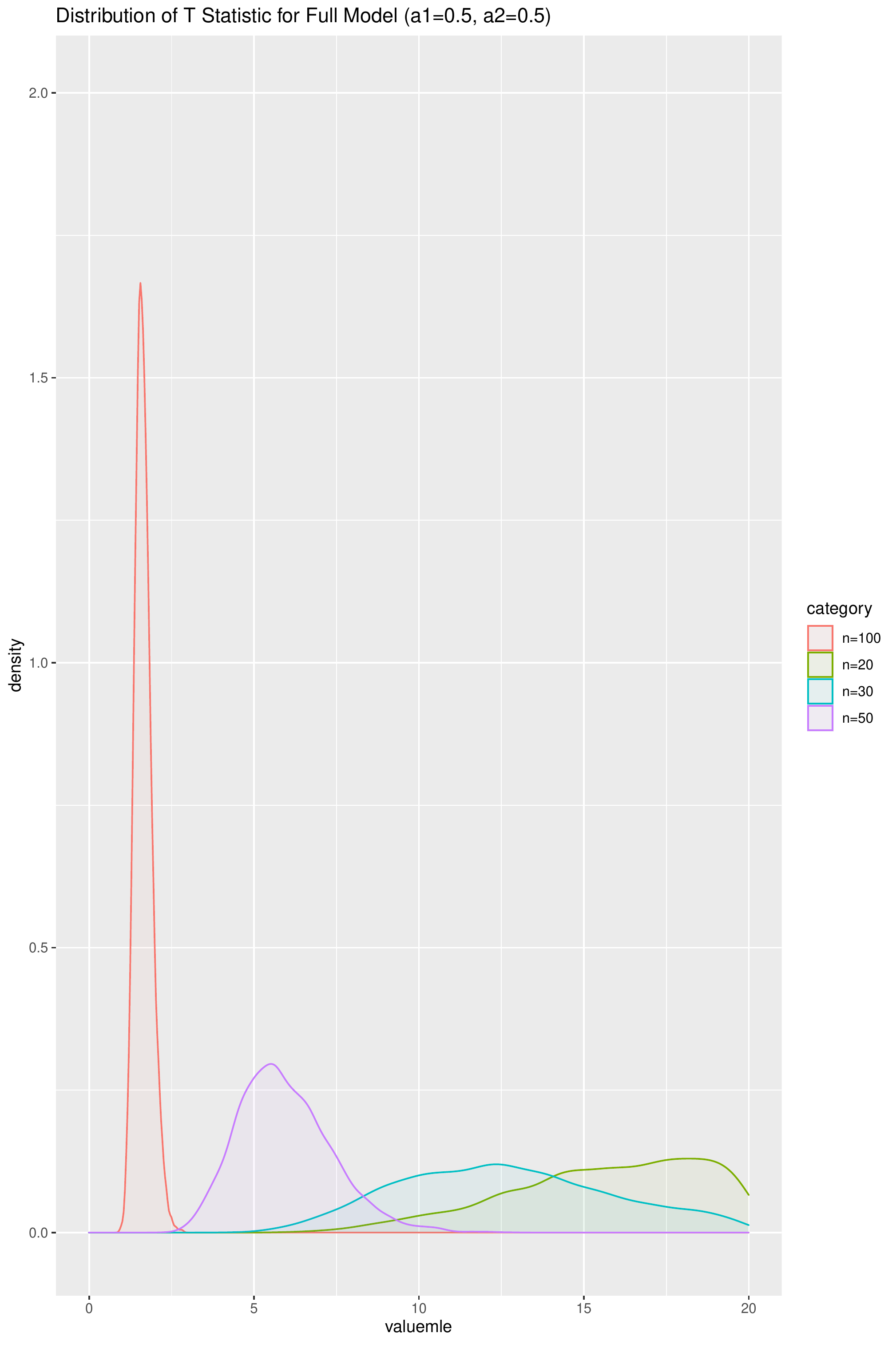}}
		\subfloat[]{\includegraphics[width=5cm,height=2.5cm]{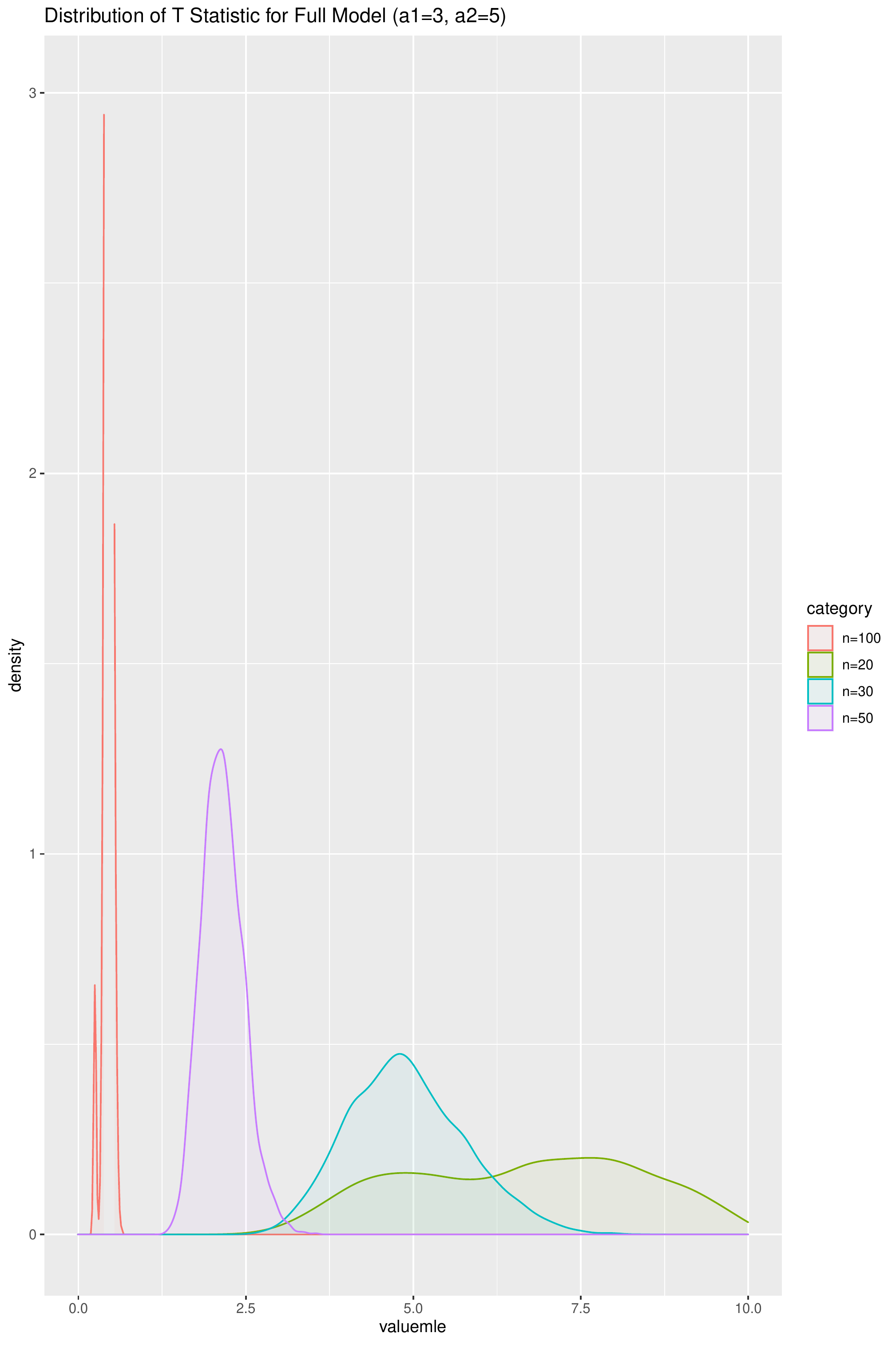}}
		\subfloat[]{\includegraphics[width=5cm,height=2.5cm]{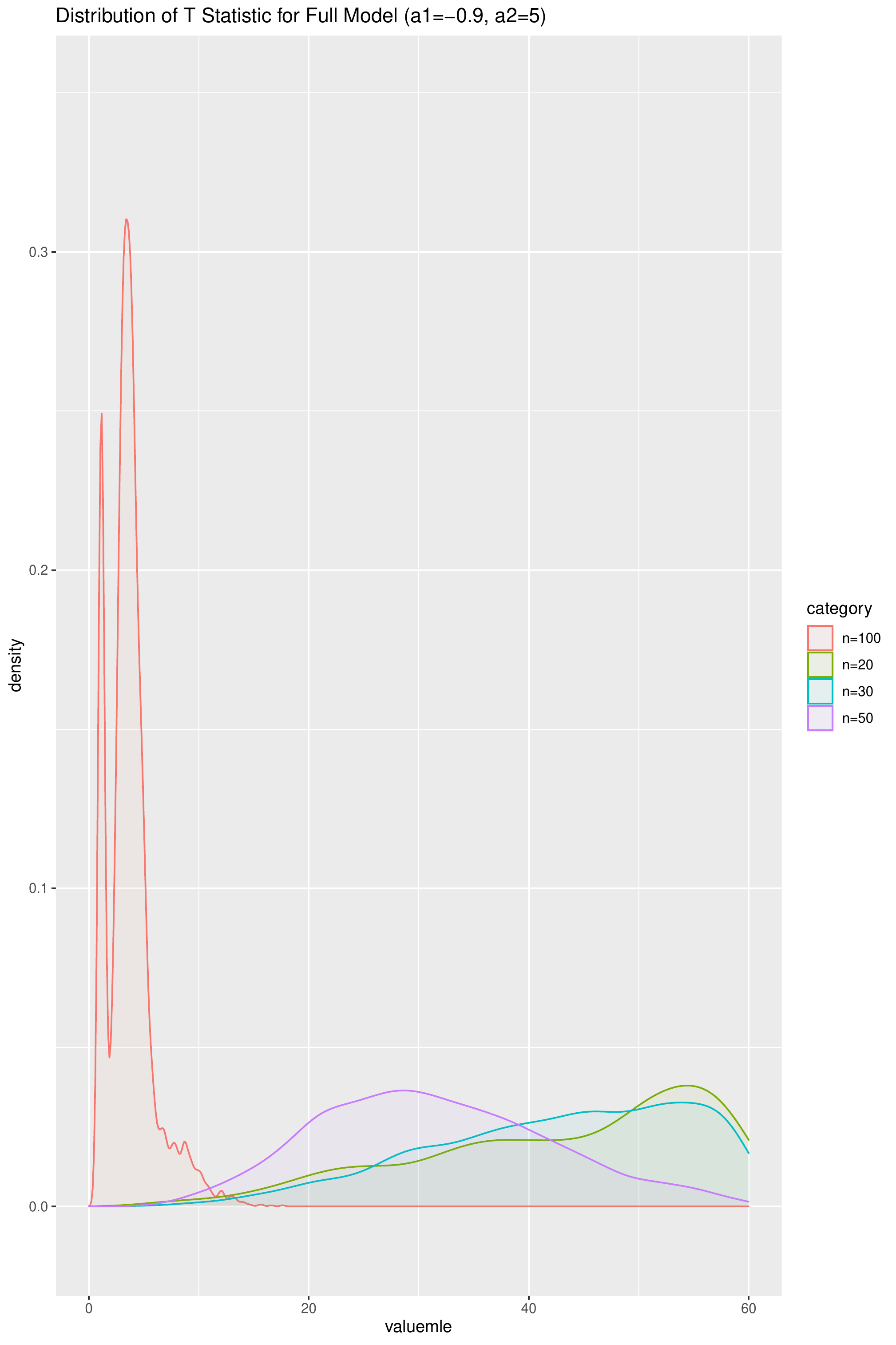}}
		\caption{$T^{(.)}_{P,n,w}$}
		\label{EX3FULL}
	\end{figure}

\begin{figure}[H]
		\subfloat[]{\includegraphics[width=5cm,height=2.5cm]{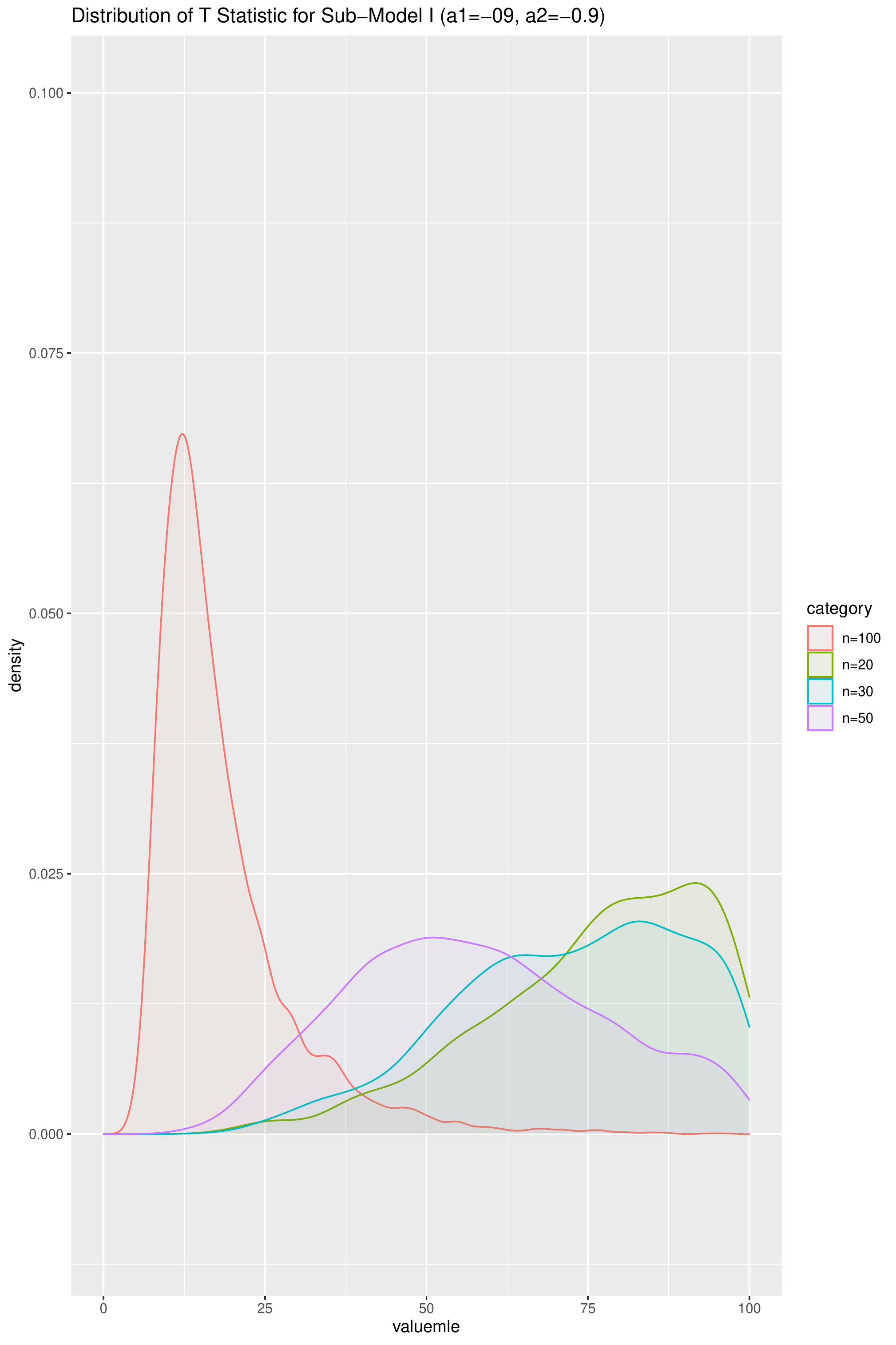}} 
		\subfloat[]{\includegraphics[width=5cm,height=2.5cm]{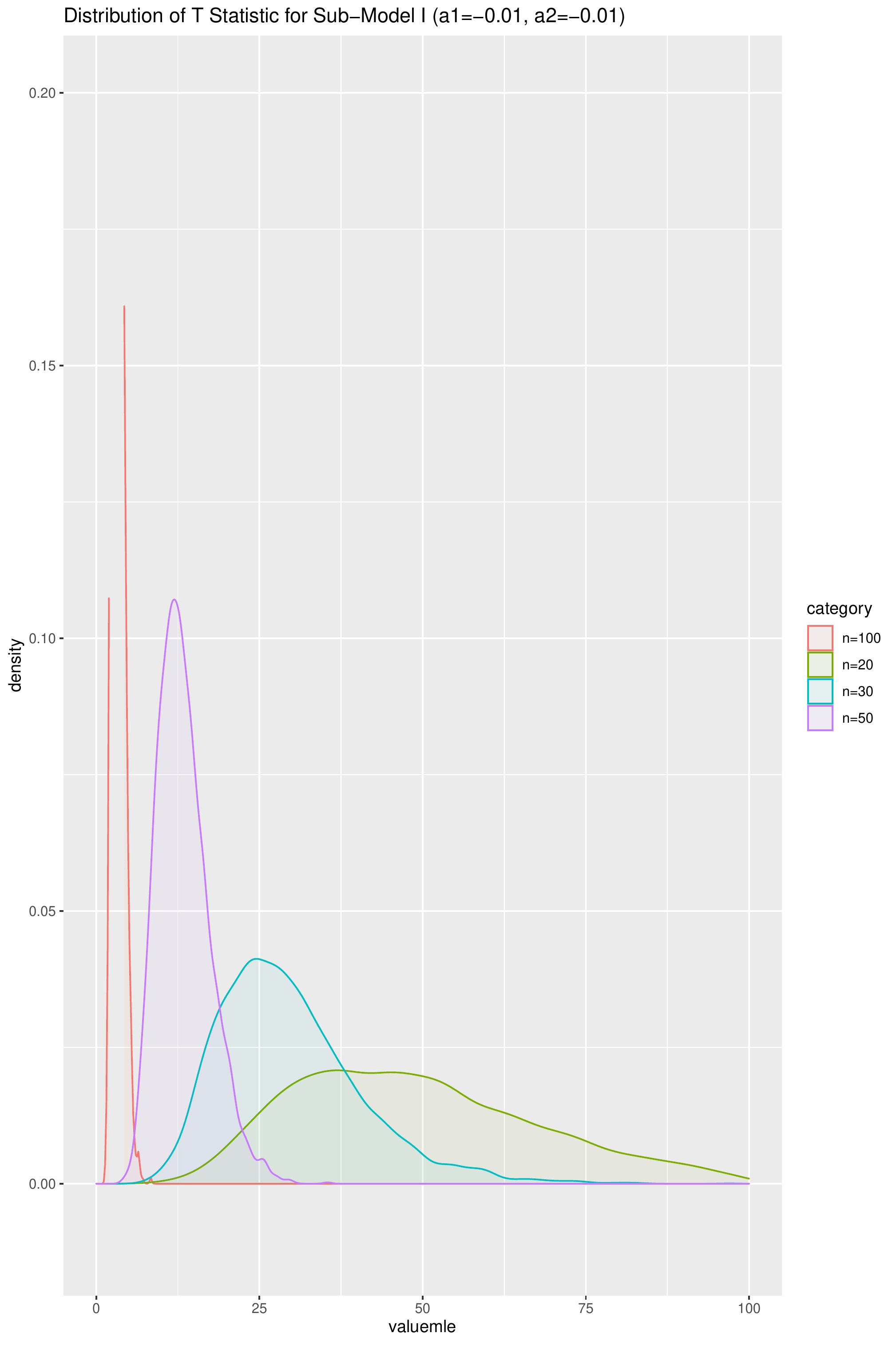}}
		\subfloat[]{\includegraphics[width=5cm,height=2.5cm]{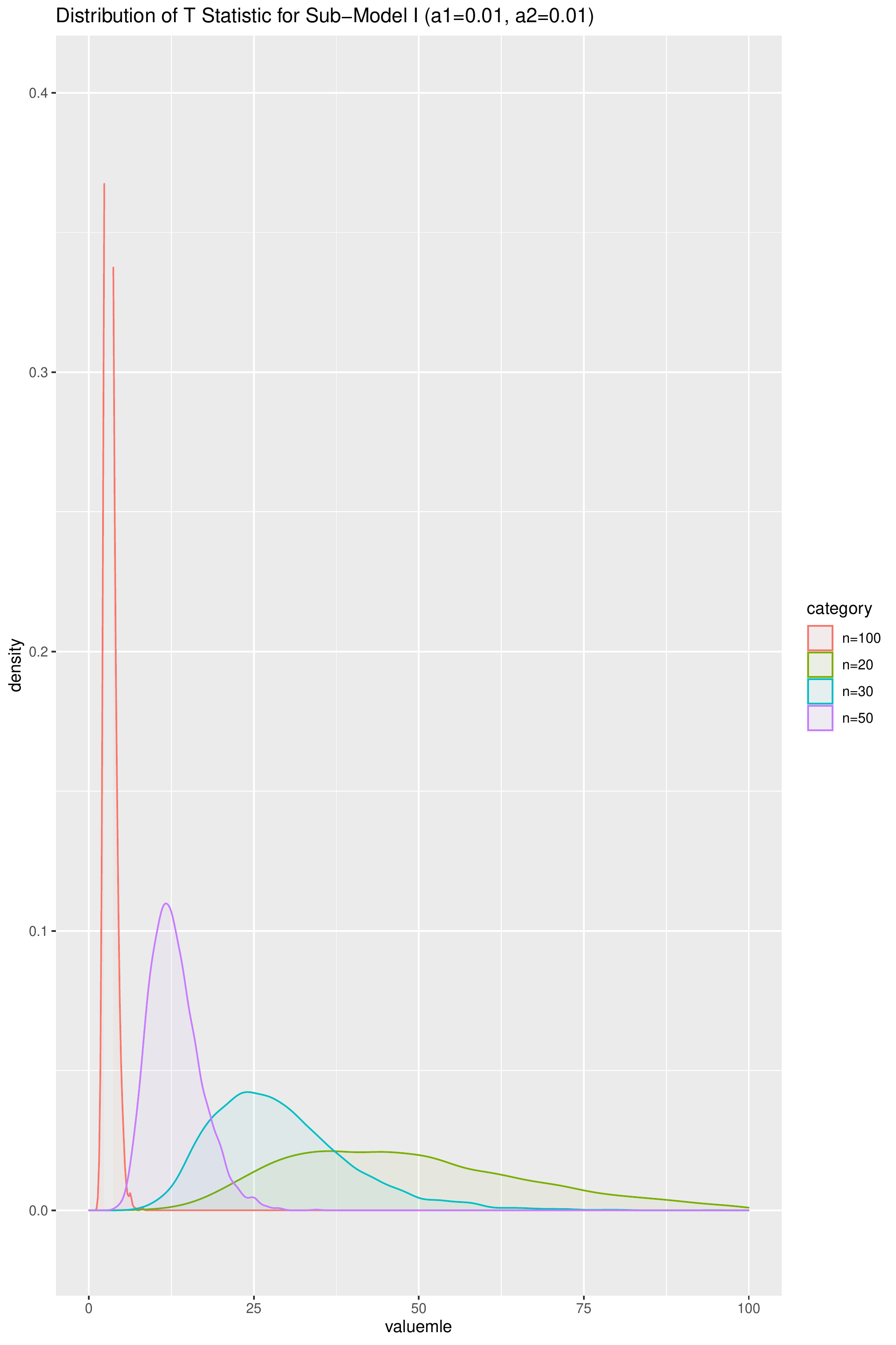}} \\
		\subfloat[]{\includegraphics[width=5cm,height=2.5cm]{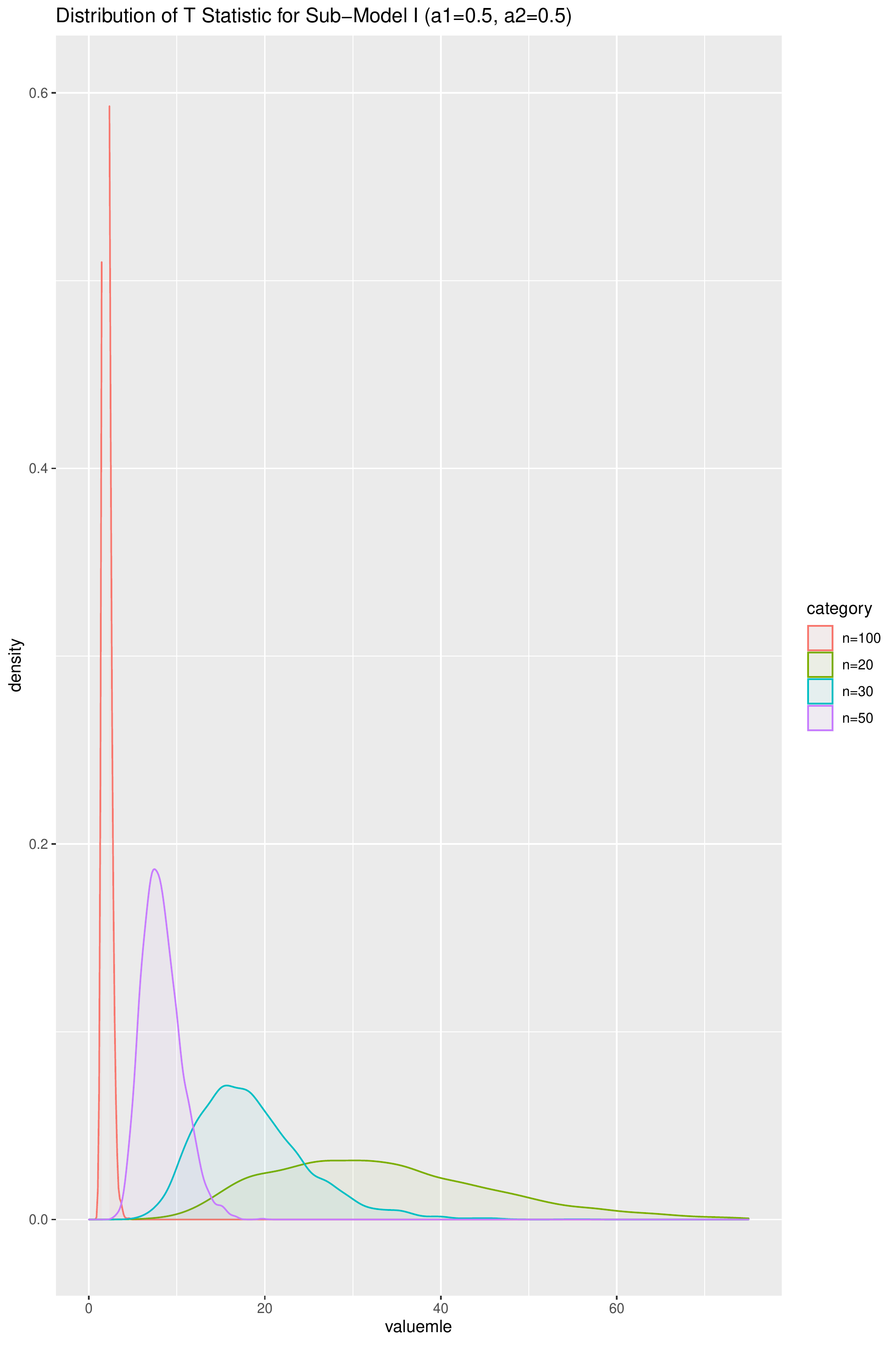}}
		\subfloat[]{\includegraphics[width=5cm,height=2.5cm]{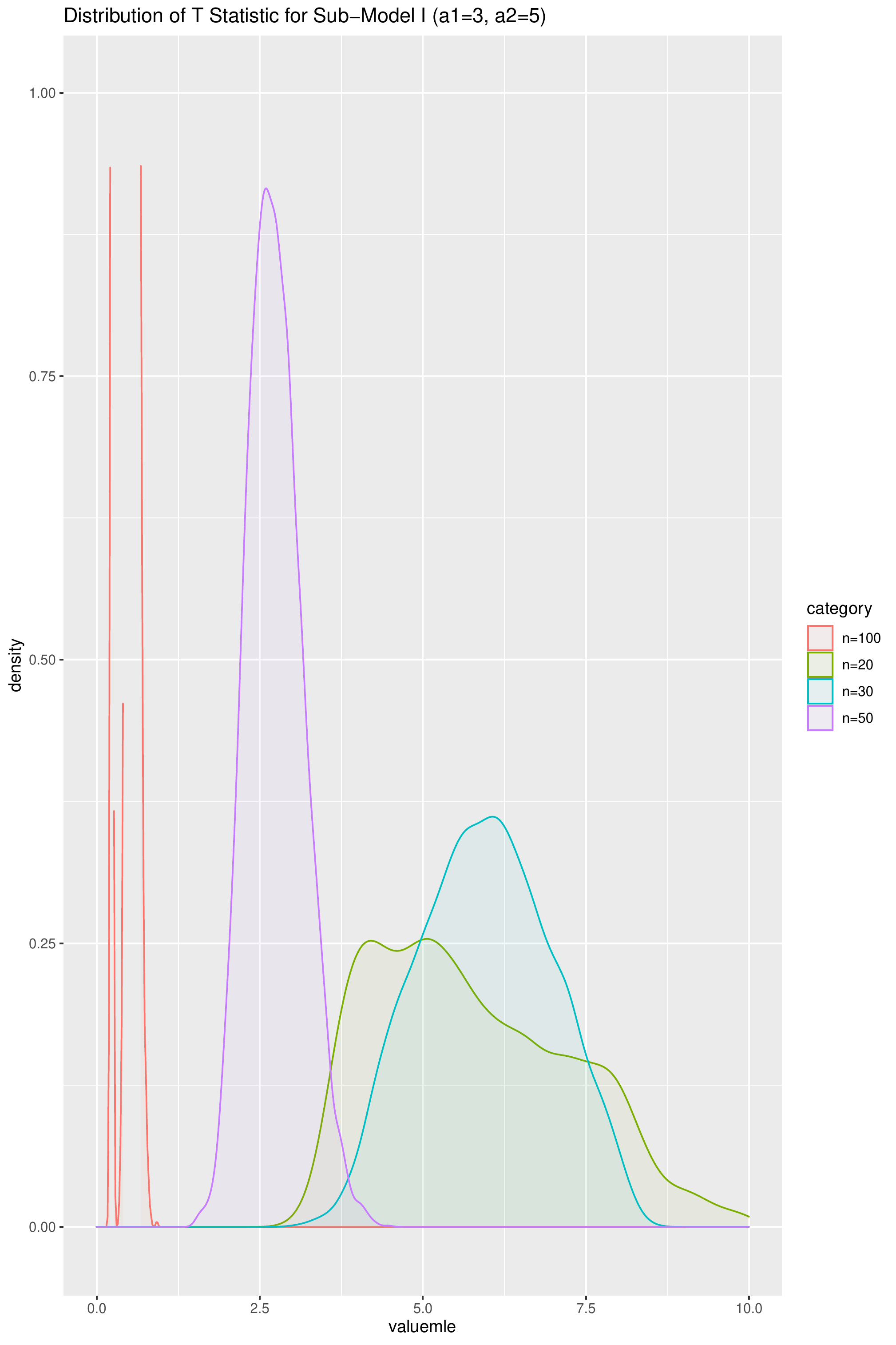}}
		\subfloat[]{\includegraphics[width=5cm,height=2.5cm]{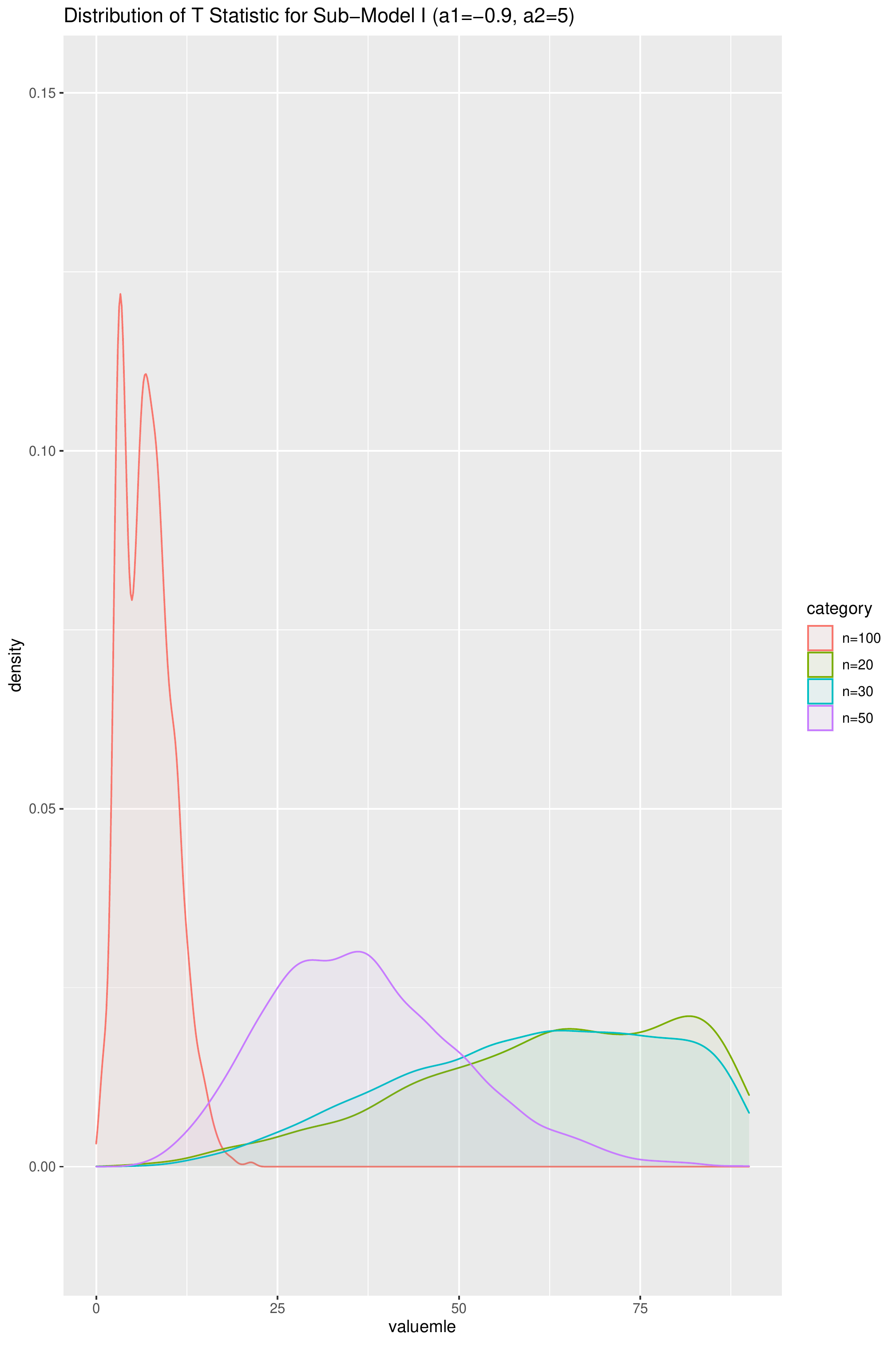}}
		\caption{$T^{(SI)}_{P,n,w}$}
		\label{EX3SI}
	\end{figure}

\begin{figure}[H]
		\subfloat[]{\includegraphics[width=5cm,height=2.5cm]{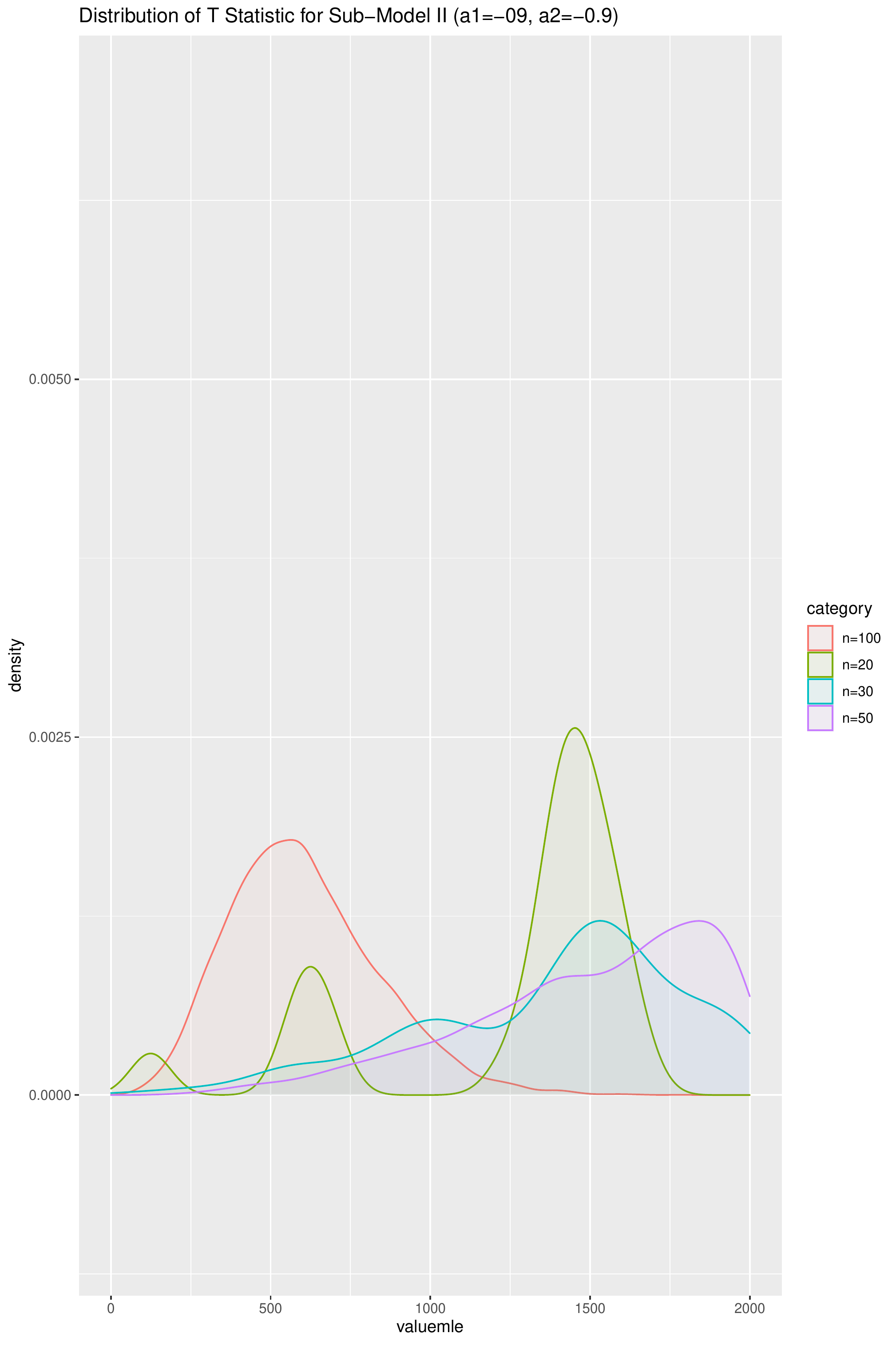}} 
		\subfloat[]{\includegraphics[width=5cm,height=2.5cm]{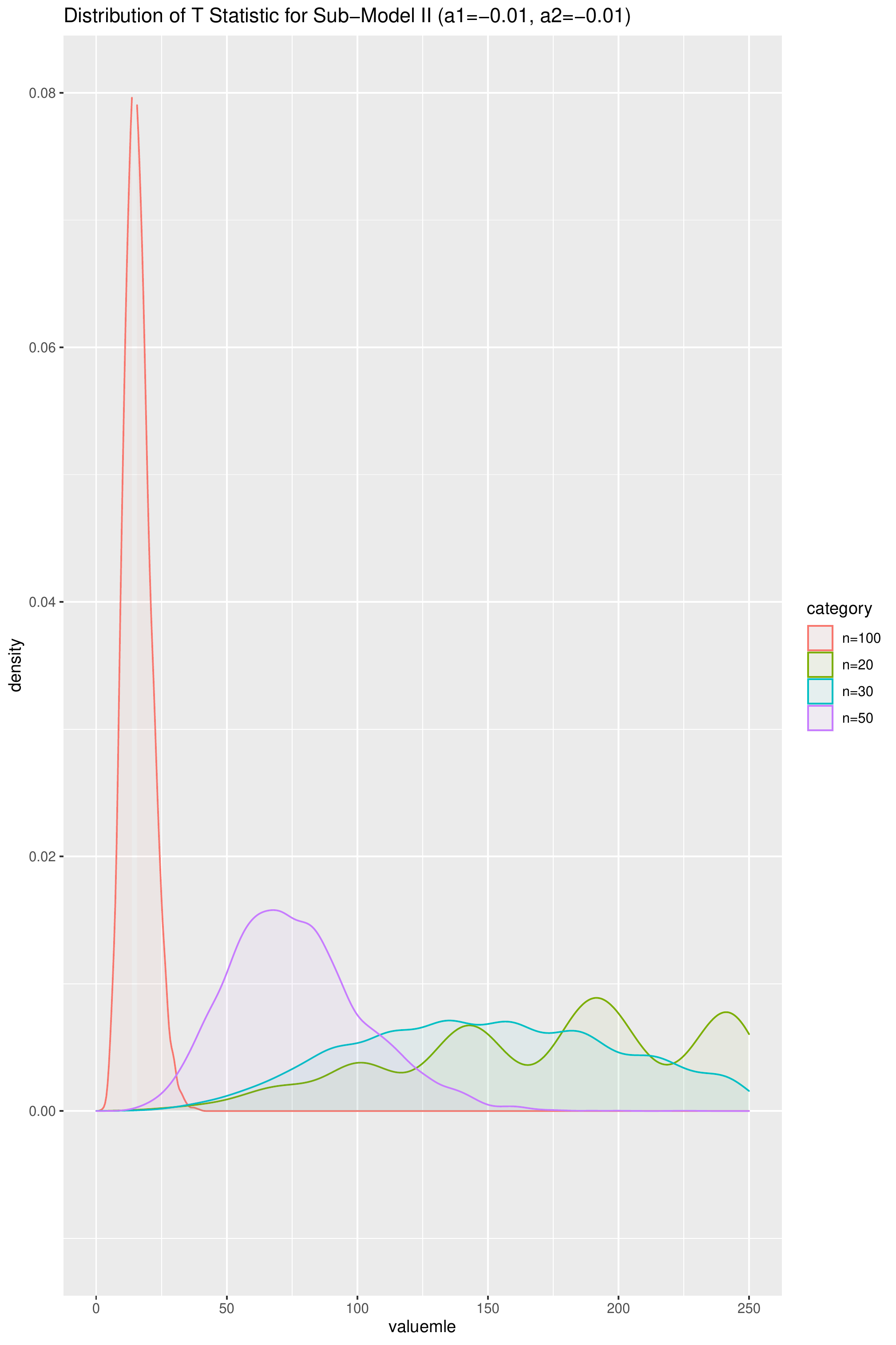}}
		\subfloat[]{\includegraphics[width=5cm,height=2.5cm]{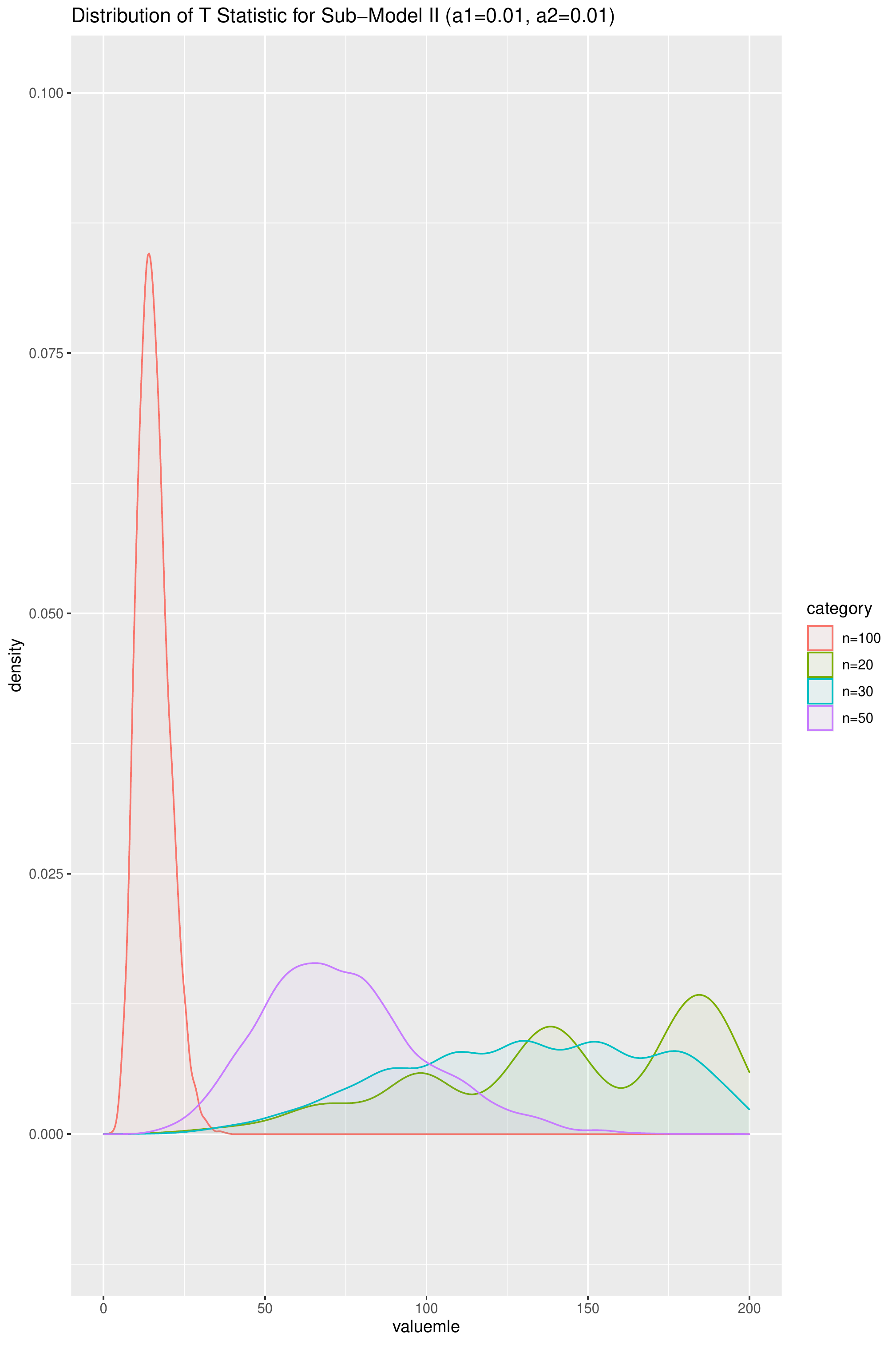}} \\
		\subfloat[]{\includegraphics[width=5cm,height=2.5cm]{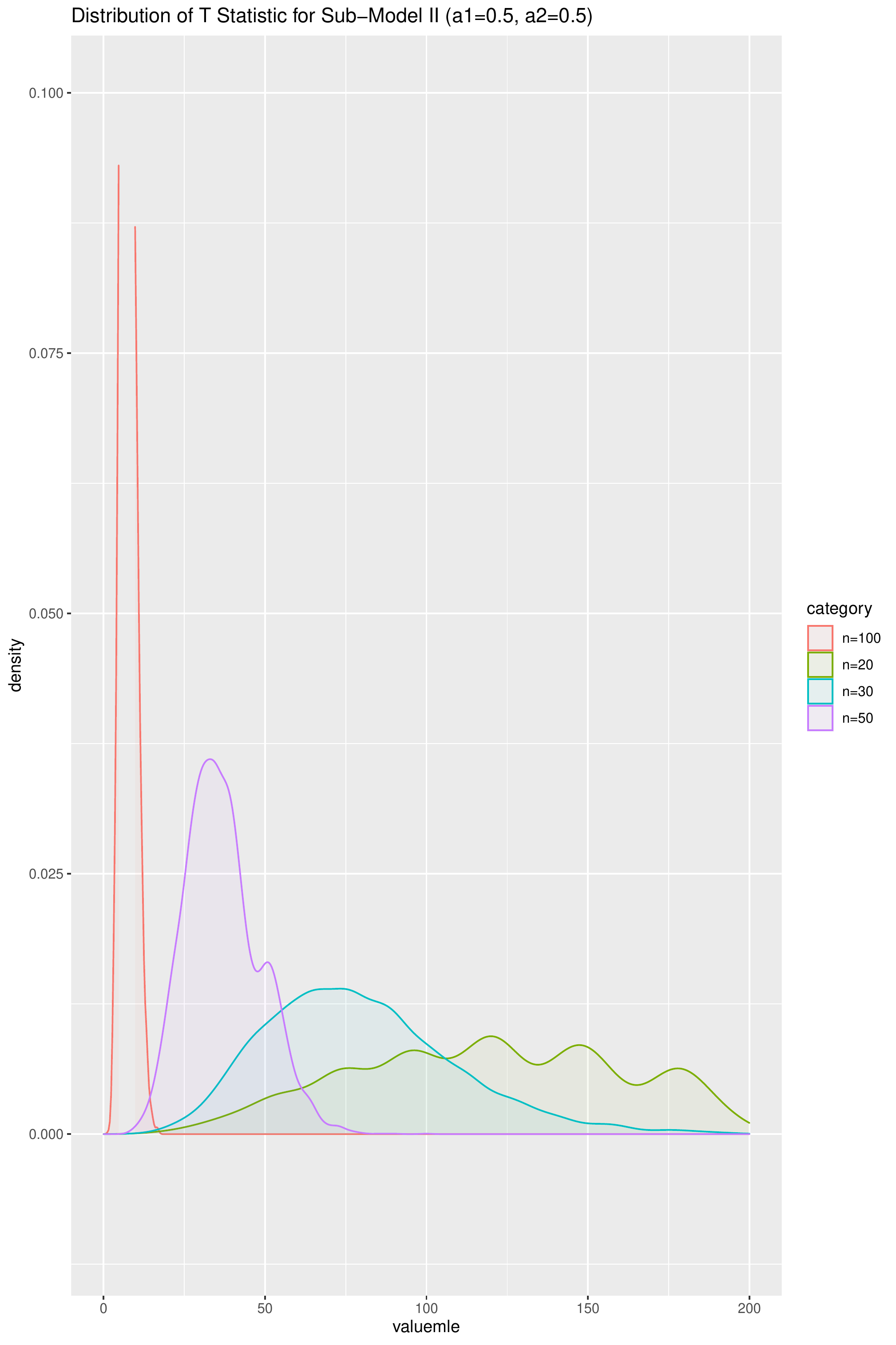}}
		\subfloat[]{\includegraphics[width=5cm,height=2.5cm]{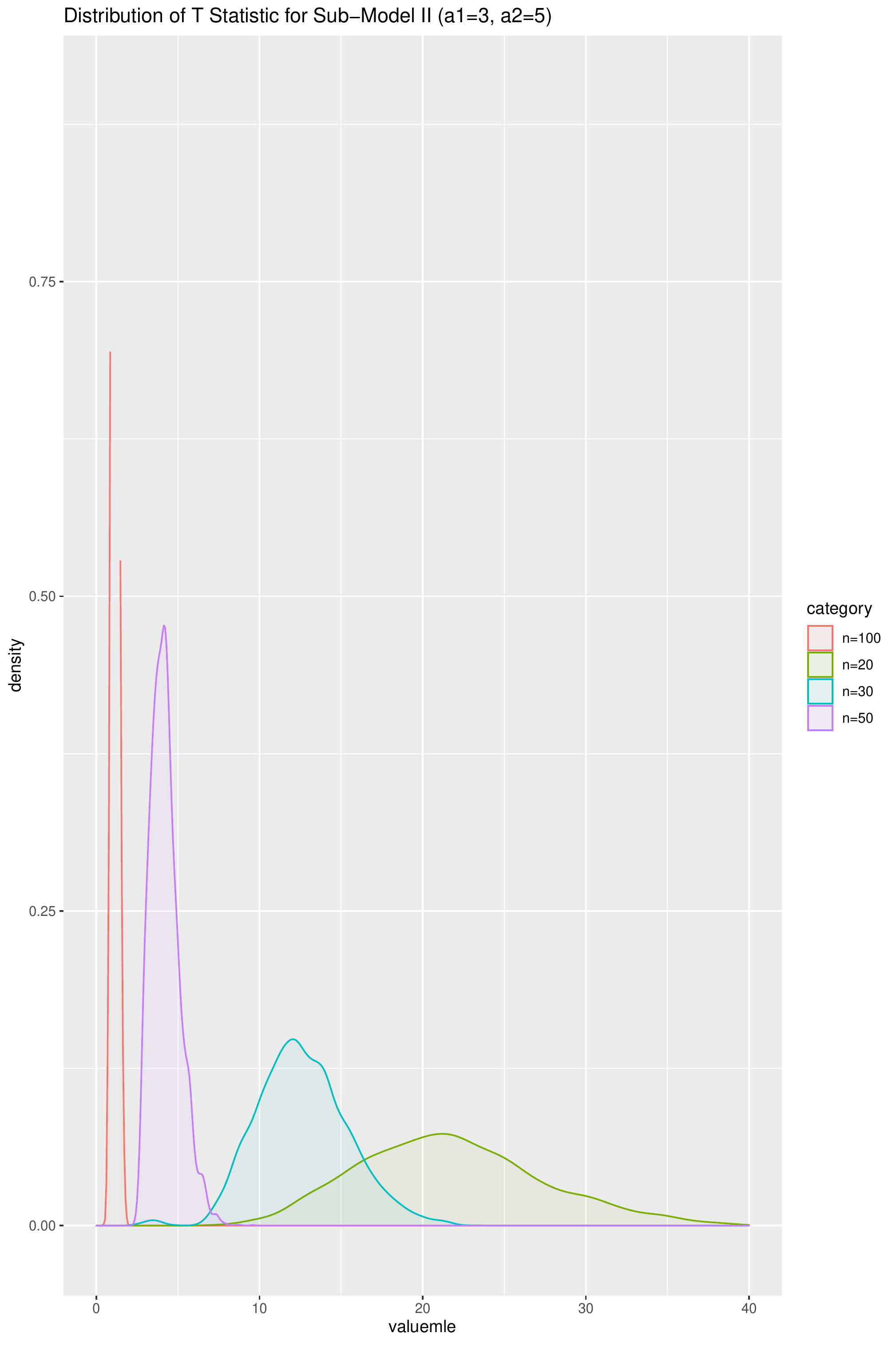}}
		\subfloat[]{\includegraphics[width=5cm,height=2.5cm]{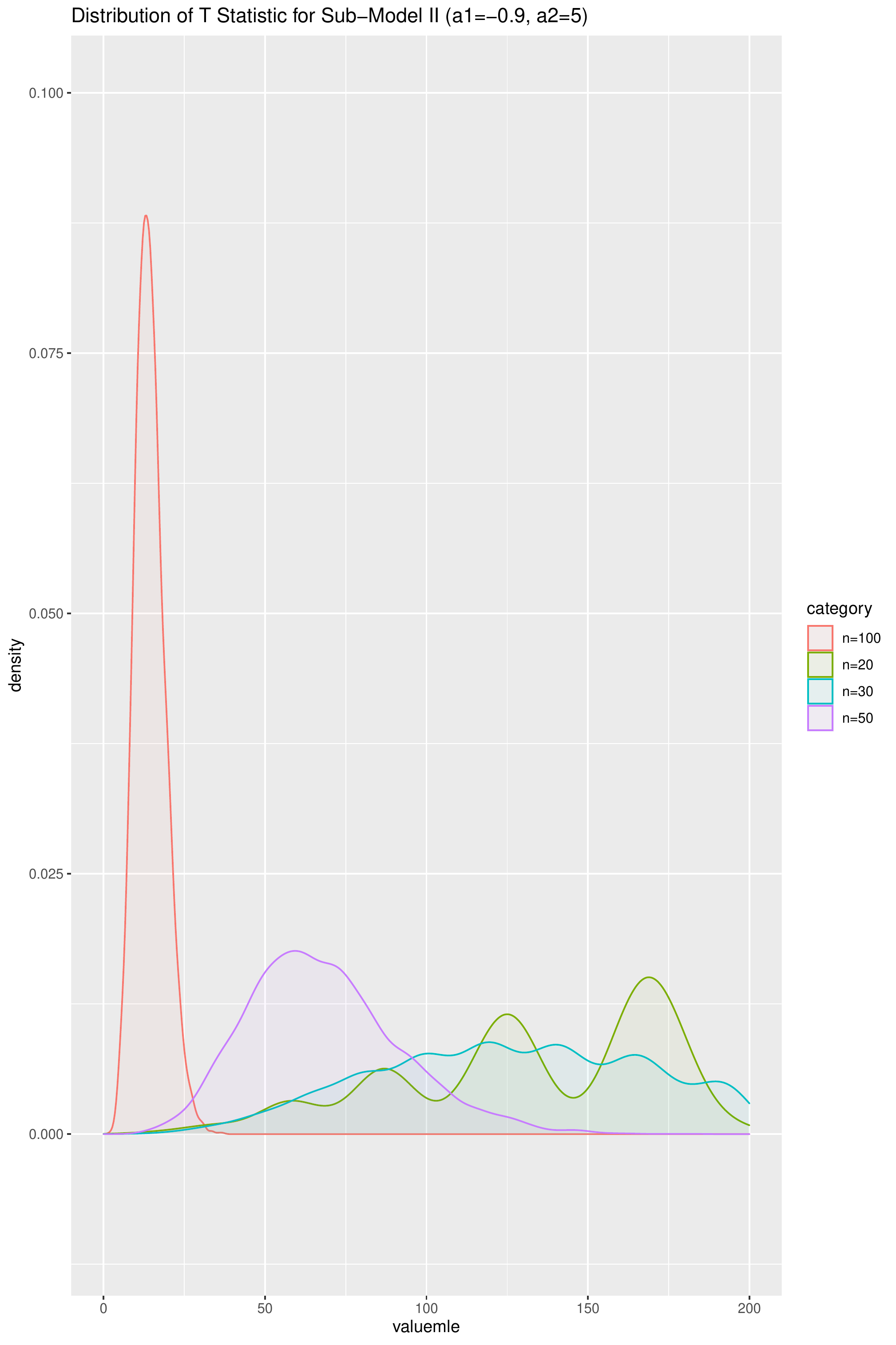}}
		\caption{$T^{(SII)}_{P,n,w}$}
		\label{EX3SII}
	\end{figure}

	Now, we consider GoF using p.g.f. (c.f. Mu\~noz and Gamero \cite{nj14}) $T^{(.)}_{P,n,w}$ with depends on the underlying weight functions. We refer to Table \ref{PGFEX1}, \ref{PGFEX2}, \ref{EX3FULL} and Figure \ref{EX1T}, \ref{EX2T}, \ref{EX3FULL},  \ref{EX3SI}, \ref{EX3SII} for small and large sample distribution of the test statistic and its quantile values for the full and its sub-models. 
	
	To better understand the behaviour of the test statistic, we examined the impact of different weights at $a_1 =-0.9,-0.5,-0.01,0.5,3$ and $a_2 =-0.9,-0.5,-0.01,0.5,5$ on the test statistic.  According to the simulation study we make a remark that irrespective of the weight choosen the test are consistent and stable for  moderetely large sample sizes. Also, note that for the increasing sample size the distribution of the test statistic are less variant and are showen to be consistent.

	\begin{sidewaystable}
		\centering
		\caption{Example 1}
		\label{PGFEX1}
		\resizebox{25cm}{!}{%
			\begin{tabular}{cc|c|c|c|c|l}
				\cline{3-6}
				& & \multicolumn{4}{ c| }{Sample size  $(0.5\% , 2.5\%, 5\%; 95\%, 97.5\%, 99.5\%  )$ } \\ \cline{3-6}
				& & $n=20$ & $n=30$ & $n=50$ & $n=100$ \\ \cline{1-6}
				\multicolumn{1}{ |c  }{\multirow{3}{*}{$T_{P,n,w}$} } &
				\multicolumn{1}{ |c| }{Full Model} & $(7.085,8.664,9.632;30.382,33.138,38.773)$ & $(5.117,5.793,6.190;15.599,16.820,20.041)$ & $(2.666,3.002,3.230;6.908,7.490,8.600)$ & $(0.885,0.960,1.005;1.635,1.720,1.911)$   \\ \cline{2-6}
				\multicolumn{1}{ |c  }{}                        &
				\multicolumn{1}{ |c| }{Sub Model I} & $(10.808,12.772,13.980;40.614,42.949,49,598)$ & $(6.560,7.770,8.455;21.579,23.463,28.388)$ & $(3.535,4.008,4.009,4.346;8.914,9.507,10.722)$ & $(0.958,1.07,1.126;2.093,2.217,2.513)$      \\ \cline{2-6}
				\multicolumn{1}{ |c  }{}                        &
				\multicolumn{1}{ |c| }{Sub Model II} & $(21.270,29.847,35.064;147.186,164.792,197.586)$ & $(16.4633,21.968,24.682;78.691,86.521,103.263)$ & $(12.906,15.543,16.812;33.672,36.303,41.303)$ & $(3.191,3.582,3.951;7.981,8.418,9.304)$  \\ \cline{1-6}
			\end{tabular}  
		}
		
		\centering
		\caption{Example 2}
		\label{PGFEX2}
		\resizebox{25cm}{!}{%
			\begin{tabular}{cc|c|c|c|c|l}
				\cline{3-6}
				& & \multicolumn{4}{ c| }{Sample size  $(0.5\% , 2.5\%, 5\%; 95\%, 97.5\%, 99.5\%  )$ } \\ \cline{3-6}
				& & $n=20$ & $n=30$ & $n=50$ & $n=100$ \\ \cline{1-6}
				\multicolumn{1}{ |c  }{\multirow{3}{*}{$T_{P,n,w}$} } &
				\multicolumn{1}{ |c| }{Full Model}& $(24.003,29.695,32.916;109.461,120.825,143.666)$ & $(18.688,21.358,23.045;56.043,60.840,72.928)$ & $(9.814,11.380,12.424;18.230,27.670,33.967)$ & $(3.340,3.758,3.964;6.958,7.362,8.115)$  \\ \cline{2-6}
				\multicolumn{1}{ |c  }{}                        &
				\multicolumn{1}{ |c| }{Sub Model I}& $(35.178,41.764,46.663;153.635,166.724,205.881)$ & $(21.55443,26.233,28.611,81.378,90.566,111.364)$ & $(12.925,14.971,16.330;36.434,39.121,44.605)$ & $(3.735,4.290,4.546;9.0899.648,11.126)$     \\ \cline{2-6}
				\multicolumn{1}{ |c  }{}                        &
				\multicolumn{1}{ |c| }{Sub Model II} & $(83.950,120.662,150.705;750.7120,854.663,1027.723)$ & $(66.689,94.637,108.678;398.689,44.015,540523)$ & $(51.414,64.302,72.087;169.210,184.953,213.675)$ & $(13.119,15.157,16.856;38.858,41.334,46.675)$  \\ \cline{1-6}
			\end{tabular}  

		}
		
		\centering
		\caption{$T^{(.)}_{P,n,w}$}
		\label{EX3FULLT}
		\resizebox{25cm}{!}{%
			\begin{tabular}{cc|c|c|c|c|l}
				\cline{3-6}
				& & \multicolumn{4}{ c| }{Sample size  $(0.5\% , 2.5\%, 5\%; 95\%, 97.5\%, 99.5\%  )$ } \\ \cline{3-6}
				& & $n=20$ & $n=30$ & $n=50$ & $n=100$ \\ \cline{1-6}
				\multicolumn{1}{ |c  }{\multirow{5}{*}{$T_{P,n,w}$} } &
				\multicolumn{1}{ |c| }{($a_1=-0.9,a_1=-0.9 $)}& $(30.918,43.773,54.367;455.914,527.123,1023.20345)$ & $(25.182,33.193,38.630;204.290,414.503)$ & $(13.956,17.766,20.112;85.184,103.305,140.462)$ & $(4.543,5.389,5.885;18.364,22.210,29.855)$  \\ \cline{2-6}
				\multicolumn{1}{ |c  }{}                        &
				\multicolumn{1}{ |c| }{($a_1=-0.01,a_1=-0.01 $)}& $(12.663,18.047,19.808;69.013,77.235,97.099)$ & $(9.193,11.206,12.488;36.494,39.859,47.084)$ & $(4.981,5.839,6.373;15.345,16.561,19.420)$ & $(1.774,1.951,2.060;3.755,3.994,4.461)$     \\ \cline{2-6}
				\multicolumn{1}{ |c  }{}                        &
				\multicolumn{1}{ |c| }{($a_1=0.01,a_1=0.01 $)} & $(10.923,13.726,15.786;54.873,61.512,75.468)$ & $(7.864,9.473,10.465;29.302,31.877,37.684)$ & $(4.216,4.899,5.315,12.289,13.215,15.395)$ & $(1.506,1.645,1.734;3.037,3.213,3.570)$  \\ \cline{2-6}
				\multicolumn{1}{ |c  }{}                        &
				\multicolumn{1}{ |c| }{($a_1=0.5,a_1=0.5 $)} & $(8.400,10.410,11.925;40.708,45.527,54.879)$ & $(5.935,7.091,7.752;20.261,21.926,25.859)$ & $(3.122,3.598,3.884;8.470,9.118,10.522)$ & $(1.120,1.213,1.271;2.100,2.210,2.445)$  \\ \cline{2-6}
				\multicolumn{1}{ |c  }{}                        &
				\multicolumn{1}{ |c| }{($a_1=3,a_1=5$)} & $(3.066,3.557,3.851;10.352,11.398,13.679)$ & $(3.032,3.353,3.566;6.451,6.745,7.330)$ & $(1.480,1.618,1.677;2.695,2.832,3.089)$ & $(0.235,0.259,0.365;0.533,0.551,0.584)$  \\ \cline{2-6}
				\multicolumn{1}{ |c  }{}                        &
				\multicolumn{1}{ |c| }{($a_1=-0.9,a_1=5 $)} & $(16.295,26.760,36.107;235.847,262.617,326.237)$ & $(14.523,22.047,27.541,124.208,137.808,165.083)$ & $(0.213,12.853,15.190;52.653,57.075,67.911)$ & $(8.825,0.945,1.028;8.050,9.502,12.366)$  \\ \cline{1-6}
			\end{tabular}  
		}
		
		\centering
		\caption{Distribution of the $FI^{(.)}_{n}$}
		\label{DIT}
		\resizebox{25cm}{!}{%
			\begin{tabular}{cc|c|c|c|c|c|l}
				\cline{3-7}
				& & \multicolumn{4}{ c| }{Sample size  $(0.5\% , 2.5\%, 5\%; 95\%, 97.5\%, 99.5\%  )$ } \\ \cline{3-7}
				& & $n=20$ & $n=30$ & $n=50$ & $n=100$ & $n=500$ \\ \cline{1-7}
				\multicolumn{1}{ |c   }{\multirow{3}{*}{$FI^{(.)}_{n}$} } &
				\multicolumn{1}{ |c| }{Full Model}& $(-7.818,-5.701,-4.665;5.033,6.900,11.231)$ & $(-7.760,-5.762,-4.698;5.084,6.703,10.598)$ & $(-7.166,-5.530,-4.748;5.034,6.544,9.666)$ & $(-7.167,-5.530,-4.748;5.034,6.544,9.666)$ & $(-6.897,-5.676,-4.841;5.047,6.063,8.20)$ \\ \cline{2-7}
				\multicolumn{1}{ |c  }{}                        &
				\multicolumn{1}{ |c| }{Sub Model I}& $(-9.537,-7.749,-6.766;9.093,11.852,17.648)$ & $(-9.947,-8.140,-7.131;8.825,11.500,15.842)$ & $(-10.477,-8.560,-7.505;8.725,10.887,15.936)$ & $(-11.018,-8.610,-7.507;9.203,11.223,15.689)$ & $(-11.759,-9.034,-7.802;8.365,10.293,13.741)$     \\ \cline{2-7}
				\multicolumn{1}{ |c  }{}                        &
				\multicolumn{1}{ |c| }{Sub Model II} & $(-15.205,-12.628,-11.181;13.710,18.740,29.800)$ & $(-15.917,-13.277,-11.695;13.924,17.255,28.748)$ & $(-17.191,-14.088,-12.162;13.748,17.947,27.665)$ & $(-18.434,-15.108,-13.076;12.997,15.976,21.976)$  & $(-20.647,-16.782,-14.478,11.629,14.497,21.093)$\\ \cline{1-7}
			\end{tabular}  

		}
	\end{sidewaystable}


    \subsubsection{K\&K Method}
    In the following we discuss finite, large and asymptotic distribution of the  test statistics $T^{(.)}_{PN}(t_1,t_2)$ and $T^{(.)}_{SPN}$  (c.f. Section 3.2.2).  
    Here, we limit our analysis to sub-models of the bivariate pseudo-Poisson model, and statistical inference or parameter estimation are well defined (see Section $7$ in Arnold and Manjunath \cite{am21}). However, in Sub-Models I and II, both the method of moments and the maximum likelihood estimators coincides. Hence, due to invariance property of the maximum likelihood estimator the defined test statistic asymptoptically follows standard normal with variance is will be inverse of Fisher information matrix.

    Now, we consider bootstrapping size of $B=5000$ with varying sample size of \\ $n=20,30,50,100,500$ at different $t_i = \pm 0.01,\pm 0.5,\pm 0.9$, $i=1,2$.
    	
\begin{figure}[H]
   	\subfloat[$t_1=-0.9,t_2=-0.9$]{\includegraphics[width=5cm,height=2.5cm]{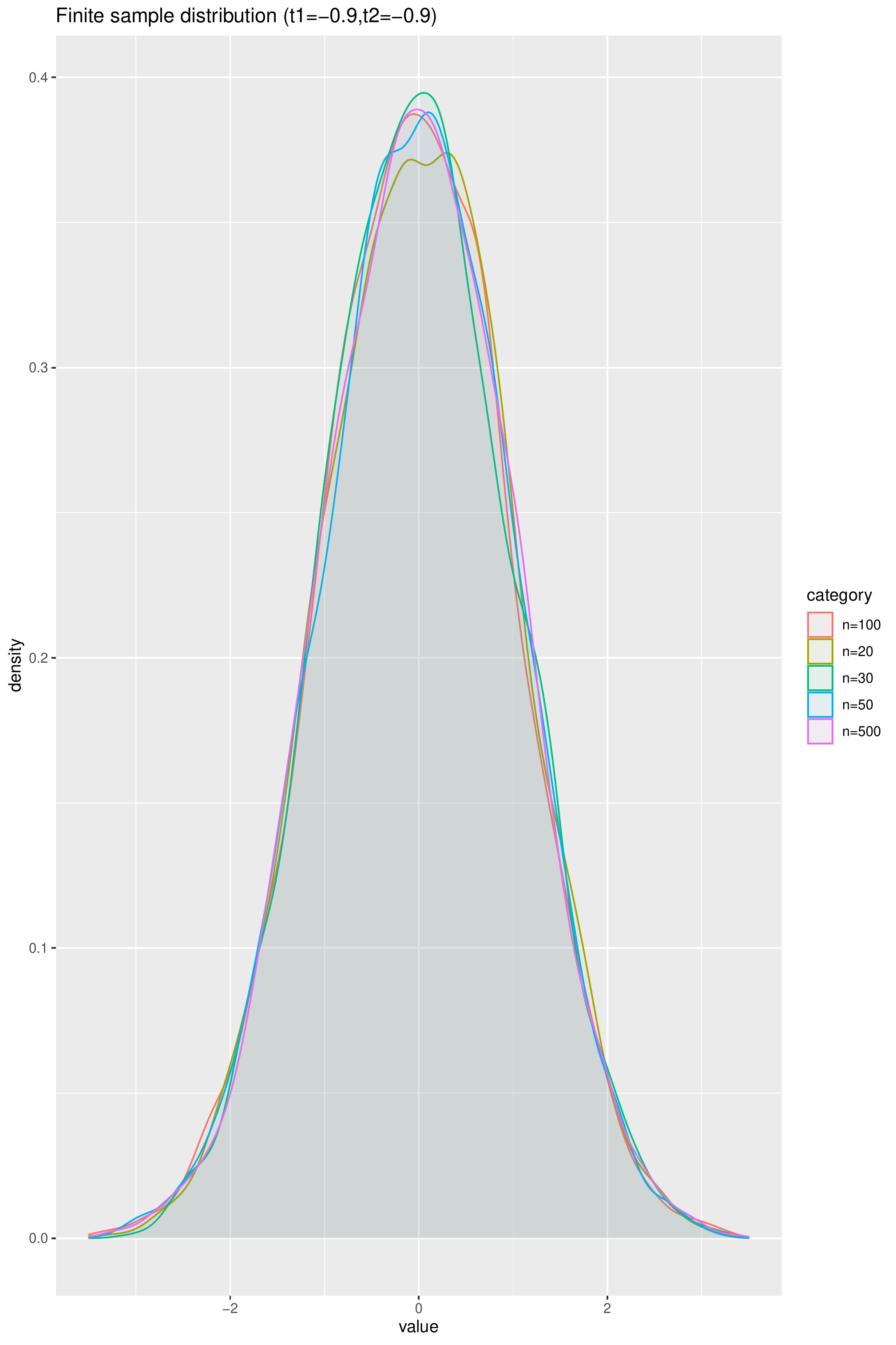}} 
   	\subfloat[$t_1=-0.5,t_2=-0.5$]{\includegraphics[width=5cm,height=2.5cm]{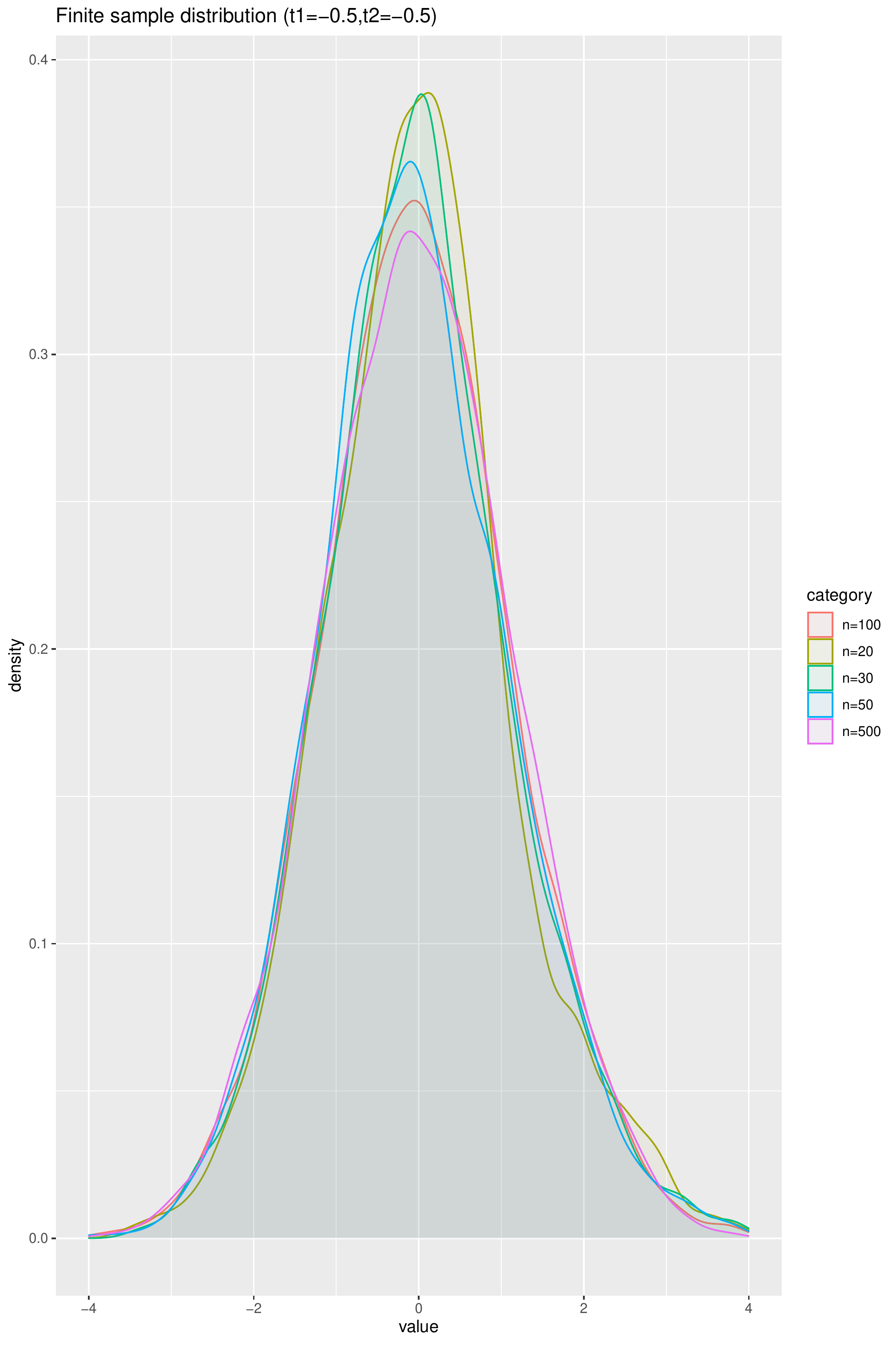}}
   	\subfloat[$t_1=-0.01,t_2=-0.01$]{\includegraphics[width=5cm,height=2.5cm]{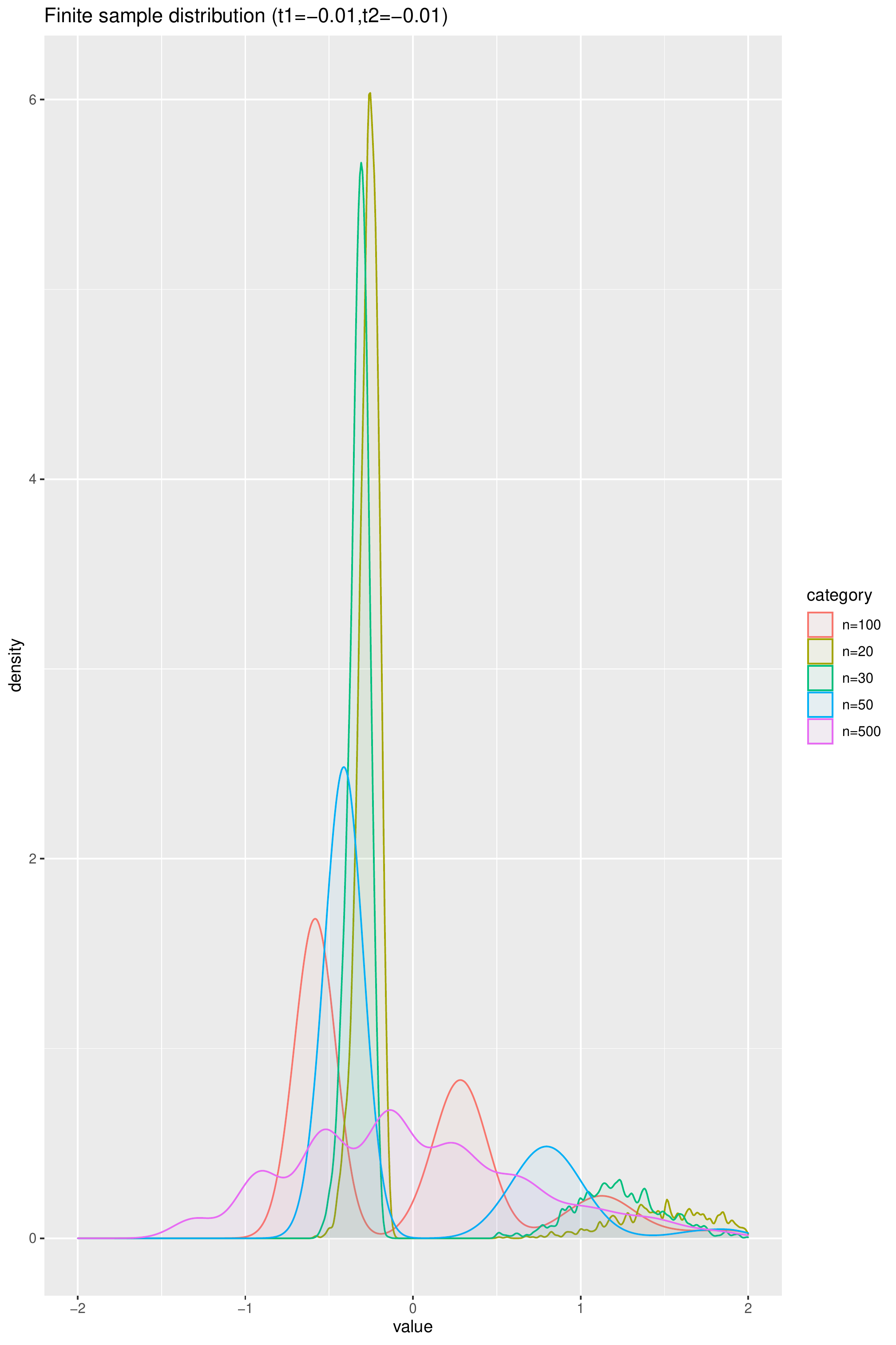}} \\
   	\subfloat[$t_1=0.01,t_2=0.01$]{\includegraphics[width=5cm,height=2.5cm]{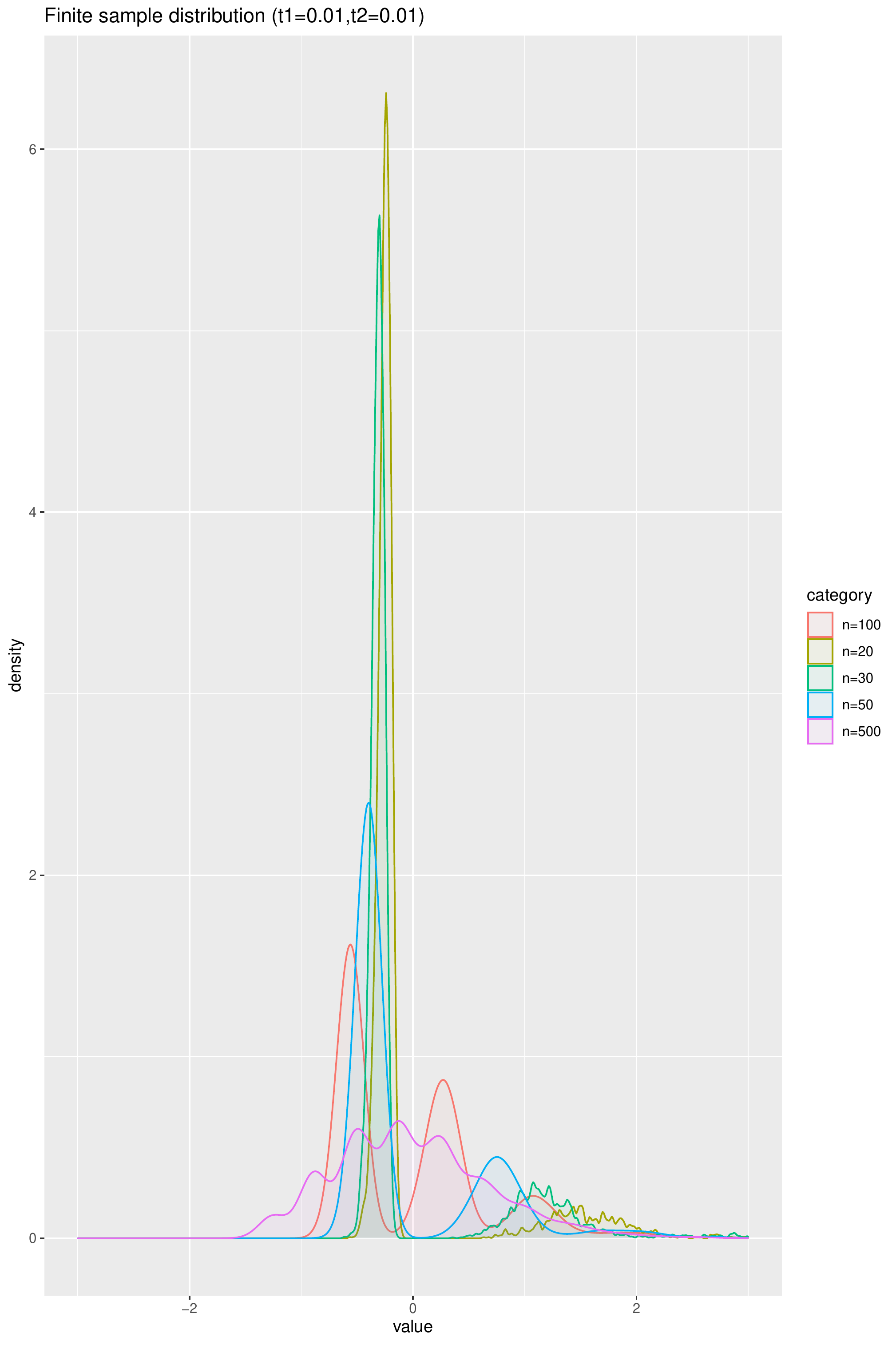}} 
   	\subfloat[$t_1=0.5,t_2=0.5$]{\includegraphics[width=5cm,height=2.5cm]{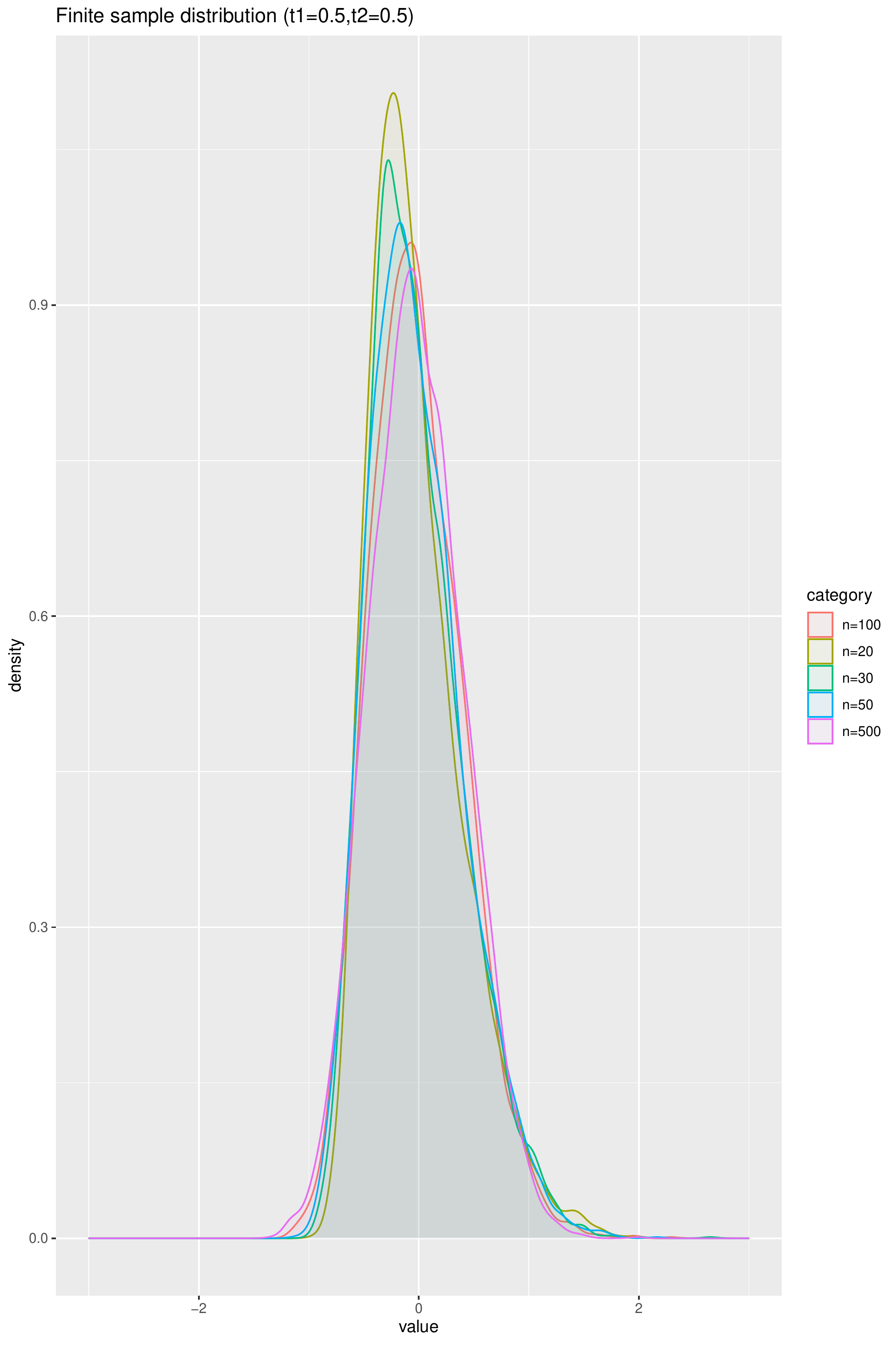}}
   	\subfloat[$t_1=0.9,t_2=0.9$]{\includegraphics[width=5cm,height=2.5cm]{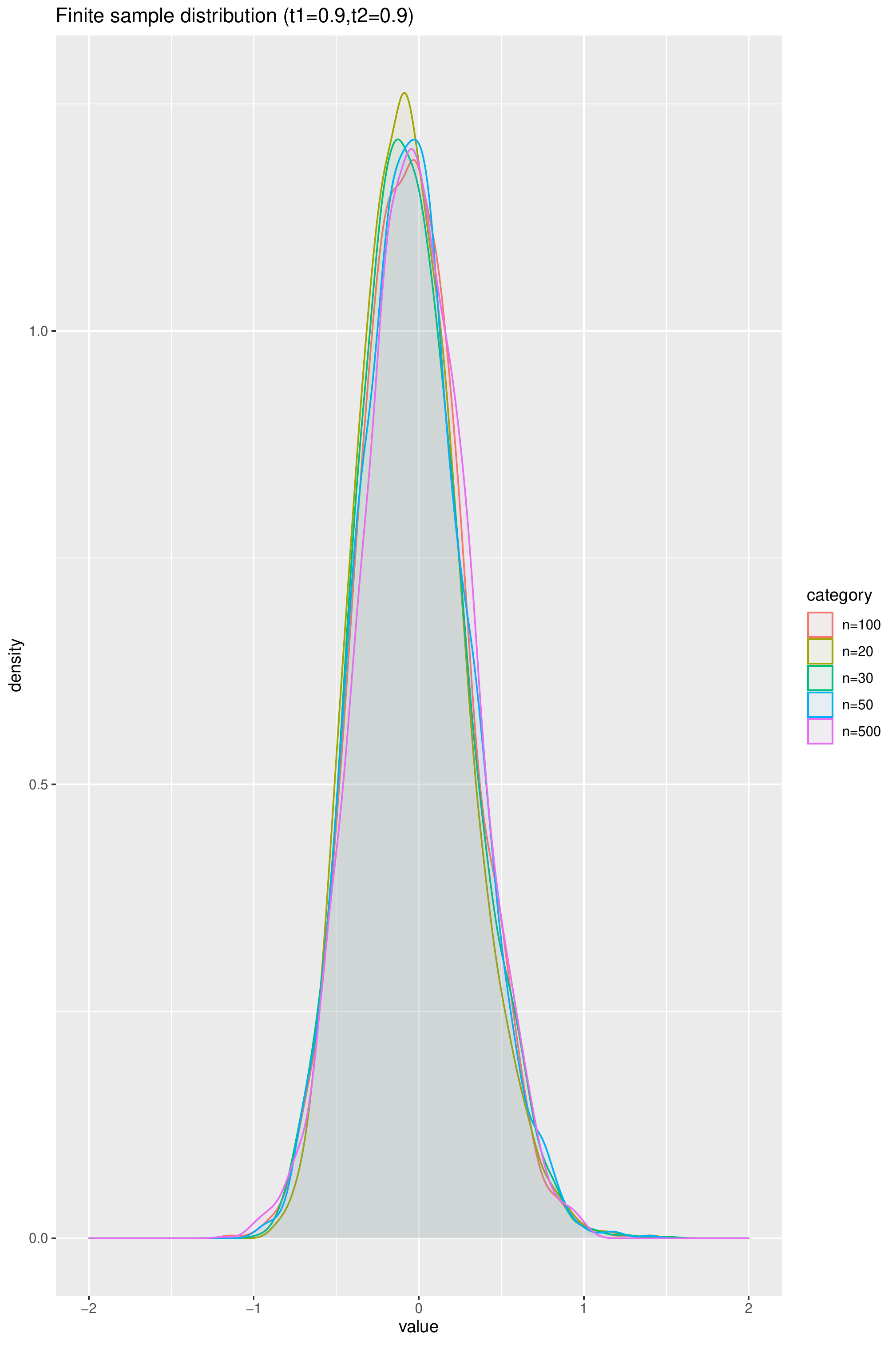}}
   	\caption{Finite sample distribution of $T_{PN}^{(SI)}$ for the Sub-Model I}
   	\label{KK Sub-Model I}

   	\subfloat[$t_1=-0.9,t_2=-0.9$]{\includegraphics[width=4cm,height=2.5cm]{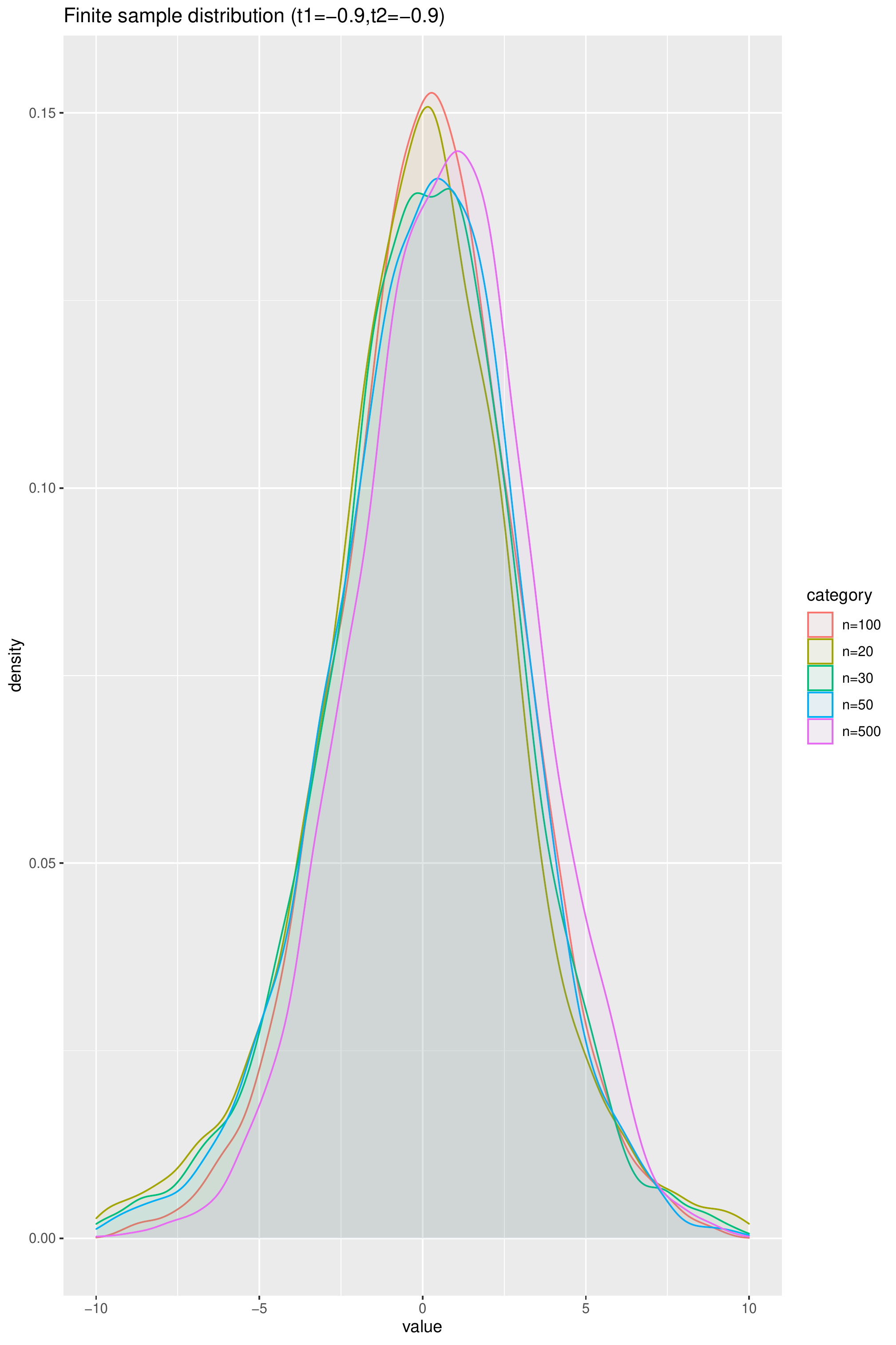}} 
   	\subfloat[$t_1=-0.01,t_2=-0.01$]{\includegraphics[width=4cm,height=2.5cm]{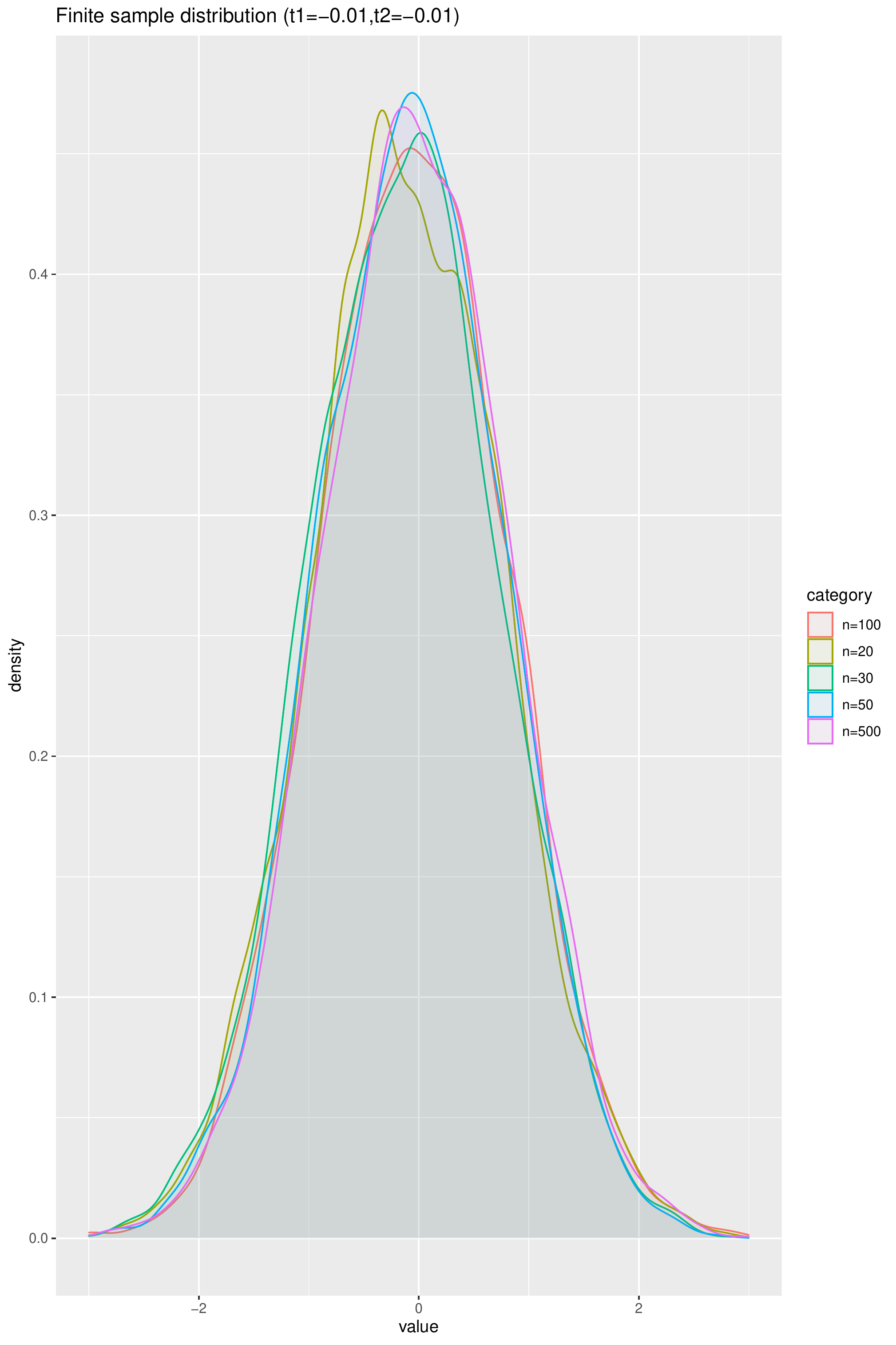}} 
   	\subfloat[$t_1=0.01,t_2=0.01$]{\includegraphics[width=4cm,height=2.5cm]{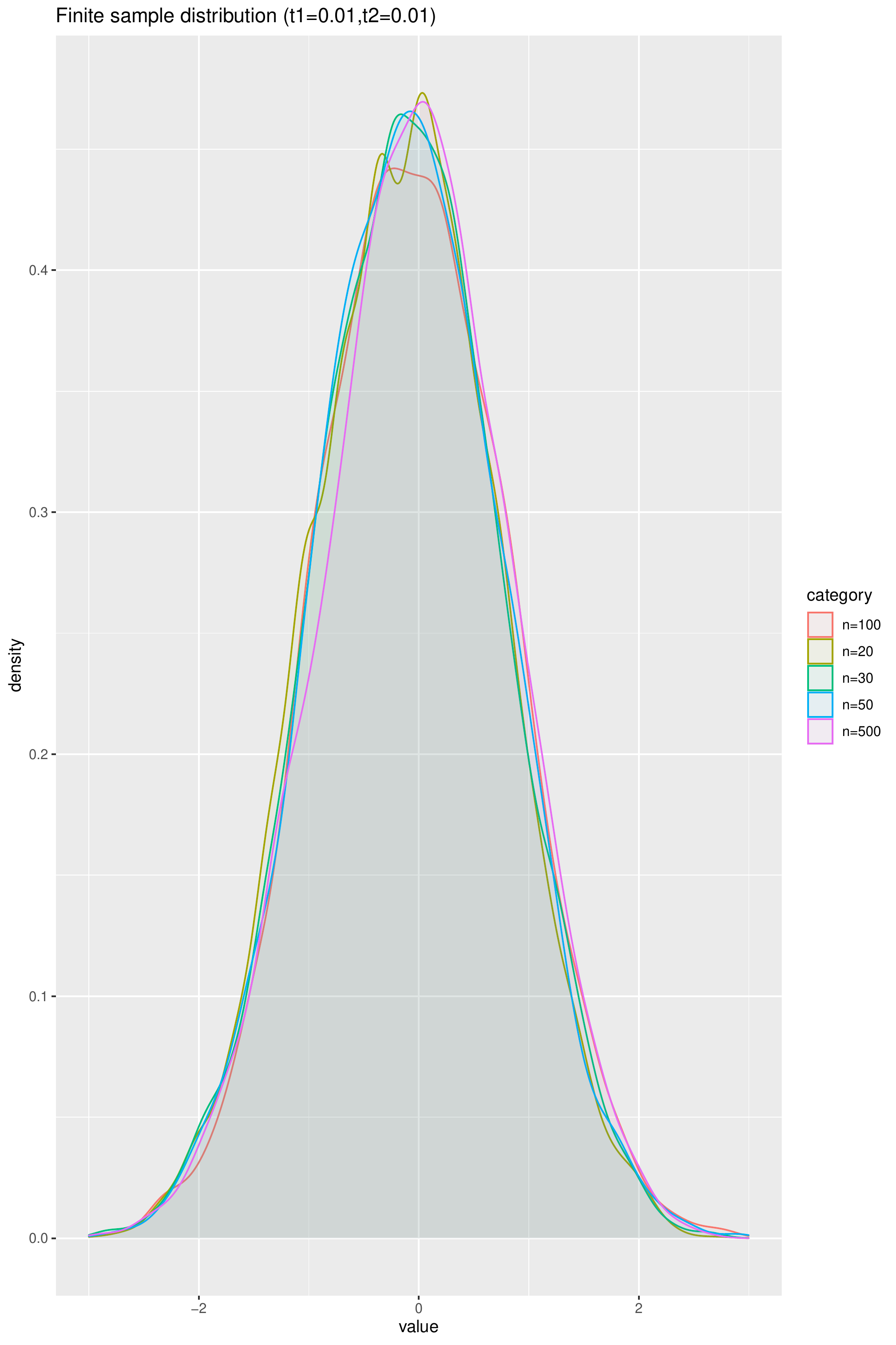}} 
   	\subfloat[$t_1=0.9,t_2=0.9$]{\includegraphics[width=4cm,height=2.5cm]{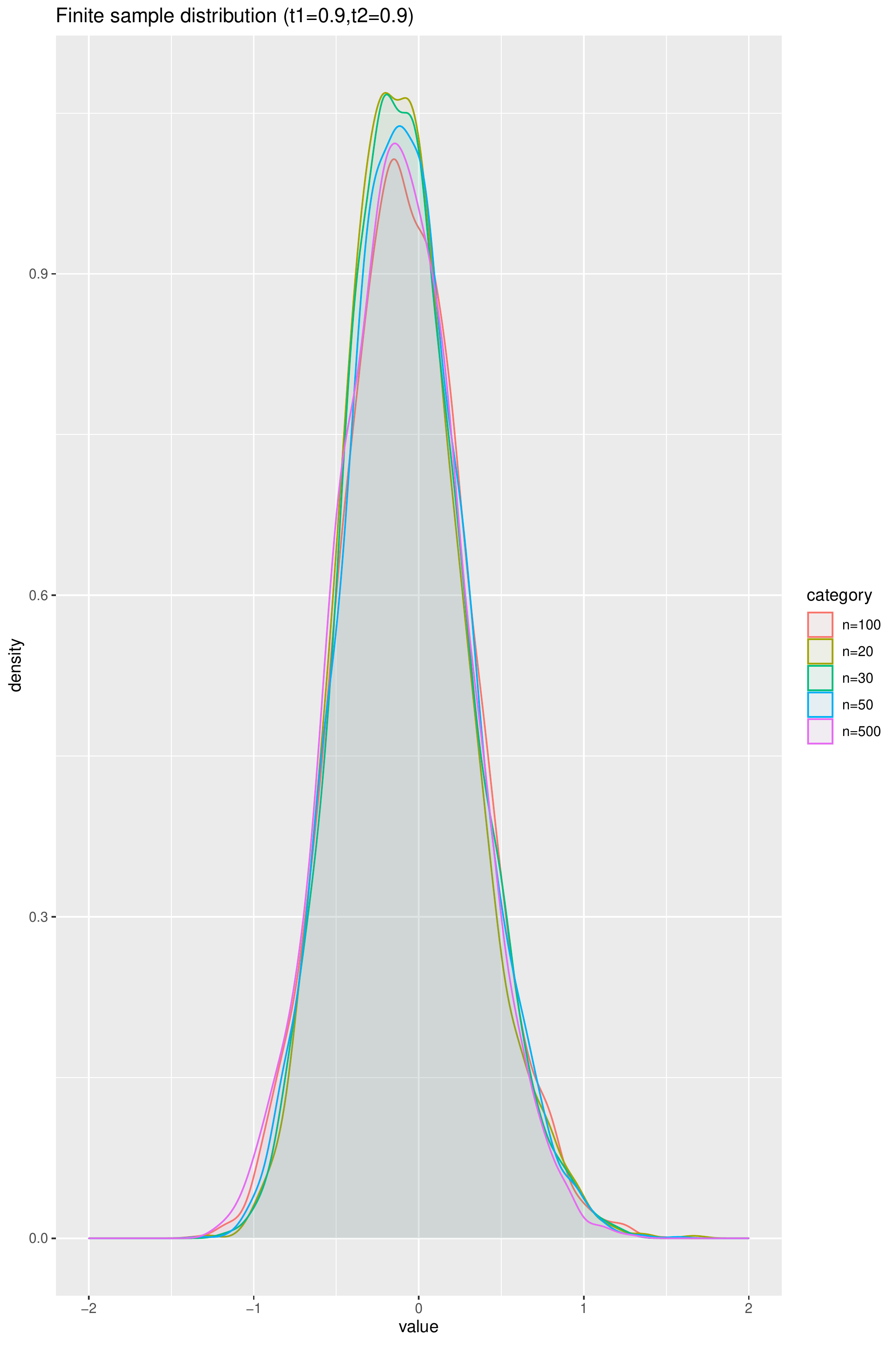}}
   	\caption{Finite sample distribution of $T_{PN}^{(SII)}$ for the Sub-Model II}
   	\label{KK Sub-Model II}
   \end{figure}

   \begin{enumerate}
   	\item[Sub-Model I] (i.e. $\lambda_2=\lambda_3$) The coresponding quantile values and density plots refer to Table \ref{KK Sub-Model I} and Figuare \ref{KKsingleSI}, respectively.
   	\item[Sub-Model II] (i.e. $\lambda_2=0$) The coresponding quantile values and density plots refer to Table \ref{KK Sub-Model II} and Figuare \ref{KKsingleSII}, respectively.
   \end{enumerate}
   
 According to the simulation study it has been observed that whenever $t_i$ is closer to zero the empirical critical points are closer to the standard normal qunatile values.  It has been recommended that the $t$ values are to be chosen  either in the neighborhood of zero or well spanned in the interval $(-1,1)$ to have consistency in the tests.

 Note that from the Table \ref{KKsingleSI} \& \ref{KKsingleSII} and also from Figure \ref{KK Sub-Model I} \& \ref{KK Sub-Model II} K\&K-method finite sample distribution depends on the selected values for $(t_1,t_2)$. In particular, at $t_1 =-0.5 (0.5)$ and $t_2 =-0.5 (0.5)$ K \& K statistic distributions are inconsistent.  Hence, we consider the test statistic $T^{(.)}_{SPN}$(defined in (3.18)) such that the test support completely deponds on  complete span of $t_i$-values, $i=1,2,$.   For an illustration of the proposed test we are analysing the finite sample distribution of the  test statistic which is computed with varying $t_1$ and $t_2$ from $-0.99$ to $0.99$ at an increment of $0.01$. 
 
 Finally, it has been argued in Feiyan Chen \cite{cf13} that such tests are robust to the choice of alternatives and that the performance of the test is better than the K\&K test because it also spanned the entire interval of $(-1,1) \times (-1,1)$.

\begin{figure}[H]
	\subfloat[$T^{(SI)}_{SPN}$]{\includegraphics[width=5cm,height=2.5cm]{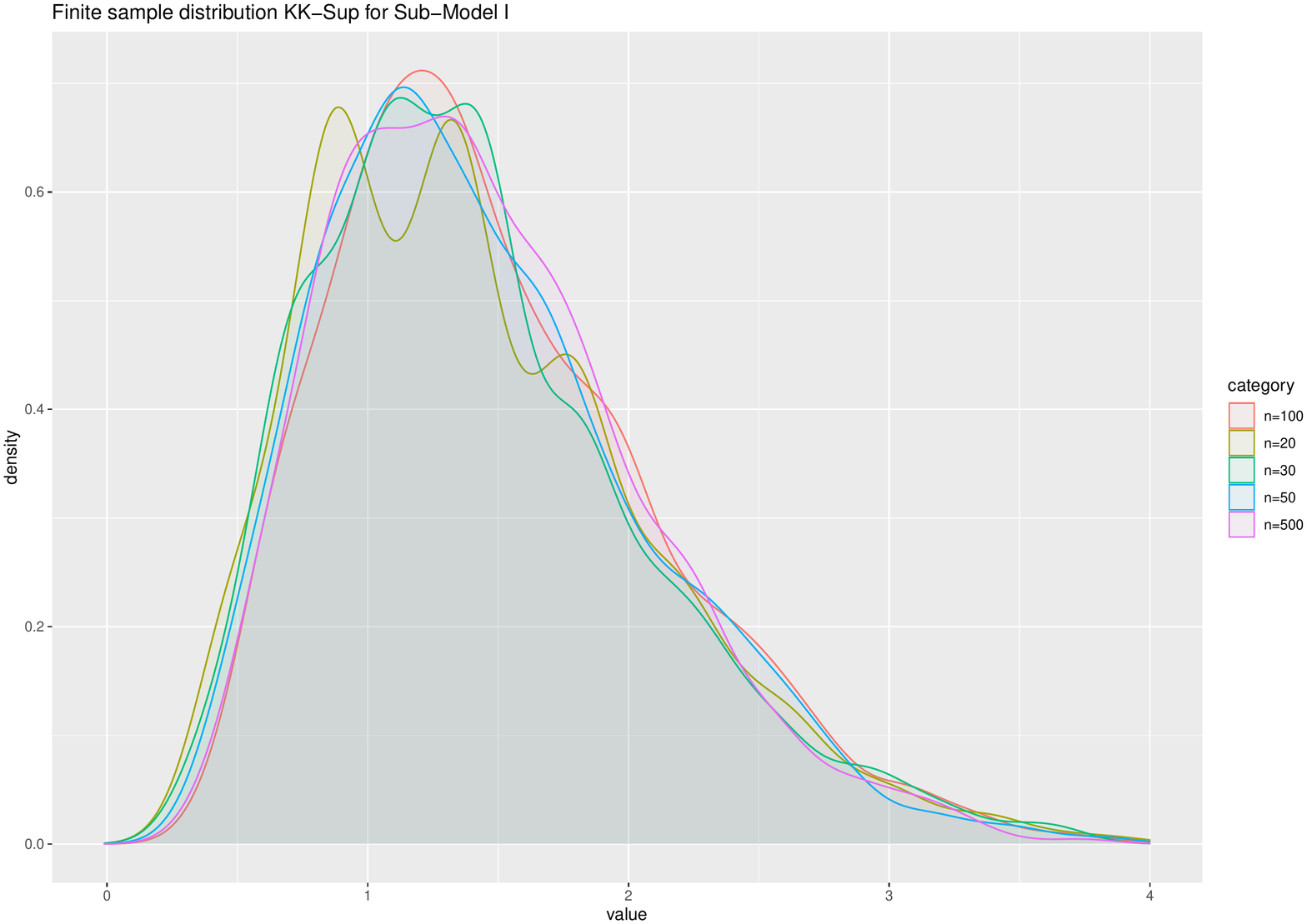}} 
	\subfloat[$T^{(SII)}_{SPN}$]{\includegraphics[width=5cm,height=2.5cm]{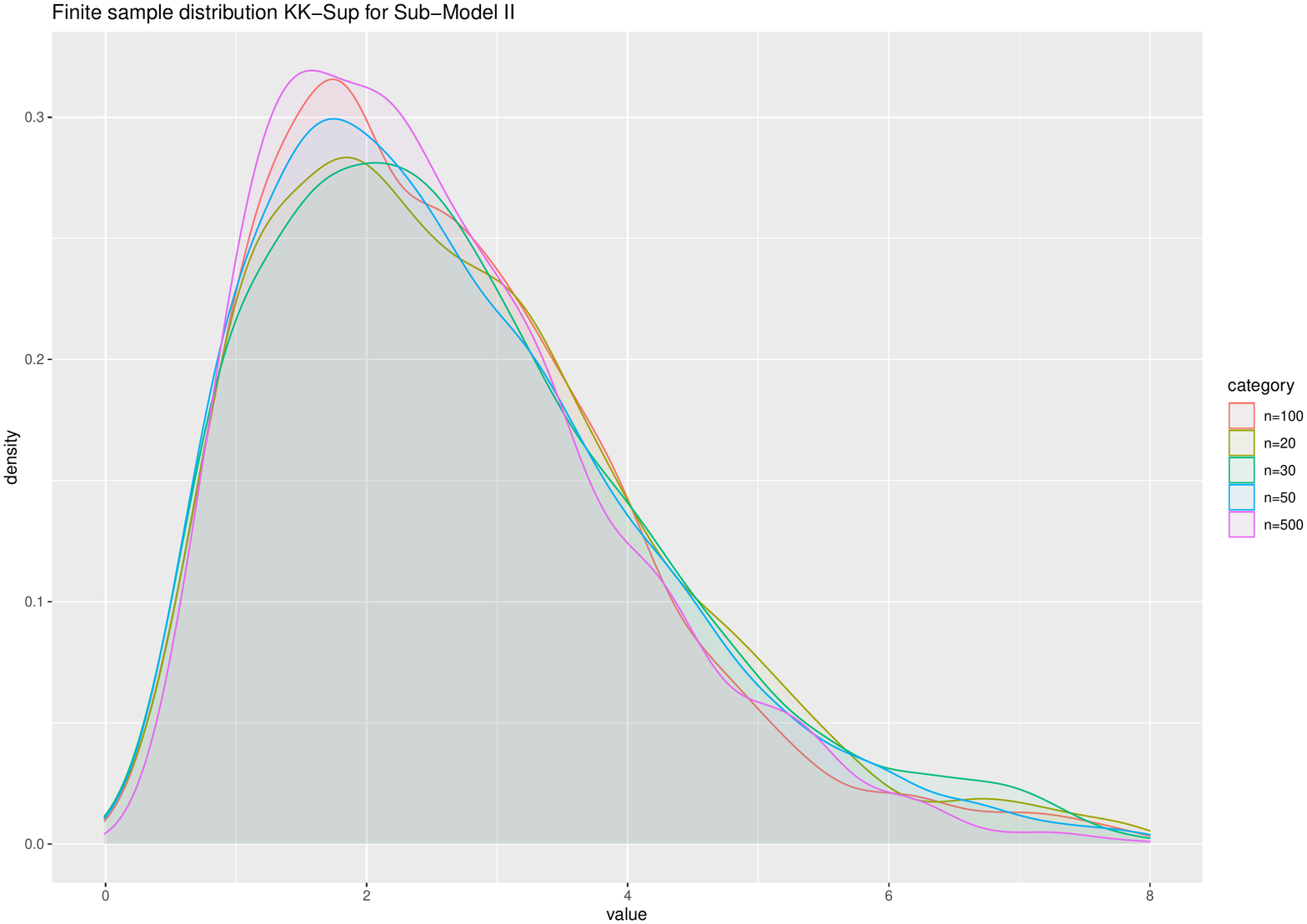}}
	\caption{Finite sample distribution of the supremum of absolute deviation for the K\&K-test statistic.}
	\label{KKSup}
\end{figure}

  We refer to Table \ref{KKSupQSI} and Figure \ref{KKSup}  for the quantile values and frequency distribution of the test statistic, respectively. The test statistic's behaviour is more stable and consistent for small and moderately large samples.

   \begin{sidewaystable}
   	\centering
   	\caption{$T_{PN}^{(SI)}$  distribution for the Sub-Model I}
   	\label{KKsingleSI}
   \resizebox{25cm}{!}{%
   \begin{tabular}{cc|c|c|c|c|c|l}
   	\cline{3-7}
   	& & \multicolumn{5}{ c| }{Sample size  $(0.5\% , 2.5\%, 5\%; 95\%, 97.5\%, 99.5\%  )$ } \\ \cline{3-7}
   	& & $n=20$ & $n=30$ & $n=50$ & $n=100$ & $n=500$\\ \cline{1-7}
   	\multicolumn{1}{ |c  }{\multirow{6}{*}{$T^{(SI)}_{PN}$} } &
   	\multicolumn{1}{ |c| }{$t_1=-0.9$, $t_2=-0.9$} & $(-2.503,-1.953,-1.652; 1.660,1.912,2.522)$ & $(-2.447,-1.886,-1.626;1.637,1.981,2.528)$ & $(-2.582,-1.975,-1.669;1.619,1.928,2.548)$ & $(-2,622,-2.011,-1.674;1.620,1.912,2.543)$ &   $(-2.582,-1.910,-1.612;1.634,1.948,2.560)$  \\ \cline{2-7}
   	\multicolumn{1}{ |c  }{}                        &
   	\multicolumn{1}{ |c| }{$t_1=-0.5$, $t_2=-0.5$} & $(-2.839,-2.187,-1.811,1.953,2.458,3.196)$ & $(-2.986,-2.168,-1.825,2.031,2.523,3.390)$ & $(-2.917,-2.189,-1.825;1.927,2.412,3.342)$ & $(-2.822,-2.198,-1.872;1.988,2.364,3.106)$   &  $(-3.021,-2.156,-1.845;1.947,2.270,3.078)$   \\ \cline{2-7}
   		\multicolumn{1}{ |c  }{}                        &
   	\multicolumn{1}{ |c| }{$t_1=-0.01$, $t_2=-0.01$} & $(-0.464,-0.402,-0.376;1.701,1.976,2.981)$ & $(-0.519,-0.456,-0.428;1.454,1.688,2.912)$ & $(-0.593,-0.535,-0.510;1.407,2.036,2.936)$ & $(-0.747,-0.695,-0.668;1.332,1.734,2.453)$ &   $(-1.388,-1.263,-0.997;1.228,1.553,2.220)$  \\ \cline{2-7}
   		\multicolumn{1}{ |c  }{}                        &
   	\multicolumn{1}{ |c| }{$t_1=0.01$, $t_2=0.01$} & $(-0.452,-0.392,-0.362;1.620,1.870,2.456)$ & $(-0.492,-0.440,-0.415;1.352,1.612,2.853)$ & $(-0.574,-0.520,0.492;1.231,1.867,2.456)$ & $(-0.713,-0.665,-0.641,1.240,1.645,2.347)$ &   $(-1.310,-1.216,-0.958;1.223,1.529,2.049)$      \\ \cline{2-7}
   		\multicolumn{1}{ |c  }{}                        &
   	\multicolumn{1}{ |c| }{$t_1=0.5$, $t_2=0.5$} & $(-0.751,-0.640,-0.581;0.771,0.977,1.423)$ & $(-0.821,-0.692,-0.619;0.760,0.966,1.281)$ & $(-0.863,-0.725,-0.634;0.753,0.9358,1.304)$ & $(-0.944,-0.753,-0.654;0.710,0.892,1.238)$ &   $(-1.008,-0.792,-0.683;0.707,0.853,1.111)$     \\ \cline{2-7}
   		\multicolumn{1}{ |c  }{}                        &
   	\multicolumn{1}{ |c| }{$t_1=0.9$, $t_2=0.9$} & $(-0.750,-0.624,-0.541;0.498,0.625,0.880)$ & $(-0.733,-0.622,-0.541;0.516,0.630,0.854)$ & $(-0.776,-0.625,-0.543;0.545,0.643,0.884)$ & $(-0.833,-0.657,-0.550;0.539,0.654,0.867)$ &   $(-0.829,-0.648,-0.552;0.554,0.669,0.842)$     \\ \cline{1-7}
   	\end{tabular}  
}

\centering
\caption{$T_{PN}^{(SII)}$  distribution for the Sub-Model II}
\label{KKsingleSII}
	\resizebox{25cm}{!}{%
		\begin{tabular}{cc|c|c|c|c|c|l}
			\cline{3-7}
			& & \multicolumn{5}{ c| }{Sample size  $(0.5\% , 2.5\%, 5\%; 95\%, 97.5\%, 99.5\%  )$ } \\ \cline{3-7}
			& & $n=20$ & $n=30$ & $n=50$ & $n=100$ & $n=500$\\ \cline{1-7}
			\multicolumn{1}{ |c  }{\multirow{6}{*}{$T^{(SII)}_{PN}$} } &
			\multicolumn{1}{ |c| }{$t_1=-0.9$, $t_2=-0.9$} & $(-17.388,-8.192,-5.956;5.168,7.046,13.129)$ & $(-13.272,-6.980,-5.286;4.862,5.972,10.505)$ & $(-9.952,-6.382,-5.020;4.653,5.737,7.871)$ & $(-7.590,-5.342,-4.345;4.612,5.568,7.522)$ &   $(-6.667,-4.788,-3.7800;5.200,5.930,7.723)$  \\ \cline{2-7}
			\multicolumn{1}{ |c  }{}                        &
			\multicolumn{1}{ |c| }{$t_1=-0.01$, $t_2=-0.01$} & $(-2.306,-1.700,-1.557;1.347,1.608,2.216)$ & $(-2.222,-1.803,-1.507;1.303,1.587,2.116)$ & $(-2.241,-1.726,-1.393;1.264,1.517,2.023)$ & $(-2.174,-1.679,-1.441;1.341,1.624,2.230)$  &   $(-2.203,-1.690,-1.390;1.362,1.588,2.173)$  \\ \cline{2-7}
			\multicolumn{1}{ |c  }{}                        &
			\multicolumn{1}{ |c| }{$t_1=0.01$, $t_2=0.01$} & $(-2.301,-1.740,-1.432;1.203,1.549,2.007)$ & $(-2.215,-1.802,-1.440;1.297,1.558,1.965)$ & $(-2.176,-1.733,-1.480;1.259,1.568,2.123)$ & $(-2.261,-1.680,-1.434;1.383,1.618,2.136)$ &   $(-2.219,-1.727,-1.436;1.363,1.624,2.065)$      \\ \cline{2-7}
	        \multicolumn{1}{ |c  }{}                        &
			\multicolumn{1}{ |c| }{$t_1=0.9$, $t_2=0.9$} & $(-0.927,-0.737,-0.647;0.580,0.750,1.014)$ & $(-0.913,-0.755,-0.657;0.572,0.714,0.986)$ & $(-0.950,-0.784,-0.669;0.590,0.723,0.981)$&$(-0.980,-0.825,-0.701,0.594,0.748,1.012)$ & $(-1.033,-0.853,-0.730;0.549,0.672,0.906)$     \\ \cline{1-7}
		\end{tabular}  
	}

\centering
\caption{$T^{(SI)}_{SPN}$ distribution for the Sub-Model I}
\label{KKSupQSI}
\resizebox{25cm}{!}{%
	\begin{tabular}{cc|c|c|c|c|c|l}
		\cline{3-7}
		& & \multicolumn{5}{ c| }{Sample size  $(0.5\% , 2.5\%, 5\%; 95\%, 97.5\%, 99.5\%  )$ } \\ \cline{3-7}
		& & $n=20$ & $n=30$ & $n=50$ & $n=100$ & $n=500$\\ \cline{1-7}
		\multicolumn{1}{ |c  }{\multirow{1}{*}{$T^{(SI)}_{SPN}$} } &
		 & $(0.380,0.450,0.528;2.647,2.987,3.005)$ & $(0.337,0.489,0.586;2.656,3.012,3.622)$ & $(0.401,0.511,0.593;2.623,2.863,3.567)$ & $(0.423,0.574,0.643;2.679,3.013,3.591)$ &   $(0.410,0.550,0.634;2.558,2.860,3.279)$  \\ \cline{1-7}
	\end{tabular}

}

\centering
\caption{$T^{(SII)}_{SPN}$ distribution for the Sub-Model II}
\label{KKSupQSII}
\resizebox{25cm}{!}{%
	\begin{tabular}{cc|c|c|c|c|c|l}
		\cline{3-7}
		& & \multicolumn{5}{ c| }{Sample size  $(0.5\% , 2.5\%, 5\%; 95\%, 97.5\%, 99.5\%  )$ } \\ \cline{3-7}
		& & $n=20$ & $n=30$ & $n=50$ & $n=100$ & $n=500$\\ \cline{1-7}
		\multicolumn{1}{ |c  }{\multirow{1}{*}{$T^{(SII)}_{SPN}$} } &
		& $(0.367,0.676,0.830;6.868,9.190,24.875)$ & $(0.422,0.651,0.771;6.121,7.016,11.323)$ & $(0.394,0.6200.783;5.559,6.506,9.039)$ & $(0.423,0.638,0.793;5.208,6.221,7.612)$ &   $(0.487,0.701,0.850;5.113,5.588,6.720)$  \\ \cline{1-7}
	\end{tabular}

}

\end{sidewaystable}

\subsubsection{GoF test free from alternative}

\begin{figure}[H]
	\subfloat[Full Model]{\includegraphics[width=5cm,height=2.5cm]{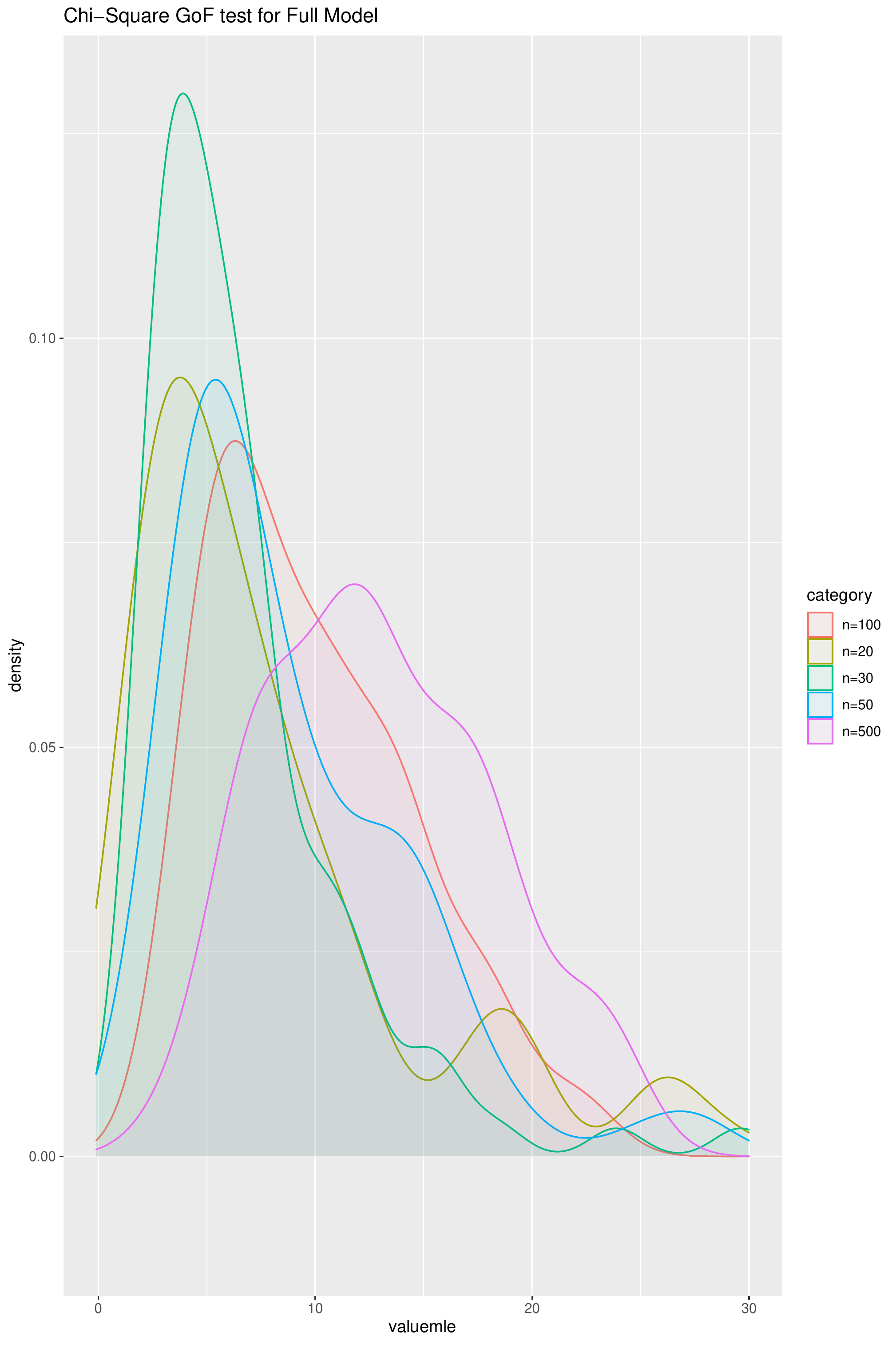}} 
	\subfloat[Sub-Model I]{\includegraphics[width=5cm,height=2.5cm]{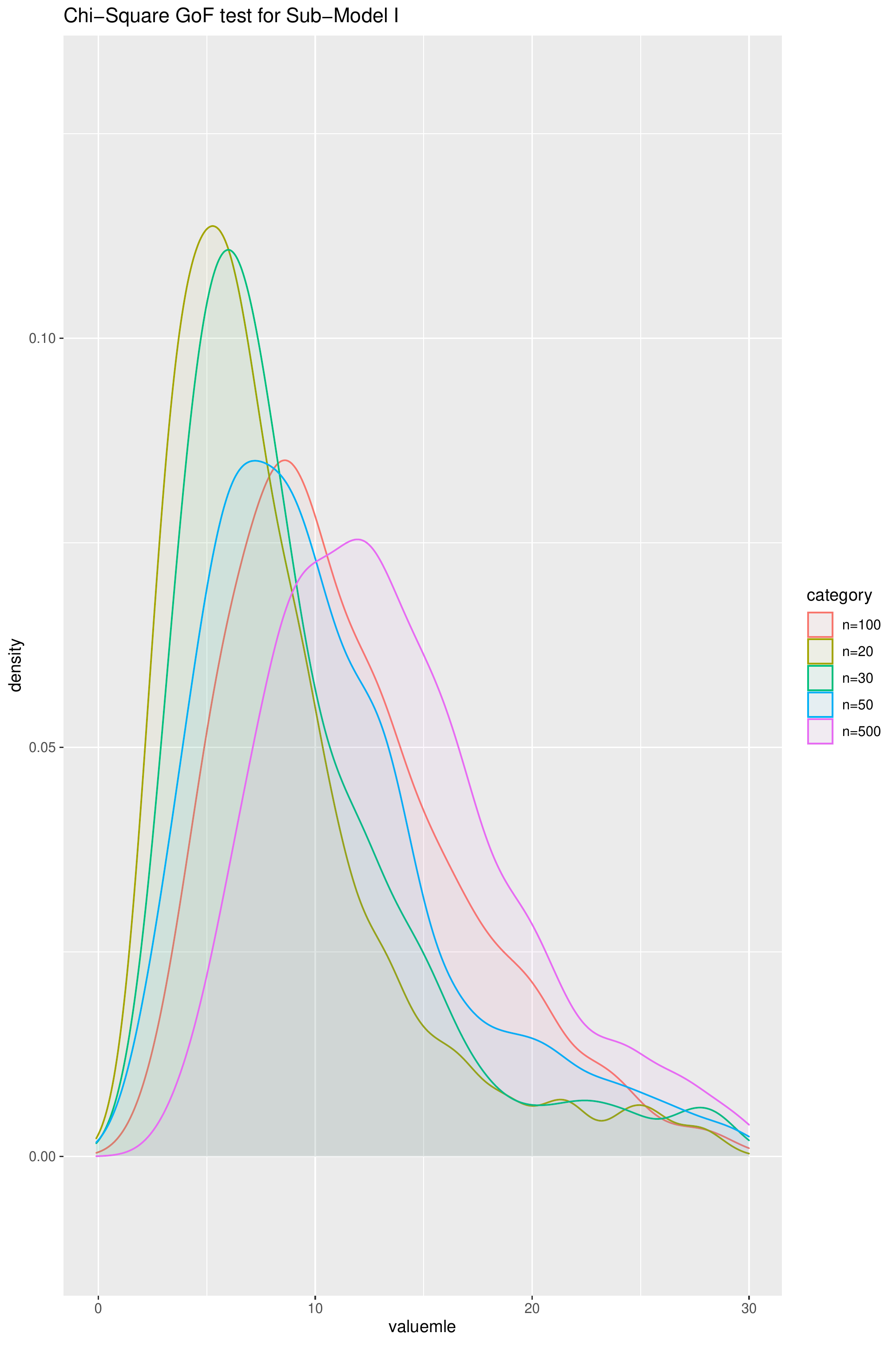}}
	\subfloat[Sub-Model II]{\includegraphics[width=5cm,height=2.5cm]{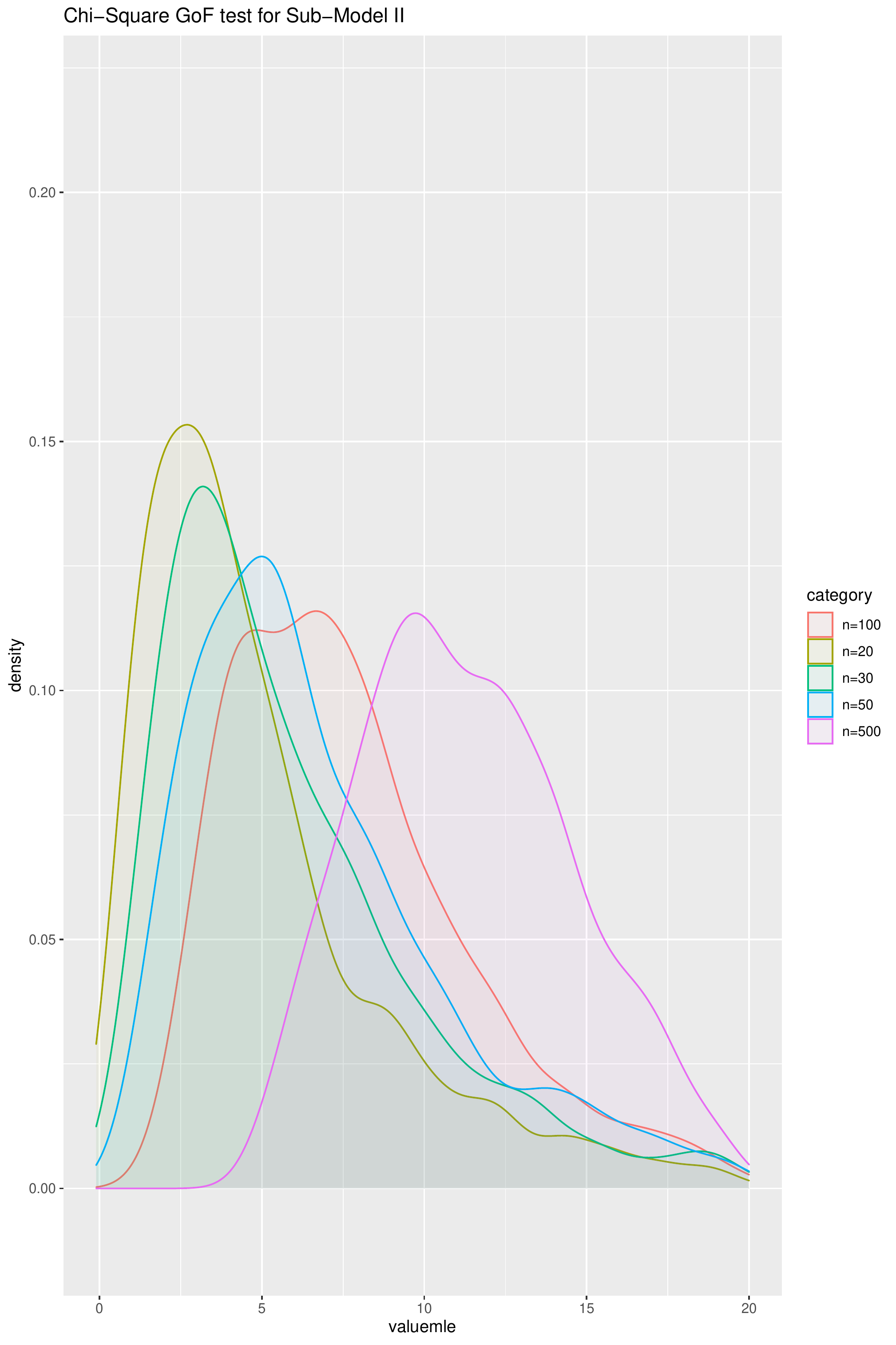}} 
	\caption{Chi-square GoF test for $k=4$.}
	\label{chis}
\end{figure}

The distribution of Chi-square GoF test statistic sample distribution  for the full and its sub-models, see Figure \ref{chis}. However, in the case of missing alternative distribution information Chi-square GoF test recommended otherwise other tests which are mentioned perform better than the Chi-square.  Also, Chi-square test dependends on the value of $k$ choosen, for an illustration we have consider $K=4$ and analysed its finite and large sample distributions. 
 
\subsubsection{Power analysis}
In the present section we will be considering classical bivariate Poisson and bivariate Com-Max Poisson distributions as alternatives  to analyse the power of each of the tests discussed above.

Hence, we simulate $n=20,30,50,100,500$ samples from $Z_i \sim Poisson (\theta_i)$, $i=1,2,3$ and taking $U= Z_1 + Z_3$ and $V=Z_2+Z_3$ the  resultant joint random variable $(U,V)$ will be $n$ observations from the classical bivariate Poisson distribution.
Nevertheless, to simulate  $n=20,30,50,100,500$ samples from the  bivariate Com-Max-Poisson, we begin with simulating an observation from the univariate Com-Max-Poisson with parameter $\theta$ and $\nu$, say $N$. Further, simulate $N$ observations from the bivariate binomial distribution with specified cell probabilities, say $(W_{1i},W_{2i})$, $i=1,2,...,N$. Then, the random vector  $(\sum_{i=1}^{N} W_{1i},\sum_{i=1}^{N}W_{2i})$ will be an observation from the bivariate Com-Max Poisson distribution. For the desired sample size, repeat the above procedure for $n$ times to have specified sample size from the bivariate Com-Mox Poisson distribution. We refer to Sellers et al. \cite{kdb16} for further discussion and an algorithm to simulate from the bivariate Com-Max Poisson using R software.

The empirical power computation is as follows

\begin{enumerate}
	\item[Step 1]  Compute GoF test statistic value for the samples from  alternative distribution, say $T_{obs}$.
	\item[Step 2]  For the given boostrapping size (say$B=5000$), compute    $T^b_{A}$ for $b \in \{ 1,2,...,B\}$.
	\item[Step 3]  Hence,  $\frac{1}{B} \{ \mbox{Total no. of $T^b_{A}$ greater than $T_{obs}$} \} $ is an empirical power of the test.
\end{enumerate}
We refer to Table \ref{power} for the each of the tests empirical powers. 
   \begin{table}[H]
 	\centering
 	\caption{Power (\% of observations) under classical bivariate Poisson (BCBP(($\theta_1 =1,\theta_2=3,\theta_3=4$))) and bivariate Com-Max Poisson (BCMP($\theta=1$,$\nu=5$, $\mu_1=0.1$,$ratio =\exp(1.5)$)) alterativies }
 	\label{power}
 	\resizebox{12cm}{!}{%
 		\begin{tabular}{cc|c|c|c|c|c|l}
 			\cline{3-7}
 			& & \multicolumn{5}{ c| }{Sample size } \\ \cline{3-7}
 			& & $n=20$ & $n=30$ & $n=50$ & $n=100$ & $n=500$\\ \cline{1-7}
 			\multicolumn{1}{ |c  }{\multirow{6}{*}{$T^{(SII)}_{PN}$} } &
 			\multicolumn{1}{ |c| }{$t_1=-0.9$, $t_2=-0.9$} & $(17.2,8.8)$ & $(21.6,4.2)$ & $(22.6,2.6)$ & $(59.2,0.03)$ &   $(90.4,0.01)$  \\ \cline{2-7}
 			\multicolumn{1}{ |c  }{}                        &
 			\multicolumn{1}{ |c| }{$t_1=-0.5$, $t_2=-0.5$} & $(0.92,82.4)$ & $(0.89,84.0)$ & $(0.91,92.4)$ & $(0.93,0.95.8)$   &  $(0.99,0.97)$   \\ \cline{2-7}
 			\multicolumn{1}{ |c  }{}                        &
 			\multicolumn{1}{ |c| }{$t_1=-0.01$, $t_2=-0.01$} & $(0.99,0.91)$ & $(0.97,0.93)$ & $(0.99,0.99)$ & $(0.99,0.97)$ &   $(0.99,0.99)$  \\ \cline{2-7}
 			\multicolumn{1}{ |c  }{}                        &
 			\multicolumn{1}{ |c| }{$t_1=0.01$, $t_2=0.01$} & $(0.99,0.99)$ & $(0.99,0.98)$ & $(0.98,0.97)$ & $(0.99,0.99)$ &   $(0.99,0.98)$      \\ \cline{2-7}
 			\multicolumn{1}{ |c  }{}                        &
 			\multicolumn{1}{ |c| }{$t_1=0.5$, $t_2=0.5$} & $(0.99,0.99)$ & $(0.91,0.99)$ & $(0.92,0.99)$ & $(0.89,0.99)$ &   $(0.81,0.99)$     \\ \cline{2-7}
 			\multicolumn{1}{ |c  }{}                        &
 			\multicolumn{1}{ |c| }{$t_1=0.9$, $t_2=0.9$} & $(0.98,0.99)$ & $(0.92,0.93)$ & $(0.94,0.96)$ & $(0.99,0.99)$ &   $(0.99,0.99)$     \\ \cline{1-7}
 			\multicolumn{2}{ c| }{}     \\ \cline{1-7}
 			\multicolumn{1}{ |c  }{\multirow{1}{*}{$T^{(SII)}_{SPN}$} } &
 			\multicolumn{1}{ |c| }{---} & $(0.95,0.92)$ & $(0.99,0.91)$ & $(0.95,0.98)$ & $(0.99,0.99)$ &   $(0.99,0.97)$     \\ \cline{1-7}
 			 \multicolumn{2}{ c| }{}     \\ \cline{1-7}
 			 \multicolumn{1}{ |c | }{\multirow{6}{*}{$T^{(.)}_{P,n,w}$} } &
 			 \multicolumn{1}{ |c| }{$a_1=-0.9,a_2=-0.9$} & $(35.2, 67.8)$ & $(56.9,82.7)$ & $(59.9,12.6)$ & $(35.0,9.2)$ &   $--$     \\ \cline{2-7}
 			 \multicolumn{1}{ |c  }{}                        &
 			 \multicolumn{1}{ |c| }{$a_1=-0.01$, $a_2=-0.01$} & $(6.0,85.4)$ & $(72.5,27.4)$ & $(24.5,15.2)$ & $(10.1,50.3)$   &  $--$   \\ \cline{2-7}
 			  \multicolumn{1}{ |c  }{}                        &
 			 \multicolumn{1}{ |c| }{$a_1=0.01$, $a_2=0.01$} & $(79.5,2.6)$ & $(20.3,16.0)$ & $(36.3,34.2)$ & $(6.1,59.8)$   &  $--$   \\ \cline{2-7}
 			  \multicolumn{1}{ |c  }{}                        &
 			 \multicolumn{1}{ |c| }{$a_1=0.5$, $a_2=0.5$} & $(19.2,37.3)$ & $(4.5,27.0)$ & $(26.1,85.4)$ & $(2.1,55.5)$   &  $--$   \\ \cline{2-7}
 			  \multicolumn{1}{ |c  }{}                        &
 			 \multicolumn{1}{ |c| }{$a_1=3$, $a_2=5$} & $(35.0,74.0)$ & $(72.7,19.0)$ & $(5.5,25.5)$ & $(1.2,10.0)$   &  $--$   \\ \cline{2-7}
 			  \multicolumn{1}{ |c  }{}                        &
 			 \multicolumn{1}{ |c| }{$a_1=-0.9$, $a_2=5$} & $(12.0,91.0)$ & $(10.8,72.0)$ & $(4.7,92.8)$ & $(0.1,43.6)$   &  $--$   \\ \cline{1-7}
 			  \multicolumn{2}{ c| }{}     \\ \cline{1-7}
 			 \multicolumn{1}{ |c | }{\multirow{6}{*}{$T^{(SI)}_{P,n,w}$} } &
 			 \multicolumn{1}{ |c| }{$a_1=-0.9,a_2=-0.9$} & $(79.3,96.5)$ & $(12.3,95.0)$ & $(11.2,21.8)$ & $(13.4,34.6)$ &   $--$     \\ \cline{2-7}
 			 \multicolumn{1}{ |c  }{}                        &
 			 \multicolumn{1}{ |c| }{$a_1=-0.01$, $a_2=-0.01$} & $(73.5,20.4)$ & $(16.9,9.5)$ & $(0.9,87.0)$ & $(0.1,13.2)$   &  $--$   \\ \cline{2-7}
 			 \multicolumn{1}{ |c  }{}                        &
 			 \multicolumn{1}{ |c| }{$a_1=0.01$, $a_2=0.01$} & $(22.4,3.8)$ & $(15.9,96.4)$ & $(2.2,93.8)$ & $(13.8,17.4)$   &  $--$   \\ \cline{2-7}
 			 \multicolumn{1}{ |c  }{}                        &
 			 \multicolumn{1}{ |c| }{$a_1=0.5$, $a_2=0.5$} & $(43.2,87.1)$ & $(27.5,4.8)$ & $(21.3,80.2)$ & $(1.1,69.4)$   &  $--$   \\ \cline{2-7}
 			 \multicolumn{1}{ |c  }{}                        &
 			 \multicolumn{1}{ |c| }{$a_1=3$, $a_2=5$} & $(40.8,43.2)$ & $(36.0,94.7)$ & $(2.1,75.8)$ & $(1.3,10.2)$   &  $--$   \\ \cline{2-7}
 			 \multicolumn{1}{ |c  }{}                       &
 			 \multicolumn{1}{ |c| }{$a_1=-0.9$, $a_2=5$} & $(54.4,87.2)$ & $(62.7,72.0)$ & $(4.7,36.8)$ & $(0.1,24.4)$   &  $--$   \\ \cline{1-7}
 			 \multicolumn{2}{ c| }{}     \\ \cline{1-7}
 			 \multicolumn{1}{ |c | }{\multirow{6}{*}{$T^{(SII)}_{P,n,w}$} } &
 			 \multicolumn{1}{ |c| }{$a_1=-0.9,a_2=-0.9$} & $(7.1,12.9)$ & $(41.9,99.9)$ & $(7.7,26.2)$ & $(60.71,64.2)$ &   $--$     \\ \cline{2-7}
 			 \multicolumn{1}{ |c  }{}                        &
 			 \multicolumn{1}{ |c| }{$a_1=-0.01$, $a_2=-0.01$} & $(71.4,74.7)$ & $(9.6,56.6)$ & $(2.9,31.6)$ & $(0.2,36.5)$   &  $--$   \\ \cline{2-7}
 			 \multicolumn{1}{ |c  }{}                        &
 			 \multicolumn{1}{ |c| }{$a_1=0.01$, $a_2=0.01$} & $(8.1,36.1)$ & $(2.5,54.4)$ & $(7.5,67.1)$ & $(0.1,37.3)$   &  $--$   \\ \cline{2-7}
 			 \multicolumn{1}{ |c  }{}                        &
 			 \multicolumn{1}{ |c| }{$a_1=0.5$, $a_2=0.5$} & $(38.9,18.9)$ & $(18.2,99.6)$ & $(5.4,78.7)$ & $(0.4,96.2)$   &  $--$   \\ \cline{2-7}
 			 \multicolumn{1}{ |c  }{}                        &
 			 \multicolumn{1}{ |c| }{$a_1=3$, $a_2=5$} & $(2.6,70.0)$ & $(2.5,1.8)$ & $(0.1,82.3)$ & $(0.0,6.0)$   &  $--$   \\ \cline{2-7}
 			 \multicolumn{1}{ |c  }{}                        &
 			 \multicolumn{1}{ |c| }{$a_1=-0.9$, $a_2=5$} & $(50.6,70.6)$ & $(55.0,36.4)$ & $(4.4,84.4)$ & $(0.0,13.2)$   &  $--$   \\ \cline{1-7}
 			 \multicolumn{2}{ c| }{}     \\ \cline{1-7}
 			 \multicolumn{1}{ |c  }{\multirow{1}{*}{$FI_n^{(.)}$} } &
 			 \multicolumn{1}{ |c| }{---} & $(58.4,34.6)$ & $(46.1,86.7)$ & $(56.0,22.9)$ & $(11.8,42.4)$ &   $(63.5,20.8)$     \\ \cline{1-7}
 			 \multicolumn{2}{ c| }{}     \\ \cline{1-7}
 			 \multicolumn{1}{ |c  }{\multirow{1}{*}{$FI_n^{(SI)}$} } &
 			 \multicolumn{1}{ |c| }{---} & $(76.9,13.5)$ & $(6.3,90.6)$ & $(83.0,40.6)$ & $(41.7,30.9)$ &   $(88.1,73.8)$     \\ \cline{1-7}
 			 \multicolumn{2}{ c| }{}     \\ \cline{1-7}
 			 \multicolumn{1}{ |c  }{\multirow{1}{*}{$FI_n^{(SII)}$} } &
 			 \multicolumn{1}{ |c| }{---} & $(69.9,94.4)$ & $(86.4,70.2)$ & $(95.0,26.6)$ & $(91.0,46.7)$ &   $(100,33.8)$     \\ \cline{1-7}
 			
 		 			\end{tabular}  
 	}
\end{table}
All tests are effective or significant in identifying from the pseudo-Poisson and Com-Max Poisson distributions, according to the power analysis. When compared to the classical bivariate Poisson, tests are moderately consistent in detecting the true population. We draw the conclusion that one needs to think about altering the parameter values and conducting additional research on the same in order to better grasp the power for the classical bivariate Poisson alternative.

	\subsection{Real-life data}
	In the following section we consider two data sets which are mentioned in Karlis and Tsiamyrtzis \cite{kt08}, Islam and Chowdhury \cite{ic17}, Leiter and Hamdani \cite{lh73} and also in Arnold and Manjunath \cite{am21}.  For empirical $p$-value computation we have simulated $5000$ observations from the pseudo-Poisson models with respective maximum likelihood values and compare it with the critical value of each of the tests.
	
	\subsection{A particular data set I}
	We consider a data sets which is mentioned in Islam and Chowdhury \cite{ic17} and also in Arnold and Manjunath \cite{am21} , the source of the data is from  the tenth wave of the Health and Retirement Study (HRS). The data represents the number of conditions ever had $(X)$ as mentioned by the doctors and utilization of healthcare services (say, hospital, nursing home, doctor and home care) $(Y)$. The Pearson correlation coefficient between $X$ and $Y$ is $0.063$. The test for independence, classical inference (m.l.e and moment estimates) and AIC values for full and its sub-models c.f.  Arnold and Manjunath \cite{am21} page 16 and 18 (Table 10). 
	
		In the following we will consider the full and its Sub-Model II. The criteria of selecting below two models are discussed in Arnold and Manjunath \cite{am21} on page 18 and Table 10. We refer to Table \ref{realI} for the critical values and its empirical p-values for the Full and Sub-Model II.

	\begin{table}[H]
		\centering
		\caption{Health and retirement study data (Full Model) and m.l.e. estimates Full model ($\hat{\lambda_1}=2.643$,$\hat{\lambda_2}=0.688$,$\hat{\lambda}_3=0.031$) for Sub-Model II ($\hat{\lambda}_3=0.031$)}
		\label{realI}
			\resizebox{10cm}{!}{%
	\begin{tabular}{cc|c|c|c|c|c|l}
				\cline{3-4}
				& & \multicolumn{2}{ c| }{$n = 5567$  } \\ \cline{3-4}
				& & Test statistic value & $p$-value \\ \cline{1-4}
				\multicolumn{1}{ |c  }{\multirow{6}{*}{$T^{(SII)}_{PN}$} } &
				\multicolumn{1}{ |c| }{$t_1=-0.9$, $t_2=-0.9$} & $151.734$ &  $0.025$ \\ \cline{2-4}
				\multicolumn{1}{ |c  }{}                        &
				\multicolumn{1}{ |c| }{$t_1=-0.5$, $t_2=-0.5$} & $-2870.383$ &  $0.891$ \\ \cline{2-4}
				\multicolumn{1}{ |c  }{}                        &
				\multicolumn{1}{ |c| }{$t_1=-0.01$, $t_2=-0.01$} & $755.821$ &  $0.901$ \\ \cline{2-4}
				\multicolumn{1}{ |c  }{}                        &
				\multicolumn{1}{ |c| }{$t_1=0.01$, $t_2=0.01$} &  $803.119$ &   $0.921$    \\ \cline{2-4}
				\multicolumn{1}{ |c  }{}                        &
				\multicolumn{1}{ |c| }{$t_1=0.5$, $t_2=0.5$} & $1713.7$ &   $0.141$   \\ \cline{2-4}
				\multicolumn{1}{ |c  }{}                        &
				\multicolumn{1}{ |c| }{$t_1=0.9$, $t_2=0.9$} & $3710.615$  & $0.164$      \\ \cline{1-4}
				\multicolumn{2}{ c| }{}     \\ \cline{1-4}
				\multicolumn{1}{ |c  }{\multirow{1}{*}{$T^{(SII)}_{SPN}$} } &
				\multicolumn{1}{ |c| }{---} &  $12.740$ &    $0.097$  \\ \cline{1-4} 
				\multicolumn{2}{ c| }{}     \\ \cline{1-4}
					\multicolumn{1}{ |c  }{\multirow{6}{*}{$T^{(.)}_{PN}$} } &
				\multicolumn{1}{ |c| }{$a_1=-0.9$, $a_2=-0.9$} & $578.674$ &  $0.01$ \\ \cline{2-4}
				\multicolumn{1}{ |c  }{}                        &
				\multicolumn{1}{ |c| }{$a_1=-0.5$, $a_2=-0.55$} & $117.940$ &  $0.99$ \\ \cline{2-4}
				\multicolumn{1}{ |c  }{}                        &
							\multicolumn{1}{ |c| }{$a_1=-0.01$, $a_2=-0.01$} & $64.179$ &  $0.8$ \\ \cline{2-4}
				\multicolumn{1}{ |c  }{}                        &
				\multicolumn{1}{ |c| }{$a_1=1$, $a_2=1$} & $67.564$ &  $0.09$ \\ \cline{2-4}
				\multicolumn{1}{ |c  }{}                        &
				\multicolumn{1}{ |c| }{$a_1=3$, $a_2=5$} & $21.739$ &  $0.12$ \\ \cline{2-4}
				\multicolumn{1}{ |c  }{}                        &
				\multicolumn{1}{ |c| }{$a_1=-0.9$, $a_2=5$} & $23.830$ &  $0.02$ \\ \cline{1-4}
				\multicolumn{2}{ c| }{}     \\ \cline{1-4}
					\multicolumn{1}{ |c  }{\multirow{3}{*}{$T^{(SII)}_{PN}$} } &
				\multicolumn{1}{ |c| }{$a_1=-0.9$, $a_2=-0.9$} & $659.816$ &  $0.99$ \\ \cline{2-4}
				\multicolumn{1}{ |c  }{}                        &
				\multicolumn{1}{ |c| }{$a_1=1$, $a_2=1$} & $71.881$ &  $0.07$ \\ \cline{2-4}
				\multicolumn{1}{ |c  }{}                        &
				\multicolumn{1}{ |c| }{$a_1=-0.9$, $a_2=5$} & $24.465$ &  $0.02$ \\ \cline{1-4}
				\multicolumn{2}{ c| }{}     \\ \cline{1-4}
				  \multicolumn{2}{ c| }{}     \\ \cline{1-4}
				  \multicolumn{1}{ |c  }{\multirow{1}{*}{$FI^{(.)}_{n}$} } &
				  \multicolumn{1}{ |c| }{---} &  $-13.532$ &    $0.987$  \\ \cline{1-4} 
				  \multicolumn{2}{ c| }{}     \\ \cline{1-4}
				  \multicolumn{1}{ |c  }{\multirow{1}{*}{$FI^{(SII)}_{n}$} } &
				  \multicolumn{1}{ |c| }{---} &  $-25.729$ &    $0.991$  \\ \cline{1-4} 
				  \multicolumn{2}{ c| }{}     \\ \cline{1-4}
				  \multicolumn{1}{ |c  }{\multirow{1}{*}{Chi-square $^{(.)}$} } &
				  \multicolumn{1}{ |c| }{---} &  $417.653$ &    $--$  \\ \cline{1-4}

			\end{tabular}  
	}
		
	\end{table}
	
		The tests $T^{(SII)}_{PN}$ on neighborhood of $0$, $T^{(SII)}_{SPN}$, $T^{(.)}_{PN}$ (large than $-0.9$), $FI^{(.)}_n$ and $FI^{(SII)}_n$ are suggests that the Health and Retirement data fits bivariate pseudo-Poisson Full and its Sub-Model II,  
		 which agree with the AIC values listed on pages 16 \& 18 of Arnold and Manjunath's \cite{am21}.

	\subsection{A particular data set II}
	
		Now, we consider a data set which is in Leiter and Hamdani \cite{lh73}, the source of the data is a 50-mile stretch of Interstate 95 in Prince William, Stafford and Spotsylvania counties in Eastern Virginia. The data represents the number of accidents categorized as fatal accidents, injury accidents or property damage accidents, along with the corresponding number of fatalities and injuries for the period 1 January 1969 to 31 October 1970.
	For  classical inference (m.l.e and moment estimates) and AIC values for full and its sub-models c.f.  Arnold and Manjunath \cite{am21} page 17 and 19 (Table 11).  The criteria of selecting below two models are discussed in Arnold and Manjunath \cite{am21} on page 19 and Table 11.  It has been emphasized  in Leiter and Hamdani \cite{lh73} and Arnold and Manjunath \cite{am21} that mirrored Sub-model II fit the data better than any other sub-models. 
	
	In the following we will consider the  two models.  We refer to Table \ref{realII} for the critical values and its empirical $p$-values for the Full and Mirrored Sub-Model II.

	\begin{table}[H]
		\centering
		\caption{Accidents and fatalities (Full Model) and m.l.e. estimates Full model ($\hat{\lambda_1}=0.058$,$\hat{\lambda_2}=0.812$,$\hat{\lambda}_3=0.867$) and for mirrored Sub-Model II ($\hat{\lambda}_1=0.862$,$\hat{\lambda}_3=0.067$)}
		\label{realII}
		\resizebox{10cm}{!}{%
			\begin{tabular}{cc|c|c|c|c|c|l}
				\cline{3-4}
				& & \multicolumn{2}{ c| }{$n = 639$  } \\ \cline{3-4}
				& & Test statistic value & $p$-value \\ \cline{1-4}
				\multicolumn{1}{ |c  }{\multirow{6}{*}{Mirrored $T^{(SII)}_{PN}$} } &
				\multicolumn{1}{ |c| }{$t_1=-0.9$, $t_2=-0.9$} & $165.966$ &  $0.054$ \\ \cline{2-4}
				\multicolumn{1}{ |c  }{}                        &
				\multicolumn{1}{ |c| }{$t_1=-0.5$, $t_2=-0.5$} & $-359.286$ &  $0.932$ \\ \cline{2-4}
				\multicolumn{1}{ |c  }{}                        &
				\multicolumn{1}{ |c| }{$t_1=-0.01$, $t_2=-0.01$} & $-135.242$ &  $0.914$ \\ \cline{2-4}
				\multicolumn{1}{ |c  }{}                        &
				\multicolumn{1}{ |c| }{$t_1=0.01$, $t_2=0.01$} &  $-133.630$ &   $0.899$    \\ \cline{2-4}
				\multicolumn{1}{ |c  }{}                        &
				\multicolumn{1}{ |c| }{$t_1=0.5$, $t_2=0.5$} & $-126.924$ &  $0.763$    \\ \cline{2-4}
				\multicolumn{1}{ |c  }{}                        &
				\multicolumn{1}{ |c| }{$t_1=0.9$, $t_2=0.9$} & $-220.890$  & $0.558$      \\ \cline{1-4}
				\multicolumn{2}{ c| }{}     \\ \cline{1-4}
				\multicolumn{1}{ |c  }{\multirow{1}{*}{Mirrored $T^{(SII)}_{SPN}$} } &
				\multicolumn{1}{ |c| }{---} &  $4.237$ &    $0.544$  \\ \cline{1-4} 
				\multicolumn{2}{ c| }{}     \\ \cline{1-4}
				\multicolumn{1}{ |c  }{\multirow{6}{*}{$T^{(.)}_{PN}$} } &
			\multicolumn{1}{ |c| }{$a_1=-0.9$, $a_2=-0.9$} & $1057.191$ &  $0.99$ \\ \cline{2-4}
			\multicolumn{1}{ |c  }{}                        &
			\multicolumn{1}{ |c| }{$a_1=-0.5$, $a_2=-0.55$} & $100.903$ &  $0.98$ \\ \cline{2-4}
			\multicolumn{1}{ |c  }{}                        &
			\multicolumn{1}{ |c| }{$a_1=-0.01$, $a_2=-0.01$} & $24.906$ &  $0.87$ \\ \cline{2-4}
			\multicolumn{1}{ |c  }{}                        &
			\multicolumn{1}{ |c| }{$a_1=1$, $a_2=1$} & $3.786$ &  $0.91$ \\ \cline{2-4}
			\multicolumn{1}{ |c  }{}                        &
			\multicolumn{1}{ |c| }{$a_1=3$, $a_2=5$} & $1.178$ &  $0.01$ \\ \cline{2-4}
			\multicolumn{1}{ |c  }{}                        &
			\multicolumn{1}{ |c| }{$a_1=-0.9$, $a_2=5$} & $152.5798$ &  $0.007$ \\ \cline{1-4}
			\multicolumn{2}{ c| }{}     \\ \cline{1-4}
				\multicolumn{2}{ c| }{}     \\ \cline{1-4}
				\multicolumn{1}{ |c  }{\multirow{3}{*}{Mirrored $T^{(SII)}_{PN}$} } &
				\multicolumn{1}{ |c| }{$a_1=-0.9$, $a_2=-0.9$} & $78.337$ &  $0.40$ \\ \cline{2-4}
				\multicolumn{1}{ |c  }{}                        &
				\multicolumn{1}{ |c| }{$a_1=1$, $a_2=1$} & $4.049$ &  $0.91$ \\ \cline{2-4}
				\multicolumn{1}{ |c  }{}                        &
				\multicolumn{1}{ |c| }{$a_1=-0.9$, $a_2=5$} & $1.438$ &  $0.20$ \\ \cline{1-4}
				\multicolumn{2}{ c| }{}     \\ \cline{1-4}
				\multicolumn{1}{ |c  }{\multirow{1}{*}{$FI^{(.)}_{n}$} } &
				\multicolumn{1}{ |c| }{---} &  $2.289$ &    $0.986$  \\ \cline{1-4} 
				\multicolumn{2}{ c| }{}     \\ \cline{1-4}
				\multicolumn{1}{ |c  }{\multirow{1}{*}{Mirrored $FI^{(SII)}_{n}$} } &
				\multicolumn{1}{ |c| }{---} &  $3.443$ &    $0.3$  \\ \cline{1-4} 
				\multicolumn{2}{ c| }{}     \\ \cline{1-4}
				\multicolumn{1}{ |c  }{\multirow{1}{*}{Chi-square $^{(.)}$} } &
				\multicolumn{1}{ |c| }{---} &  $586$ &    $--$  \\ \cline{1-4}

			\end{tabular}  
		}
		
	\end{table}

	The tests $T^{(SII)}_{PN}$ on neighborhood of $0$, $T^{(SII)}_{SPN}$, $T^{(.)}_{PN}$ (large than $-0.9$), $FI^{()}_n$ and $FI^{(SII)}_n$ suggests that the Accidents and Fatalities data fits well the bivariate pseudo-Poisson Full and its mirrored Sub-Model II,  which is inline with the AIC values listed on pages 16 \& 18 of Arnold and Manjunath's \cite{am21}.
	  
	\section{Conclusion}
The GoF tests for the bivariate pseudo-Poisson and its sub-models were the main emphasis of the current note. Based on p.g.f, moments, and Chi-square tests, we proposed  a few GoF tests. The test based on the  bivariate Fisher index of dispersion-based GoF test is a new contribution to the bivariate count variables.  The supremum of the absolute difference between the calculated p.g.f. and its empirical equivalent is the robust  GoF test, i.e., robust to the choice of the alternative distributions. Additionally, we took into account a few existing tests that depend on the estimated p.g.f and its empirical results, such as K\&K, Munoz, and Gamero approaches. Finally, the Chi-square GoF test results for the pseudo-Poisson data were also examined.

A finite sample, a fairly large sample, and asymptotic distributions of test statistics are examined for each of the tests discussed. In addition, we looked at the power and efficacy of each statistical test using the bivariate Com-Max-Poisson and the bivariate Classical Bivariate (BCP) as alternative distributions. It has been demonstrated that a test based on the supremum and index of dispersion is reliable, consistent, and satisfying. Particularly, the supremum-based test proved to be more robust to the choice of alternative distributions.  Additionally, we suggest utilising the Munoz and Gamero (M\&G) test for moderately small samples and the supremum (robust) and dispersion tests for moderately large samples. Due to the asymptotic distribution of the test statistic, we also recommend K\&K and dispersion tests for sufficiently large data sets. Also, due to its robust property, we suggest considering the supremum and Chi-square GoF tests if there are no reasonable alternatives to the hypothesis.

The bivariate pseudo-Poisson distribution has been highly advised as the primary choice when modelling bivariate count data whenever the marginals exhibit equal- and over-dispersed, see Arnold and Manjunath \cite{am21}. The GoF tests that have been suggested will unquestionably add yet another tool for evaluating the compatibility of the bivariate count data. Briefly said, writers are working on a R package that covers fitting (classical and Bayesian analysis) and testing for the bivariate pseudo-Poisson model. The developed package will merit a spot in the toolkit of contemporary modellers because of its simple structure and fast computation.
 	
\section{Acknowledgment(s)}
We thank Prof. B.C. Arnold for useful suggestions and feedback in every step of the   successful realization of the article.

\vspace{0.2cm}
The first author’s research was sponsored by the Institution of Eminence (IoE), University of Hyderabad (UoH-IoE-RC2-21-013).

	\appendix
	\section{Appendices}
	\subsection{Marginal probability of $Y$}
	For the marginal distribution of $Y$,  the probability that $Y=0$ can be computed as 
	\begin{eqnarray}
		P(Y=0)= G_{Y}(0) 
		= e^{- \lambda_2} e^{\lambda_1 (e^{-\lambda_3}-1)}  .
	\end{eqnarray}

	For the probability that $Y=1$ we have
	\begin{eqnarray*}
		\frac{d}{dt}G_{Y}(t_2) 
		= G_{Y}(t_2) \left [\lambda_1 \lambda_3 e^{\lambda_3(t_2-1)} + \lambda_2 \right]
	\end{eqnarray*}
	
	\begin{eqnarray}
		P(Y=1)= \frac{ \frac{d}{dt}G_{Y}(t_2)  |_{t_2=0}}{1\underline!}= G_{Y}(0) \left [\lambda_1 \lambda_3 e^{-\lambda_3} + \lambda_2 \right].
	\end{eqnarray}
	
	Similarly, $P(Y=2)$ is given as
	
	\begin{eqnarray*}
		\frac{d^2}{dt^2}G_{Y}(t_2) 
		= G_{Y}(t_2) \left[ \left( \lambda_1 \lambda_2 e^{\lambda_2(t_2-1)} + \lambda_2 \right)^2  + \lambda_1 \lambda^2_3 e^{\lambda_3(t_2-1)}\right]
	\end{eqnarray*}
	
	\begin{eqnarray}
		P(Y=2)= \frac{ \frac{d}{dt}G_{Y}(t_2)  |_{t_2=0}}{2!}= \frac{G_{Y}(0)}{2!} \left[ \left( \lambda_1 \lambda_2 e^{-\lambda_2} + \lambda_2 \right)^2  + \lambda_1 \lambda^2_3 e^{-\lambda_3}\right]
	\end{eqnarray}
	and finally $P(Y=3)$ ismodels
	\begin{eqnarray}
		\frac{d^3}{dt^3}G_{Y}(t_2)  
		= && G_{Y}(t_2) \left [\lambda_1  \lambda_3 ( \left(\lambda_1 \lambda_3 e^{\lambda_3 (t_2-1)} + \lambda_2\right)^2 + \lambda_3  \right. \nonumber\\
		&&
		\left. (2   \left(\lambda_1 \lambda_3 e^{\lambda_3 (t_2-1)}+ \lambda_2\right) + \lambda_3 (1 + \lambda_1 e^{\lambda_3 (t_2-1)}))) e^{\lambda_3 (t_2-1)}+ \right. \nonumber \\ && 
		\left.   \lambda_2  (    \left(\lambda_1 \lambda_3 e^{\lambda_3 (t_2-1)}+ \lambda_2\right)^2 + 
		\lambda_1 \lambda_3^2 e^{\lambda_3 (t_2-1)}) \right]
	\end{eqnarray}

	\begin{eqnarray}
		P(Y=3)= \frac{1}{3!}\frac{d^3}{dt^3}G_{X_2}(0)  |_{t_2=0}
		= && \frac{G_{Y}(0)}{6} \left [\lambda_1  \lambda_3 ( \left(\lambda_1 \lambda_3 e^{-\lambda_3} + \lambda_2\right)^2 + \lambda_3  \right. \nonumber \\
		&&
		\left. (2   \left(\lambda_1 \lambda_3 e^{-\lambda_3 }+ \lambda_2\right) + \lambda_3 (1 + \lambda_1 e^{-\lambda_3 }))) e^{-\lambda_3 }+ \right. \nonumber \\ && 
		\left.   \lambda_2  (    \left(\lambda_1 \lambda_3 e^{-\lambda_3 }+ \lambda_2\right)^2 + 
		\lambda_1 \lambda_3^2 e^{-\lambda_3 }) \right].
	\end{eqnarray}
On similar line one can extend the above procedure to get albeit complicated values for the probability that $Y$ assumes any positive value. 

	\vspace{0.5cm}

	\subsection{Other conditional distribution of the bivariate pseudo-Poisson}
	In the following we are deriving other conditional distribution, i.e., conditional distribution of $X$ given $Y=y$ by induction for the Sub-model II.  Consider the joint mass function of pseudo-Poisson Sub-model II
	
	\[ f_{X,Y}(x,y) = \begin{cases} 
		\frac{e^{-\lambda_1 \lambda_1^x}}{x!} \frac{e^{-\lambda_3 x} (\lambda_3 x)^y}{y!} &  x= 1,2,...; y=0,1,2,...\\
		e^{-\lambda_1} & (x,y)=(0,0). 
	\end{cases}
	\]
	Now, consider the case in which $y=0$ then for each $x=0,1,2,...$ the conditional mass function will be 
	
	\begin{eqnarray}
		f_{x|Y}(x|0) &=& \frac{P(X=x,Y=0)}{P(Y=0)} \nonumber \\
		&=& \frac{e^{-\lambda_1 e^{-\lambda_3}} (\lambda_1 e^{-\lambda_3})^x}{x!}
	\end{eqnarray}
	Indeed the above conditional mass function is a Poisson distribution with mean equal to $\lambda_1 e^{-\lambda_3}$.

	Next, consider the case with $y=1$. For each $x=1,2,...$ we have
	
	\begin{eqnarray}
		f_{x|Y}(x|1) &=& \frac{P(X=x,Y=1)}{P(Y=1)} \nonumber \\
		&=& \frac{e^{-\lambda_1 e^{-\lambda_3}} (\lambda_1 e^{-\lambda_3})^{x-1}}{(x-1)!}
	\end{eqnarray}
	which is recognizable as the distribution of $1$ plus a Poisson($\lambda_1 e^{-\lambda_3}$).

	For $y \geq 1$ and for each $x=1,2,...$  we have
	a
	\begin{eqnarray}
		f_{X|Y}(x|y) &=& \frac{P(X=x,Y=y)}{P(Y=y)} \nonumber \\
		&=& \frac{ \frac{e^{-\lambda_1 e^{-\lambda_3}} (\lambda_1 e^{-\lambda_3})^{x-1} x^y}{(x-1)!}}{\mu_y}
	\end{eqnarray}
	
	where $\mu_y$ is the $y$th moment of a Poisson($\lambda_1 e^{-\lambda_3}$) variable.  Note that the expression $\mu_y$ can also be expressed in terms of factorial moments and the $y$th factorial moment is $(\lambda_1 e^{-\lambda_3})^y$. Thus we have 
	
	\begin{eqnarray}
		\mu_y = \sum_{j=0}^{y} S(y,j) (\lambda_1 e^{-\lambda_3})^y
	\end{eqnarray}	
	where $S(y,j)$ is a Stirling number of the second kind. Also note that if $y \geq 1$ then $S(y,0)=0$.
	\subsection{Examples}
	Consider the following examples: 
	
	\begin{example}
		define   $w_1(t_1,t_2 ) =   c_1 + c_2  t_1 t_2 + c_3  t^2_1 t_2^{2}$, $(t_1,t_2)^T \in [0,1]^2$,  $c_1, c_2, c_3 \in \mathbb{R}$  and $T_{n,w_1}(\hat{\lambda}_1,\hat{\lambda}_2, \hat{\lambda}_3) $ is
		
		\begin{eqnarray}
			&& c_1 \Biggl\{   \frac{1}{n} \sum_{i=1}^{n} \sum_{j=1}^{n} \Bigg( \fraca{1}{(X_i + X_j +1)(Y_i + Y_j +1)}\Bigg) + \nonumber \\
			&+ & \sum_{k=0}^{\infty} \sum_{l=0}^{\infty}  \sum_{m=0}^{\infty} \sum_{n=0}^{\infty} \Bigg ( \mathscr{P}(k ; \hat{\lambda}_1) \mathscr{P}(l;\hat{\lambda}_2 + k \hat{\lambda}_3) \mathscr{P}(m ; \hat{\lambda}_1) \mathscr{P}(n;\hat{\lambda}_2 + m \hat{\lambda}_3)  \nonumber \\&& \int_{0}^{1} \int_{0}^{1}   t_1^{k+ m} t^{l+ n}_2   dt_1 dt_2  \Bigg )  - \nonumber \\
			&& -  2  \sum_{i=1}^{n}   \sum_{x=0}^{\infty} \sum_{y=0}^{\infty} \Bigg ( \mathscr{P}(x ; \hat{\lambda}_1) \mathscr{P}(y;\hat{\lambda}_2 + x\hat{\lambda}_3) \int_{0}^{1} \int_{0}^{1}   t_1^{x+X_i } t^{y + Y_i }_2   dt_1 dt_2  \Bigg ) \nonumber \Biggr\} + \nonumber   \\
			&&c_2 \Biggl\{  \frac{1}{n} \sum_{i=1}^{n} \sum_{j=1}^{n} \Bigg( \fraca{1}{(X_i + X_j +2)(Y_i + Y_j +2)}\Bigg) + \nonumber \\
			&+ & \sum_{k=0}^{\infty} \sum_{l=0}^{\infty}  \sum_{m=0}^{\infty} \sum_{n=0}^{\infty} \Bigg ( \mathscr{P}(k ; \hat{\lambda}_1) \mathscr{P}(l;\hat{\lambda}_2 + k \hat{\lambda}_3) \mathscr{P}(m ; \hat{\lambda}_1) \mathscr{P}(n;\hat{\lambda}_2 + m \hat{\lambda}_3)  \nonumber \\&& \int_{0}^{1} \int_{0}^{1}   t_1^{k+ m+ 1} t^{l+ n+ 1}_2   dt_1 dt_2  \Bigg )  - \nonumber \\
			&& -  2  \sum_{i=1}^{n}   \sum_{x=0}^{\infty} \sum_{y=0}^{\infty} \Bigg ( \mathscr{P}(x ; \hat{\lambda}_1) \mathscr{P}(y;\hat{\lambda}_2 + x\hat{\lambda}_3) \int_{0}^{1} \int_{0}^{1}   t_1^{x+X_i +1 } t^{y + Y_i +1 }_2   dt_1 dt_2  \Bigg ) \nonumber \Biggr\} +  \\
			&& c_3\Biggl\{  \frac{1}{n} \sum_{i=1}^{n} \sum_{j=1}^{n} \Bigg( \fraca{1}{(X_i + X_j + 3)(Y_i + Y_j +3)}\Bigg) + \nonumber \\
			&+ & \sum_{k=0}^{\infty} \sum_{l=0}^{\infty}  \sum_{m=0}^{\infty} \sum_{n=0}^{\infty} \Bigg ( \mathscr{P}(k ; \hat{\lambda}_1) \mathscr{P}(l;\hat{\lambda}_2 + k \hat{\lambda}_3) \mathscr{P}(m ; \hat{\lambda}_1) \mathscr{P}(n;\hat{\lambda}_2 + m \hat{\lambda}_3)  \nonumber
		\end{eqnarray}
		\begin{eqnarray}
			&& \int_{0}^{1} \int_{0}^{1}   t_1^{k+ m+ 2} t^{l+ n+ 2}_2   dt_1 dt_2  \Bigg )  - \nonumber \\
			&& -  2  \sum_{i=1}^{n}   \sum_{x=0}^{\infty} \sum_{y=0}^{\infty} \Bigg ( \mathscr{P}(x ; \hat{\lambda}_1) \mathscr{P}(y;\hat{\lambda}_2 + x\hat{\lambda}_3) \int_{0}^{1} \int_{0}^{1}   t_1^{x+X_i + 2} t^{y + Y_i + 2}_2   dt_1 dt_2  \Bigg ) \Biggr\} 
		\end{eqnarray}
	\end{example}
	
	where $\mathscr{P}(i;. \hat{\lambda})$ is a Poisson probability at $i$ for the estimated parameter $\hat{\lambda}$.
	
	Further simplication gives us
	
	\begin{eqnarray}
		T_{n,w_1} (\hat{\lambda}_1,\hat{\lambda}_2, \hat{\lambda}_3)	&=&\frac{c_1}{n} \sum_{i=1}^{n} \sum_{j=1}^{n} \Bigg( \fraca{1}{(X_i + X_j +1)(Y_i + Y_j +1)}\Bigg) + \nonumber \\ &&\frac{c_2}{n} \sum_{i=1}^{n} \sum_{j=1}^{n} \Bigg( \fraca{1}{(X_i + X_j +2)(Y_i + Y_j +2)}\Bigg) + \nonumber \\
		&&\frac{c_3}{n} \sum_{i=1}^{n} \sum_{j=1}^{n} \Bigg( \fraca{1}{(X_i + X_j + 3)(Y_i + Y_j +3)}\Bigg) + \nonumber  \\
		&&c_1 \sum_{k=0}^{\infty} \sum_{l=0}^{\infty}  \sum_{m=0}^{\infty} \sum_{n=0}^{\infty} \Bigg (  
		\frac{\mathscr{P}(k ; \hat{\lambda}_1) \mathscr{P}(l;\hat{\lambda}_2 + k \hat{\lambda}_3) \mathscr{P}(m ; \hat{\lambda}_1) \mathscr{P}(n;\hat{\lambda}_2 + m \hat{\lambda}_3)}{(k+m+1)(l+n+1)} \Bigg ) +  \nonumber \\ 
		&&c_2 \sum_{k=0}^{\infty} \sum_{l=0}^{\infty}  \sum_{m=0}^{\infty} \sum_{n=0}^{\infty} \Bigg (  
		\frac{\mathscr{P}(k ; \hat{\lambda}_1) \mathscr{P}(l;\hat{\lambda}_2 + k \hat{\lambda}_3) \mathscr{P}(m ; \hat{\lambda}_1) \mathscr{P}(n;\hat{\lambda}_2 + m \hat{\lambda}_3)}{(k+m+2)(l+n+2)} \Bigg ) +  \nonumber \\
		&&c_3 \sum_{k=0}^{\infty} \sum_{l=0}^{\infty}  \sum_{m=0}^{\infty} \sum_{n=0}^{\infty} \Bigg (  
		\frac{\mathscr{P}(k ; \hat{\lambda}_1) \mathscr{P}(l;\hat{\lambda}_2 + k \hat{\lambda}_3) \mathscr{P}(m ; \hat{\lambda}_1) \mathscr{P}(n;\hat{\lambda}_2 + m \hat{\lambda}_3)}{(k+m+3)(l+n+3)} \Bigg ) +  \nonumber \\
		&& -  2  c_1 \sum_{i=1}^{n}   \sum_{x=0}^{\infty} \sum_{y=0}^{\infty} \Bigg (  \frac{\mathscr{P}(x ; \hat{\lambda}_1) \mathscr{P}(y;\hat{\lambda}_2 + x\hat{\lambda}_3)}{(x+X_i +1 )(y+Y_i+1)} \Bigg ) \nonumber \\
		&&-2 c_2  \sum_{i=1}^{n}   \sum_{x=0}^{\infty} \sum_{y=0}^{\infty} \Bigg (  \frac{\mathscr{P}(x ; \hat{\lambda}_1) \mathscr{P}(y;\hat{\lambda}_2 + x\hat{\lambda}_3)}{(x+X_i +2 )(y+Y_i+2)} \Bigg )
		\nonumber  \\ 
		&& -2 c_3 \sum_{i=1}^{n}   \sum_{x=0}^{\infty} \sum_{y=0}^{\infty} \Bigg (  \frac{\mathscr{P}(x ; \hat{\lambda}_1) \mathscr{P}(y;\hat{\lambda}_2 + x\hat{\lambda}_3)}{(x+X_i +3 )(y+Y_i+3)}  \Bigg ).
	\end{eqnarray}
	
	 We refer to Table \ref{PGFEX1} and Figuare \ref{EX1T} for the quantile values and frequency distribution for  $a_1=1$ and $a_2=1$, respectively.

	\begin{example}
		For a general form of $w(.,.)$, consider  $w_2(t_1,t_2 ) =     t^{a_1}_1 t^{a_2}_2$, $(t_1,t_2)^T \in [0,1]^2$, $a_1, a_2 \in (-1,\infty)$, which allows us to include a negative powers as well,  then the $T_{n,w_2}$ is
	\end{example}
	
	\begin{eqnarray}
		&& \frac{1}{n} \sum_{i=1}^{n} \sum_{j=1}^{n} \Bigg( \fraca{1}{(X_i +\nonumber X_j + a_1+1)(Y_i + Y_j + a_2 +1)}\Bigg) + \nonumber \\
		&+ & \sum_{k=0}^{\infty} \sum_{l=0}^{\infty}  \sum_{m=0}^{\infty} \sum_{n=0}^{\infty} \Bigg ( \mathscr{P}(k ; \hat{\lambda}_1) \mathscr{P}(l;\hat{\lambda}_2 + k \hat{\lambda}_3) \mathscr{P}(m ; \hat{\lambda}_1) \mathscr{P}(n;\hat{\lambda}_2 + m \hat{\lambda}_3)  \nonumber \\&& \int_{0}^{1} \int_{0}^{1}   t_1^{k+ m+ a_1} t^{l+ n+ a_2}_2   dt_1 dt_2  \Bigg )  - \nonumber \\
		&& -  2  \sum_{i=1}^{n}   \sum_{x=0}^{\infty} \sum_{y=0}^{\infty} \Bigg ( \mathscr{P}(x ; \hat{\lambda}_1) \mathscr{P}(y;\hat{\lambda}_2 + x\hat{\lambda}_3) \int_{0}^{1} \int_{0}^{1}   t_1^{x+X_i + a_1} t^{y + Y_i + a_2}_2   dt_1 dt_2  \Bigg ). \nonumber
	\end{eqnarray}
	
	Now, further simplification will give us closed form expression for the statistic 
	
	\begin{eqnarray}
		T_{n,w_2}  (\hat{\lambda}_1,\hat{\lambda}_2 \hat{\lambda}_3)&=& \frac{1}{n} \sum_{i=1}^{n} \sum_{j=1}^{n} \Bigg( \fraca{1}{(X_i + X_j + a_1+1)(Y_i + Y_j + a_2 +1)}\Bigg) + \nonumber \\
		&+ & \sum_{k=0}^{\infty} \sum_{l=0}^{\infty}  \sum_{m=0}^{\infty} \sum_{n=0}^{\infty}  \Bigg (
		\fraca{ \mathscr{P}(k ; \hat{\lambda}_1) \mathscr{P}(l;\hat{\lambda}_2 + k \hat{\lambda}_3) \mathscr{P}(m ; \hat{\lambda}_1) \mathscr{P}(n\mathscr{P}(x ; \hat{\lambda}_1) }{(k+m+a_1+1)(l+n+a_2+1)}  \Bigg) -
		\nonumber \\
		&&	- 2  \sum_{i=1}^{n}   \sum_{x=0}^{\infty} \sum_{y=0}^{\infty} \Bigg (  \frac{ \mathscr{P}(x ; \hat{\lambda}_1) \mathscr{P}(y;\hat{\lambda}_2 + x \hat{\lambda}_3)}{(x+X_i + a_1+1)(y + Y_i + a_2+1)}
		\Bigg ). 
	\end{eqnarray}
	 We refer to Table \ref{PGFEX2} and Figuare \ref{EX2T} for the quantile values and frequency distribution for  $a_1=1$ and $a_2=1$, respectively. Also, with varying $a_1$  and $a_2$ values  finite sample distribution of the statistics, see Table \ref{EX3FULLT} and Figuares \ref{EX3FULL}, \ref{EX3SI} and \ref{EX3SII}.


\begin{thebibliography}{}
		
		
		
		\bibitem{acb99} 
		Arnold, B.C., Castillo, E., and Sarabia, J.M., Conditional Specification of Statistical Models, Springer Series in Statistics, New York (1999).
		
		
		\bibitem{am21}
		Arnold, B.C., and Manjunath, B.G., (2021), Statistical inference for distributions with one Poisson conditional, Journal of Applied Statistics, 48:12306--2325, DOI: 10.1080/02664763.2021.1928017.
		
		
		\bibitem{amsv22}
		Arnold, B.C., Veeranna, B.,  Manjunath, B.G. and Shobha, B. (2022),  Bayesian inference for Pseudo-Poisson data, J. of statist. comput. and simul. (Accepted)
		
		
		\bibitem{bl00}
		Batsidis, A. and Lemonte, A.J., On goodness-of-fit tests for the Neyman type A distribution, 
		REVSTAT-Statistical Journal, \url{https://www.ine.pt/revstat/pdf/Ongoodness-of-fittestsfortheNeymantypedistribution.pdf}
		
		
		\bibitem{br97}
		Best, D.J. and Rayner, J.C.W. (1997), Crockett's  test fit for the bivariate Poisson, Biometrical Journal, 39(4): 423--430.
		
		\bibitem{cf13}
		Chen, Feiyan, The goodness-of-fit tests for geometric models, (2013). Dissertations. 350.
		\url{https://digitalcommons.njit.edu/dissertations/350}. 
		
			
		
		\bibitem{gd75}
		Gleeson, A.C. and Douglas, J.B. (1975),  Quadrat sampling and the estimation of Neyman Type A and Thomas distributional parameters, Austral. J. Statist., 17:2, 103--113.
		
		\bibitem{gh20}
		G\"urtler, N. and Henze, N., (2000). Recent and classical goodness-of-fit tests for the Poisson distribution, J. of Stat. Planning and Inference,  90, 207--225.
		
		\bibitem{ic17}
		Islam, M.A. and Chowdhury, R.I. Analysis of Repeated Measures Data, Springer Nature, Singapore, (2017).
		
		\bibitem{jkk05}
		Johnson, N.L., Kemp, A.W. and Kotz, S., Univariate Discrete Distributions, John Wiley \& Sons, New Jersey (2005).
		
		\bibitem{kt08}	
		Karlis, D., and Tsiamyrtzis, P. (2008), Exact Bayesian modeling for bivariate Poisson data and extensions. Statistics and Computing, 18:1, 27--40.
		
		\bibitem{kk92}
		Kocherlakota, S. and Kocherlakota, K. Bivariate discrete distributions, Marcel Dekker Inc., New York (1992). 
		
		\bibitem{kp18}
		Kokonendji, C. C. and Puig, P. (2018), Fisher dispersion index for multivariate count distributions: A review and a new proposal, J. Multivar. Anal., 165, 180--193.
		
		\bibitem{lh73}
		Leiter, R.E. and M.A. Hamdan, M.A. (1973), Some bivariate probability models applicable to traffic accidents and fatalities, Int. Stat. Rev., 41, 87--100.
		
		\bibitem{m07}
		Meintanis, S.G. (2007),  A new goodness-of-fit test for certain bivariate distributions applicable to traffic accidents, Statistical Methodology, 4, 22--34.
		
		\bibitem{m16}
		Meintanis, S.G. (2016), A review of testing procedures based on the empirical characteristic function, The South African Stat. Journal, 50, 1--14.
		
		\bibitem{mv20}
		Mijburgh, P.A. and Visagie, H. J. I. (2020), An overview of goodness-of-fit tests for the  Poisson distribution, The South African Stat. Journal, 54:2, 207--230.
		
		\bibitem{n19}
		Nikitin, Y.Y. (2019), Tests based on characterizations and their efficiencies: a survey,
		Acta et Commentationes Universitatis Tartuensis de Mathematica, 21:1, 3--24. 
		
		\bibitem{n21}
		Novoa-Mu\~{n}oz, F. (2021), Goodness-of-fit tests for the bivariate Poisson distribution, Communications in Statistics--Simulation and Computation, 50:7, 1998--2014.
		
		\bibitem{nj14}
		Novoa-Mu\~{n}oz, F. and Jim\'{e}nez-Gamero, M.D. (2014),  Testing for the bivariate Poisson distribution,  Metrika, 77, 771--793.
		
		\bibitem{nj16}
		Novoa-Mu\~{n}oz, F. and Jim\'{e}nez-Gamero, M.D. (2014), A goodness-of-fit test for the multivariate Poisson distribution, SORT, 40:1, 113--138.
		
		\bibitem{uc91}
		Rueda, R., O'Reilly, F., and Pérez-Abreu, V. (1991), Goodness of Fit for the Poisson Distribution Based on the Probability Generating Function, Communications in Statistics - Theory and Methods, 20:10, 3093--3110. 
		
		\bibitem{kdb16} 
		Sellers, K.F., Morris, S.D. and  Balakrishnan, N. (2016), Bivariate Conway-Maxwell-Poisson distribution: Formulation, properties, and inference, J. Multi. Analysis, 150, 152--168. 
		
	\end{thebibliography}
\end{document}